\documentclass[A4]{article}
\usepackage[dvips]{graphicx}
\begin{document}
\newcommand{\Kstarr}{\mathrm{K^{*\pm}}}
\newcommand{\Kstarm}{\mathrm{K^{*-}}}
\newcommand{\Kstarmn}{\mathrm{K^{*-}}}
\newcommand{\Km}{\mathrm{K^{-}}}
\newcommand{\Dsl}{{\mathrm{ D_s}}\ell}
\newcommand{\bt}{\begin{tabular}}
\newcommand{\et}{\end{tabular}}
\newcommand{\Mecqt}{{$\mathrm{MeV/c^2}$}}
\newcommand{\Mecqm}{{\mathrm{MeV/c^2}}}
\newcommand{\Gecqt}{{$\mathrm{GeV/c^2}$}}
\newcommand{\Gecqm}{{\mthrm{GeV/c^2}}}
\newcommand{\DB}{\Delta B}
\newcommand{\as}{\alpha_{ s}}
\newcommand{\ep}{\varepsilon}
\newcommand{\rp}{\tau({\rm B^+})/\tau({\rm B^0_d})}
\newcommand{\rs}{\tau({\rm B^0_s})/\tau({\rm B^0_d})}
\newcommand{\rl}{\tau({\rm \Lambda_b})/\tau({\rm B^0_d})}
\newcommand{\dgs}{\Delta \Gamma_{\Bs}}
\newcommand{\dgbs}{\Delta \Gamma_{\rm \Bs}/\Gamma_{\rm \Bs}}
\newcommand{\tbs}{\tau_{\Bs}}
\newcommand{\tbd}{\tau_{\Bd}}
\newcommand{\Gs}{\Gamma_{{\rm B^0_s}}}
\newcommand{\Gd}{\Gamma_{{\rm B^0_d}}}
\newcommand{\rhobar}{\overline {\rho}}
\newcommand{\etabar}{\overline{\eta}}
\newcommand{\epsilonk}{\left|\varepsilon_K \right|}
\newcommand{\vubovcb}{\left | \frac{V_{ub}}{V_{cb}} \right |}
\newcommand{\vubsvcb}{\left | V_{ub}/V_{cb}  \right |}
\newcommand{\vtdovts}{\left | \frac{V_{td}}{V_{ts}} \right |}
\newcommand{\epsp}{\frac{\varepsilon^{'}}{\varepsilon}}
\newcommand{\dmd}{\Delta m_d}
\newcommand{\dms}{\Delta m_s}
\newcommand{\Do}{{\rm D}^0}
\newcommand{\pimora}{\pi^{\ast}}
\newcommand{\Bstar}{{\rm B}^{\ast}}
\newcommand{\Bstarstar}{{\rm B}^{\ast \ast}}
\newcommand{\Dstar}{{\rm D}^{\ast}}
\newcommand{\Dstars}{{\rm D}^{\ast +}_s}
\newcommand{\Dstaro}{{\rm D}^{\ast 0}}
\newcommand{\Dstarp}{{\rm D}^{\ast +}}
\newcommand{\Dstarstar}{{\rm D}^{\ast \ast}}
\newcommand{\pistar}{\pi^{\ast}}
\newcommand{\pisstar}{\pi^{\ast \ast}}
\newcommand{\bptre}{\rm b^{+}_{3}}
\newcommand{\Vcb}{\left | {\rm V}_{cb} \right |}
\newcommand{\Vub}{\left | {\rm V}_{ub} \right |}
\newcommand{\Vcs}{\left | {\rm V}_{cs} \right |}
\newcommand{\Vtd}{\left | {\rm V}_{td} \right |}
\newcommand{\Vts}{\left | {\rm V}_{ts} \right |}
\newcommand{\bp}{\rm b^{+}_{1}}
\newcommand{\bo}{\rm b^0}
\newcommand{\bos}{\rm b^0_s}
\newcommand{\bss}{\rm b^s_s}
\newcommand{\qq}{\rm q \overline{q}}
\newcommand{\cc}{\rm c \overline{c}}
\newcommand{\BsDmX}{{B_{s}^{0}} \rightarrow D \mu X}
\newcommand{\BsDsm}{{B_{s}^{0}} \rightarrow D_{s} \mu X}
\newcommand{\BsDsX}{{B_{s}^{0}} \rightarrow D_{s} X}
\newcommand{\BDsX}{B \rightarrow D_{s} X}
\newcommand{\BDomX}{B \rightarrow D^{0} \mu X}
\newcommand{\BDpmX}{B \rightarrow D^{+} \mu X}
\newcommand{\Dsfmn}{D_{s} \rightarrow \phi \mu \nu}
\newcommand{\Dsfipi}{D_{s} \rightarrow \phi \pi}
\newcommand{\DsfX}{D_{s} \rightarrow \phi X}
\newcommand{\DpfX}{D^{+} \rightarrow \phi X}
\newcommand{\DofX}{D^{0} \rightarrow \phi X}
\newcommand{\DfX}{D \rightarrow \phi X}
\newcommand{\DsD}{B \rightarrow D_{s} D}
\newcommand{\DsmX}{D_{s} \rightarrow \mu X}
\newcommand{\DmX}{D \rightarrow \mu X}
\newcommand{\Zbb}{Z^{0} \rightarrow \rm b \overline{b}}
\newcommand{\Zcc}{Z^{0} \rightarrow \rm c \overline{c}}
\newcommand{\Rbb}{\frac{\Gamma_{Z^0 \rightarrow \rm b \overline{b}}}
{\Gamma_{Z^0 \rightarrow Hadrons}}}
\newcommand{\Rcc}{\frac{\Gamma_{Z^0 \rightarrow \rm c \overline{c}}}
{\Gamma_{Z^0 \rightarrow Hadrons}}}
\newcommand{\bb}{b \overline{b}}
\newcommand{\str}{\rm s \overline{s}}
\newcommand{\Bs}{\rm{B^0_s}}
\newcommand{\Bsb}{\overline{\rm{B^0_s}}}
\newcommand{\Bp}{\rm{B^{+}}}
\newcommand{\Bm}{\rm{B^{-}}}
\newcommand{\Bo}{\rm{B^{0}}}
\newcommand{\Bob}{\overline{\rm{B^{0}}}}
\newcommand{\Bd}{\rm{B^{0}_{d}}}
\newcommand{\Bdb}{\overline{\rm{B^{0}_{d}}}}
\newcommand{\Lb}{\Lambda^0_b}
\newcommand{\Lbb}{\overline{\Lambda^0_b}}
\newcommand{\Kstar}{\rm{K^{\star 0}}}
\newcommand{\phim}{\rm{\phi}}
\newcommand{\Ds}{\rm{D}_s}
\newcommand{\Dsp}{\rm{D}_s^+}
\newcommand{\Dsm}{\rm{D}_s^-}
\newcommand{\Dp}{\rm{D}^+}
\newcommand{\Dn}{\rm{D}^0}
\newcommand{\Dsb}{\overline{\rm{D}_s}}
\newcommand{\Dm}{\rm{D}^-}
\newcommand{\Dnb}{\overline{\rm{D}^0}}
\newcommand{\Lc}{\Lambda_c}
\newcommand{\Lcb}{\overline{\Lambda_c}}
\newcommand{\Dstarm}{\rm{D}^{\ast -}}
\newcommand{\Dsstarp}{\rm{D}_s^{\ast +}}
\newcommand{\Pb}{P_{b-baryon}}
\newcommand{\KKpi}{\rm{ K K \pi }}
\newcommand{\GeV}{\rm{GeV}}
\newcommand{\MeV}{\rm{MeV}}
\newcommand{\nb}{\rm{nb}}
\newcommand{\Zzero}{{\rm Z}^0}
\newcommand{\MZ}{\rm{M_Z}}
\newcommand{\MW}{\rm{M_W}}
\newcommand{\GF}{\rm{G_F}}
\newcommand{\Gm}{\rm{G_{\mu}}}
\newcommand{\MH}{\rm{M_H}}
\newcommand{\MT}{\rm{m_{top}}}
\newcommand{\GZ}{\Gamma_{\rm Z}}
\newcommand{\Afb}{\rm{A_{FB}}}
\newcommand{\Afbs}{\rm{A_{FB}^{s}}}
\newcommand{\sigmaf}{\sigma_{\rm{F}}}
\newcommand{\sigmab}{\sigma_{\rm{B}}}
\newcommand{\NF}{\rm{N_{F}}}
\newcommand{\NB}{\rm{N_{B}}}
\newcommand{\Nnu}{\rm{N_{\nu}}}
\newcommand{\RZ}{\rm{R_Z}}
\newcommand{\rhob}{\rho_{eff}}
\newcommand{\Gammanz}{\rm{\Gamma_{Z}^{new}}}
\newcommand{\Gammani}{\rm{\Gamma_{inv}^{new}}}
\newcommand{\Gammasz}{\rm{\Gamma_{Z}^{SM}}}
\newcommand{\Gammasi}{\rm{\Gamma_{inv}^{SM}}}
\newcommand{\Gammaxz}{\rm{\Gamma_{Z}^{exp}}}
\newcommand{\Gammaxi}{\rm{\Gamma_{inv}^{exp}}}
\newcommand{\rhoZ}{\rho_{\rm Z}}
\newcommand{\thw}{\theta_{\rm W}}
\newcommand{\swsq}{\sin^2\!\thw}
\newcommand{\swsqmsb}{\sin^2\!\theta_{\rm W}^{\overline{\rm MS}}}
\newcommand{\swsqbar}{\sin^2\!\overline{\theta}_{\rm W}}
\newcommand{\cwsqbar}{\cos^2\!\overline{\theta}_{\rm W}}
\newcommand{\swsqb}{\sin^2\!\theta^{eff}_{\rm W}}
\newcommand{\ee}{{e^+e^-}}
\newcommand{\eeX}{{e^+e^-X}}
\newcommand{\gaga}{{\gamma\gamma}}
\newcommand{\mumu}{\ifmmode {\mu^+\mu^-} \else ${\mu^+\mu^-} $ \fi}
\newcommand{\eeg}{{e^+e^-\gamma}}
\newcommand{\mumug}{{\mu^+\mu^-\gamma}}
\newcommand{\tautau}{{\tau^+\tau^-}}
\newcommand{\qqb}{{q\overline{q}}}
\newcommand{\eegg}{e^+e^-\rightarrow \gamma\gamma}
\newcommand{\eeggg}{e^+e^-\rightarrow \gamma\gamma\gamma}
\newcommand{\eeee}{e^+e^-\rightarrow e^+e^-}
\newcommand{\eeeeee}{e^+e^-\rightarrow e^+e^-e^+e^-}
\newcommand{\eeeeg}{e^+e^-\rightarrow e^+e^-(\gamma)}
\newcommand{\eeeegg}{e^+e^-\rightarrow e^+e^-\gamma\gamma}
\newcommand{\eeeg}{e^+e^-\rightarrow (e^+)e^-\gamma}
\newcommand{\eemumu}{e^+e^-\rightarrow \mu^+\mu^-}
\newcommand{\eetautau}{e^+e^-\rightarrow \tau^+\tau^-}
\newcommand{\eehad}{e^+e^-\rightarrow {\rm hadrons}}
\newcommand{\eettg}{e^+e^-\rightarrow \tau^+\tau^-\gamma}
\newcommand{\eell}{e^+e^-\rightarrow l^+l^-}
\newcommand{\Ztopig}{{\rm Z}^0\rightarrow \pi^0\gamma}
\newcommand{\Ztogg}{{\rm Z}^0\rightarrow \gamma\gamma}
\newcommand{\Ztoee}{{\rm Z}^0\rightarrow e^+e^-}
\newcommand{\Ztoggg}{{\rm Z}^0\rightarrow \gamma\gamma\gamma}
\newcommand{\Ztomumu}{{\rm Z}^0\rightarrow \mu^+\mu^-}
\newcommand{\Ztotautau}{{\rm Z}^0\rightarrow \tau^+\tau^-}
\newcommand{\Ztoll}{{\rm Z}^0\rightarrow l^+l^-}
\newcommand{\Ztocc}{{\rm Z^0\rightarrow c \overline c}}
\newcommand{\Lamp}{\Lambda_{+}}
\newcommand{\Lamm}{\Lambda_{-}}
\newcommand{\Pt}{\rm P_{t}}
\newcommand{\Gee}{\Gamma_{ee}}
\newcommand{\Gpig}{\Gamma_{\pi^0\gamma}}
\newcommand{\Ggg}{\Gamma_{\gamma\gamma}}
\newcommand{\Gggg}{\Gamma_{\gamma\gamma\gamma}}
\newcommand{\Gmumu}{\Gamma_{\mu\mu}}
\newcommand{\Gtautau}{\Gamma_{\tau\tau}}
\newcommand{\Ginv}{\Gamma_{\rm inv}}
\newcommand{\Ghad}{\Gamma_{\rm had}}
\newcommand{\Gnu}{\Gamma_{\nu}}
\newcommand{\GnuSM}{\Gamma_{\nu}^{\rm SM}}
\newcommand{\Gll}{\Gamma_{l^+l^-}}
\newcommand{\Gff}{\Gamma_{f\overline{f}}}
\newcommand{\Gtot}{\Gamma_{\rm tot}}
\newcommand{\Rb}{\mbox{R}_b}
\newcommand{\Rc}{\mbox{R}_c}
\newcommand{\al}{a_l}
\newcommand{\vl}{v_l}
\newcommand{\af}{a_f}
\newcommand{\vf}{v_f}
\newcommand{\ael}{a_e}
\newcommand{\ve}{v_e}
\newcommand{\amu}{a_\mu}
\newcommand{\vmu}{v_\mu}
\newcommand{\atau}{a_\tau}
\newcommand{\vtau}{v_\tau}
\newcommand{\ahatl}{\hat{a}_l}
\newcommand{\vhatl}{\hat{v}_l}
\newcommand{\ahate}{\hat{a}_e}
\newcommand{\vhate}{\hat{v}_e}
\newcommand{\ahatmu}{\hat{a}_\mu}
\newcommand{\vhatmu}{\hat{v}_\mu}
\newcommand{\ahattau}{\hat{a}_\tau}
\newcommand{\vhattau}{\hat{v}_\tau}
\newcommand{\vtildel}{\tilde{\rm v}_l}
\newcommand{\avsq}{\ahatl^2\vhatl^2}
\newcommand{\Ahatl}{\hat{A}_l}
\newcommand{\Vhatl}{\hat{V}_l}
\newcommand{\Afer}{A_f}
\newcommand{\Ael}{A_e}
\newcommand{\Aferb}{\overline{A_f}}
\newcommand{\Aelb}{\overline{A_e}}
\newcommand{\AVsq}{\Ahatl^2\Vhatl^2}
\newcommand{\Iwk}{I_{3l}}
\newcommand{\Qch}{|Q_{l}|}
\newcommand{\roots}{\sqrt{s}}
\newcommand{\pT}{p_{\rm T}}
\newcommand{\mt}{m_t}
\newcommand{\Rechi}{{\rm Re} \left\{ \chi (s) \right\}}
\newcommand{\up}{^}
\newcommand{\abscosthe}{|cos\theta|}
\newcommand{\sint}{\mbox{$\sin\theta$}}
\newcommand{\cost}{\mbox{$\cos\theta$}}
\newcommand{\mcost}{|\cos\theta|}
\newcommand{\epair}{\mbox{$e^{+}e^{-}$}}
\newcommand{\mupair}{\mbox{$\mu^{+}\mu^{-}$}}
\newcommand{\taupair}{\mbox{$\tau^{+}\tau^{-}$}}
\newcommand{\gamgam}{\mbox{$e^{+}e^{-}\rightarrow e^{+}e^{-}\mu^{+}\mu^{-}$}}
\newcommand{\fullskip}{\vskip 16cm}
\newcommand{\halfskip}{\vskip  8cm}
\newcommand{\quarskip}{\vskip  6cm}
\newcommand{\abitskip}{\vskip 0.5cm}
\newcommand{\ba}{\begin{array}}
\newcommand{\ea}{\end{array}}
\newcommand{\bc}{\begin{center}}
\newcommand{\ec}{\end{center}}
\newcommand{\be}{\begin{eqnarray}}
\newcommand{\eeq}{\end{eqnarray}}
\newcommand{\bes}{\begin{eqnarray*}}
\newcommand{\ees}{\end{eqnarray*}}
\newcommand{\Kz}{\ifmmode {\rm K^0_s} \else ${\rm K^0_s} $ \fi}
\newcommand{\Zz}{\ifmmode {\rm Z^0} \else ${\rm Z^0 } $ \fi}
\newcommand{\qqbar}{\ifmmode {\rm q\overline{q}} \else ${\rm q\overline{q}} $ \fi}
\newcommand{\ccbar}{\ifmmode {\rm c\overline{c}} \else ${\rm c\overline{c}} $ \fi}
\newcommand{\bbbar}{\ifmmode {\rm b\overline{b}} \else ${\rm b\overline{b}} $ \fi}
\newcommand{\xxbar}{\ifmmode {\rm x\overline{x}} \else ${\rm x\overline{x}} $ \fi}
\newcommand{\rphi}{\ifmmode {\rm R\phi} \else ${\rm R\phi} $ \fi}
\renewcommand\topfraction{1.}
\newcommand{\BK}{B_K}
\newcommand{\nubar}{\overline{\nu_{\ell}}}
\newcommand{\snb}{\sin{(2\beta)}}
\newcommand{\sna}{\sin{(2\alpha)}}
\newcommand{\vcb}{\left | {\rm V}_{cb} \right |}
\newcommand{\vub}{\left | {\rm V}_{ub} \right |}
\newcommand{\vus}{\left | V_{us} \right |}
\newcommand{\vud}{\left | V_{ud} \right |}
\newcommand{\vtd}{\left | {V}_{td} \right |}
\newcommand{\vts}{\left | { V}_{ts} \right |}
\newcommand{\fbdsqbd}{f_{B_d} \sqrt{\hat B_{B_d}}}
\newcommand{\fbssqbs}{f_{B_s} \sqrt{\hat B_{B_s}}}

\vspace{15mm}

\begin{center}
{\LARGE\bf Current Status of the CKM Matrix and\\
\vspace*{0.2cm}
 the CP Violation}
\vspace*{2cm} 

{\Large \bf Achille Stocchi}

\vspace*{0.5cm}
{\bf\large Laboratoire de l'Acc\'el\'erateur Lin\'eaire,}\\
 IN2P3-CNRS et Universit\'e de Paris-Sud, B\^at.\ 200, BP 34, 
 91898 Orsay Cedex, France\\
\end{center}

\vspace*{4cm}

\begin{abstract}

\noindent
These lectures give an introduction and the current status of flavour physics in the quark sector, 
with special attention to the CKM matrix and CP violation. We describe
the measurements which contribute to the determination of the CKM matrix 
elements and how, together with important theoretical developments, 
they have significantly improved our 
knowledge on the flavour sector of the Standard Model.
These lectures are complemented by the seminar of U. Mallik 
(see these proceedings) which describes in more details the most 
recent CP-violating related measurements by the B-factories. \\
The results presented are up-to-date till winter 2004.

\end{abstract}
\vspace*{3cm}

\noindent
{\it keywords :}\\
{\it CKM matrix, CP violation, Beauty (B) hadrons, B decays, Unitarity Triangle}
\newpage
\tableofcontents
\newpage

\section{Introduction}

Accurate studies of the production and decay properties of beauty 
and charm hadrons are exploiting a unique laboratory for testing 
the Standard Model in the fermion sector, for studying QCD 
in the non-perturbative regime and for searching for New Physics 
through virtual processes. 

In the Standard Model, weak interactions among quarks are encoded
in a 3 $\times$ 3 unitary matrix: the CKM matrix. The existence of 
this matrix conveys the fact that quarks,
in weak interactions, act as linear combinations 
of mass eigenstates \cite{ref:cabi,ref:km}.

The CKM matrix can be parametrized in terms of four free 
parameters which are measured in several physics processes.

In a frequently used parametrization, these parameters are named: $\lambda$, A, 
$\overline{\rho}$ and $\overline{\eta}$\footnote{ $ \overline{\rho} = \rho ( 1-\frac{\lambda^2}{2} ) ~~~;~~~ \overline{\eta} = 
\eta ( 1-\frac{\lambda^2}{2} )$\cite{ref:BLO}.}.
The Standard Model predicts relations between the different processes 
which depend upon these parameters. CP violation is accommodated in the CKM
matrix and its existence is related to $\overline{\eta} \neq 0$.
The unitarity of the CKM matrix can be visualized as a triangle in the
($\overline{\rho},\overline{\eta}$) plane. Several quantities, depending upon $\overline{\rho}$
and $\overline{\eta}$ can be measured and they must define compatible values for
the two parameters, if the Standard Model is the correct description of these
phenomena. Extensions of the Standard Model can provide different predictions
for the position of the apex of the triangle, given by the 
$\overline{\rho}$ and $\overline{\eta}$ coordinates.
The most precise determination of these parameters is obtained using 
B decays, $\Bo-\Bob$ oscillations and CP asymmetry in the B and in the K sectors.

Many additional measurements of B meson properties (mass, branching fractions, lifetimes...) 
are necessary to constrain the Heavy Quark theories [Operator Product Expansion (OPE) /Heavy Quark 
Effective Theory (HQET) /Lattice QCD (LQCD)] to allow for precise extraction of the CKM parameters.  
In addition, to be able to extract the Standard Model parameters, it is also 
necessary to control and measure the backgrounds, and to acquire a detailed understanding of 
the experimental apparatus. 
All these aspects are important because they propagate as 
systematic errors attached to the values extracted for the CKM parameters. 
For instance, the values and the uncertainties of the B hadron lifetimes 
enter in many important quantities.
Experimental progress in various B physics measurements has been crucial in
 the determination of the CKM matrix elements. These last aspects are not treated in these lectures.

\section{Short story: from strangeness to the CKM Matrix}
\label{sec:storia}

The discovery of the strange particle in 1947 was totally unexpected and can be seen as the 
beginning of a new era in particle physics which has not ended yet. Just after the pion
discovery by C.M.G. Lattes, H. Muirhead, C.F. Powell
and G.P. Occhialini \cite{ref:lattes47}, in 1947, the same year, C.C. Butler 
and G.D. Rochester \cite{ref:rochester47} reveal, 
having exposed a cloud chamber to cosmic rays, the existence of a still-heavier
unstable particle 
decaying in a typical V-topology; this earlier name could be ascribed to the characteristic 
topology of the tracks which were produced when a neutral particle decays into two 
charged particles. At the same time there were also events in which
 a charged particle trajectory had a sharp break
indicating a decay (V$^{\pm}$) (corresponding to the decay 
$K^+ \rightarrow \mu^+ \nu_{\mu}$).

In fact, the first example of  such  particles was reported by 
L. Leprince-Ringuet 
and M. Lh\'eritier in 1944 \cite{ref:leprince}. They observed a 
secondary cosmic ray particle, in a cloud chamber placed at the 
Laboratoire de l'Argenti\`ere 
(Hautes-Alpes), producing a recoil electron (energetic delta ray). From the measured curvatures 
of the ongoing and outgoing particles and using the value of the scattering
angle of the electron it was possible to 
determine the mass of the incident particle which was found to be of ~495 $\pm$ 60 MeV/c$^2$. 
It is today clear that this particle is the charged Kaon, nevertheless this discovery came too 
early, since even the pion was not discovered at that time!

It took two years to confirm the result of Rochester and Butler.
These experiments were continued at higher altitude and with high degree 
of precision \cite{ref:thompson}. The results unambiguously established the existence 
of two states:
$ \Lambda \rightarrow p \pi^- ~~{\rm and} ~~ K^0    \rightarrow \pi^+ \pi^- .$
In 1953 it became possible to produce those V-particles in accelerators\footnote{The Brookhaven 3 GeV Cosmotron was the first accelerator delivering
 strange particles, followed by the Berkeley 6 GeV Bevatron.} and it was then
clear that they were produced in reactions mediated by the strong interaction;
 furthermore
those particles were always produced in pairs ($associated~ production$). 
On the
other hand their typical lifetime was of about 10$^{-10}$ s which is 
a typical time scale of the weak interaction\footnote {The scattering cross section of 
events like $\pi^- p \rightarrow K^0 \Lambda$ corresponds to the geometrical cross section
of hadrons ($\simeq$ 10$^{-13}$ cm)$^2$ which indicates that the $\Lambda$ and
the K$^0$ are produced through strong interactions. The natural lifetime 
of the strong interaction can be estimated using the relationship $\tau_{st} = (had. radius)/c \simeq
10^{-23} s.$}. These particles are then ``strange'', as they are produced through the strong
interaction whereas they decay through weak interaction processes. The solution was 
proposed, after several unfruitful tentative, by M. Gell-Mann \cite{ref:gellman53}, to introduce  
a new additive quantum number: the $strangeness$\footnote{The observation of events such as
$\Xi^- \rightarrow \Lambda \pi^-$, the so-called cascade events, and the non-observation 
of events such as $\Xi^- \rightarrow n \pi^-$, closed up the option that the 
$strangeness$ could be a multiplicative quantum number (a kind of ``strange parity'') being +1 for 
strange particles and -1 for non-strange ones. It was in fact the first indication of the 
existence of double strange particles.}.
The strangeness was assigned to be -1 for the $\Lambda$, the 
$\overline{\rm K}^0$ and the K$^-$ ( and +1 for the
corresponding antiparticles ), -2 for the $\Xi^-$ and 0 for all non-strange particles and making the 
hypothesis that this new quantum number is conserved by strong 
and electromagnetic interactions and is not conserved by the
weak  interaction. This allows to ``explain'', a posteriori, why 
strange particles are always produced in pairs (by strong interactions $\Delta$S=0) 
and have a relatively long lifetime (decay through weak interactions $\Delta$S=1).

In the decay, the strangeness changes by one unit and these transitions
 were classified as $\Delta S$ = 1.
An intense experimental activity on strange particles
shown, in the fifties, that the absolute decay rate for these transitions 
was suppressed by a factor of about 20 as compared with 
the corresponding rate measured in $\Delta S$ =0 decays. 

In the early 60's the existence of new constituents of matter was postulated:
they were called quarks. 
They were independently introduced by M. Gell-Mann\cite{ref:gellman64} and G. Zweig \cite{ref:zweig64} 
and they should transform according to the fundamental representation of SU(3). They were supposed 
to have spin 1/2 and to exist in three varieties: the quark $u$ with charge +2/3, the quarks 
$d$ and $s$  with charge -1/3. By analogy with leptons it was suggested that the quarks were 
also organized into doublets and the existence of a new quark of charge 2/3 was proposed 
\cite{ref:bjor}.

In 1963, N. Cabibbo proposed \cite{ref:cabi} a model to account for the suppression
of $\Delta$S=1 transitions.
In this model
the $d$ and $s$ quarks, involved in weak processes, are rotated by a mixing angle 
$\theta_{c}$: the Cabibbo angle. The quarks are organized in a doublet: \\
\begin{equation}
\begin{array}{ccc}
\left( \begin{array}{c} u \\ d_c \end{array} \right)~=~
&
\left(  \begin{array}{c} u \\ d ~cos \theta_c + s ~sin \theta_c \end{array} \right)
\end{array}
\label{eq:dc}
\end{equation}
the small value of sin~$\theta_c$ ($\simeq$ 0.22) is responsible for the suppression of
strange particle decays (the coupling being proportional to $sin^{2}\theta_{c}$). 
In this picture the slight suppression of $ n \rightarrow p~ e^{-} \overline{\nu_{e}}$ 
with respect to the rate of $ \mu^{-} \rightarrow e^{-} \nu_{\mu} \overline {\nu_{e}} $ 
is also explained by the fact that the coupling in the neutron decay is proportional 
to $cos^{2}\theta_{c}$.

In this model, the neutral current coupling can be written\footnote{More formally. The charged currents are described by the operators 
$J_{\mu}^{+(-)}$. The existence of the neutral current is needed to complete the group 
algebra (obtained by commuting the operators $J_{\mu}^{+}$ and $J_{\mu}^{-}$) and 
necessarily contains $\Delta$S=$\pm$ 1 terms.}:
\begin{equation}
u \overline{u} + d \overline{d} ~cos^2 \theta_c + s \overline{s} ~sin^2 \theta_c + (s \overline{d} +
d \overline{s} ) ~cos \theta_c sin \theta_c.
\label{eq:neutral}
\end{equation}

The presence of the $(s \overline{d} + d \overline{s})$ term implies the existence 
of a flavour changing neutral current (FCNC). This was a serious problem for the Cabibbo 
model, since these couplings would produce contributions to $\Delta m_K$ and $K_L \to \mu^+ \mu^-$ 
decays which are larger by several order of magnitude.

In 1970 S. Glashow, J. Iliopoulos and L. Maiani \cite{ref:gim} (GIM) proposed the introduction of a
new quark, named $c$, of charge 2/3 and the introduction of a new doublet of quarks formed by 
the $c$ quark and by a combination of the $s$ and $d$ quarks orthogonal to $d_c$ (eq. \ref{eq:dc}):
\begin{equation}
\begin{array}{ccc}
\left( \begin{array}{c} c \\ s_c \end{array} \right)~=~
&
\left( \begin{array}{c} c \\ s ~cos \theta_c - d ~sin \theta_c  \end{array} \right).
\end{array}
\end{equation}
In this way the $(s \overline{d}+d \overline{s})$ term (in Eq. \ref{eq:neutral}), in the neutral current, is cancelled.

The discovery of the charm quark in the form of $c \overline{c}$ bound states
\cite{ref:jpsi} and the observation of charmed particles decaying into 
strange particles \cite{ref:cleo} (the $c \overline{s}$ transitions which are proportional
to $cos^{2}\theta_c$ dominate over the $c \overline{d}$ transitions which are 
proportional to $sin^{2} \theta_c$) represent a tremendous triumph of
this picture.

It should be reminded that a candidate event for the decay of a charm hadron was 
first observed in 1971, in Japan, in an emulsion detector exposed to cosmic rays \cite{ref:niu}: 
$X^{\pm} \to h^{\pm} \pi^0$. The lifetime of $h^{\pm}$ and its mass were found to be
~10$^{-14}$ sec and 1.8 GeV respectively ! (see \cite{ref:bigicharm} and \cite{ref:niu2} for more details).

The charge current, mediated by the emission of a W boson, can then be written: 
\begin{equation}
( \overline{u}  \overline{c} )  \gamma^{\mu} (1-\gamma_5) V 
\left( \begin{array}{c} d \\ s \end{array} \right)
\end{equation}
where $ \gamma^{\mu} (1-\gamma_5)$ is the $V-A$ current, which accounts also for parity violation, 
$u,d,s,c$ are the mass eigenstates and V is defined as:
\begin{equation}
V =
\left(
\begin{array}{cc}
cos \theta_c   ~~~sin \theta_c  \\
-sin \theta_c  ~~~cos \theta_c 
\end{array}
\right).
\end{equation}
V is the Cabibbo unitary matrix which specifies the quark states which are involved in
weak interactions. 
In 1975 the Mark I group at SPEAR discovered the third charged lepton: the $\tau$ \cite{ref:tau}. 
Two years later the fifth quark, the $b$, was found at FNAL \cite{ref:fnal}. The indirect 
existence for the top quark $t$ from the observation of ${\rm B}_d^0-\overline{{\rm B}_d^0}$ 
oscillations \cite{ref:bdbd} 
suggested the existence of an heavier version of the doublets ($u$,$d$) and ($c$,$s$)\footnote{Another indirect piece of evidence for the existence of top was the measurement of 
the $Z^0$ coupling to a $b \overline{b}$ pairs, which shown that the $b$ quark is a member of 
an doublet with partner that cannot be a $u$ or a $c$ quark.}. 
The $t$ quark was finally discovered in 1995 at Fermilab \cite{ref:top} in $p \overline{p}$ collisions. 

The existence of three quark doublets was already proposed by M. Kobayashi and K. Maskawa in 1973 
\cite{ref:km} as a possible explanation for CP violation. Their proposal is a 
generalization of the Cabibbo rotation and implies that the weak flavour changing transitions
are described by a $3\times 3$ unitary matrix:
\begin{equation}
\begin{array}{ccccccccc}
\left ( \begin{array}{c} u \\ c \\ t \end{array} \right )
&
\rightarrow
&
V
&
\left ( \begin{array}{c} d \\ s \\ b \end{array} \right )
&
,
&
V =
&
\left ( \begin{array}{ccc} 
V_{ud} ~~ V_{us} ~~ V_{ub} \\
V_{cd} ~~ V_{cs} ~~ V_{cb} \\
V_{td} ~~ V_{ts} ~~ V_{tb}
\end{array} \right )
\end{array}.
\end{equation}

This matrix conveys the fact that there is an arbitrary rotation, usually applied to 
the -1/3 charged quarks, which is due to the mismatch between the strong and the 
weak eigenstates.
This matrix can be parame\-trized using three real parameters and
one phase which cannot be removed by redefining the quark field phases. This phase leads 
to the violation of the CP symmetry. In fact since CPT is a good symmetry for all quantum
field theories, the complexity of the Hamiltonian implies that the time reversal invariance
T and thus CP is violated\footnote
{The time is an anti-linear operator: 
$T(\lambda_1|\psi_1>+\lambda_2|\psi_2>)=\lambda_1^*T|\psi_1>+\lambda_2^*T| \psi_2>$.
It can be simply understood, recalling that $\psi(x,t)$ and $\psi^*(x,-t)$ (and
not $\psi(x,-t)$) obey to the same Schrodinger equation.
If the operator T is applied to the Standard Model Lagrangian and thus to the CKM
matrix: $T~V(CKM)|..> =V^*(CKM)~T|...>$. If $V(CKM)$ is complex, $V(CKM) \ne V(CKM)^*$.
In this case the Hamiltonian does not commute with T, thus T is not conserved and, 
since CPT is conserved, CP is violated.}. 
In this picture the Standard Model includes CP violation in a simple way.

\section{The Standard Model in the fermion sector and the CKM matrix}
\label{sec:SMfermions}

The Standard Model is based on the $SU(2)_L \times U(1)_Y$ gauge symmetry,
where the index $L$ stands for left, since only the $left$ handed particles
are implied in the charged weak current.
%\begin{table*}[htb]
%\label{tab:smheli}
%\begin{center}
%\begin{tabular}{|c|c|c|c|c|c|}
% \hline
%           &          &  $I$   &  $I_3$  &   $Q$  &   $Y$   \\
% \hline    
%doublet L  & $u_L$    & 1/2  &    1/2  &  2/3 &   1/3 \\
%           & $d_L$    & 1/2  &   -1/2  & -1/3 &   1/3 \\     
% \hline
%singlet R  & $u_R$    &  0  &      0   &  2/3 &   4/3 \\
%           & $d_R$    &  0  &      0   & -1/3 &  -2/3 \\
%\hline
%\hline
%doublet L  & $\nu_e$  & 1/2  &    1/2  &   0  &   -1  \\
%           & $e_L^-$  & 1/2  &   -1/2  &  -1  &   -1  \\     
% \hline
%singlet R  & $e_R^-$  &  0  &      0   &  -1  &   -2  \\
% \hline
%\end{tabular}
%\caption{\it {Quantum numbers of the leptons and quarks in the Standard
%Model, the same holds for the other families. $I$ and $I_3$ are the weak 
%isospin and its third component, respectively, and $Y$ is the weak hypercharge.}}
%\end{center}
%\end{table*}
In a Lagrangian the mass term of a fermionic field is of the type\footnote{The  Euler-Lagrange equation implies the following 
correspondence between motion equations and Lagrangians:\\
${\cal L} = \delta^{\mu}\psi\delta_{\mu}\psi -1/2m^2\psi^2~ 
     \rightarrow  (\delta_0 \delta^0 + m^2)\psi ~
     \rightarrow  (E^2 = p^2 + m^2)~ (\rm Einstein~ equation) $.\\
${\cal L} = i \psi \gamma_0 \delta^0 \psi - m\overline{\psi} \psi~ 
     \rightarrow   (i \gamma^0 -m)\psi=0~ (\rm Dirac~ equation)$. }
\begin{equation}
m  \overline{\psi}\psi = m (\overline{\psi}_R \psi_L + \overline{\psi}_L \psi_R).
\label{eq:mass}
\end{equation}

Thus the mass implies a $left-right$ coupling\footnote{It simply follows from the properties of 
projection operators and using the equalities $\overline{\psi}_R=\overline{\psi}P_L$
and $\overline{\psi}_L=\overline{\psi}P_R$. It follows that: \\
$ m  \overline{\psi}\psi = 
  m \overline{\psi} (P_L+ P_R)\psi  =
  m \overline{\psi} (P_L P_L+ P_R P_R)\psi  =
  m [(\overline{\psi}P_L)(P_L \psi) + (\overline{\psi}P_R)(P_R \psi)] =  
  m (\overline{\psi}_R \psi_L + \overline{\psi}_L \psi_R)
$} which is not gauge invariant.
%$\psi_L$ ($\psi_R$) a doublet (singlet) under $SU(2)_L$ transformation.\\
An economical approach for introducing fermion masses, in a gauge-invariant way, is 
to consider Yukawa couplings in which contributes a weak iso-doublet field $\phi$:
\begin{equation}
 \overline{\psi}_L \phi \psi_R ~;~ \rm {with} \quad~~
\phi =
\left(\begin{array}{c} 
             \phi^+  \\  
             \phi^0  
\end{array}\right)\ ,~~
I_{\phi} = 1/2, ~ Y_{\phi} = 1~,
\label{eq:higgs}
\end{equation}
where the quantum numbers of the new doublet ($\phi$) are exactly those needed 
to restore the gauge invariance of the interaction vertex giving the mass to fermions.

In the Standard Model, the Lagrangian 
corresponding to charged weak interactions can be written as:
\begin{equation}
{\cal {L}}_W = \frac{g}{2} \overline{Q}^{Int.}_{L_{i}} \gamma^{\mu} \sigma^a Q^{Int.}_{L_{i}} W_{\mu}^{a} ~;~
Q^{Int.}_{L_{i}} = \left(\begin{array}{c} u_{L_{i}}    \\  d_{L_{i}}    \end{array}\right),
L^{Int.}_{L_{i}} = \left(\begin{array}{c} \nu_{L_{i}}  \\  \ell_{L_{i}} \end{array}\right),
\label{eq:smweak}
\end{equation}
the index $Int.$ indicates the weak interaction basis, $\sigma^a$ are the Pauli matrices (a=1,2,3), 
$W_{\mu}^a$ are the SU(2)$_L$ gauge bosons and $i$ is the quark index.
It can be noted that: $\overline{Q}^{Int.}_{L_{i}} {Q}^{Int.}_{L_{i}}=
                        \overline{Q}^{Int.}_{L_{i}} {\bf 1}_{ij} {Q}^{Int.}_{L_{j}}$.
The charged weak interactions are family blind (the quantum numbers of the Standard
Model are $I_3$ and $Y$ which do not ``feel'' the family index).

In the interaction basis the Yukawa interaction is:
\begin{equation}
 {\cal {L}}_Y = Y^d_{ij}      \overline{Q}^{Int.}_{L_{i}} \phi d^{Int.}_{R_{j}} +
              Y^u_{ij}      \overline{Q}^{Int.}_{L_{i}} \tilde{\phi} u^{Int.}_{R_{j}} +
              Y^{\ell}_{ij} \overline{L}^{Int.}_{L_{i}} \phi {\ell}^{Int.}_{R_{j}}
 ~+ ~H.C. ~,
\label{eq:yuka}
\end{equation}
where $\tilde{\phi} = i \sigma_2 \phi^{\ast }$.
In the most general case the matrices $Y_{ij}$ are complex.\\
The presence of two independent matrices $Y_{ij}$, for the $up$-type and $down$-type quark, is due 
to the behaviour of the Yukawa coupling itself.

After spontaneous symmetry breaking (SSB):
\begin{equation}
\phi =\frac{1}{\sqrt{2}}
\left(\begin{array}{c} 
             0 \\  
             v  
\end{array}\right) ~,\nonumber
\end{equation}
% $ <\phi^0> = v/\sqrt(2)~;~Re(\phi^0)\rightarrow (v+H^0)/\sqrt(2)$
and the Yukawa interaction can be written:
\begin{equation}
 {\cal {L}}_M =     \overline{d}^{Int.}_{L_{i}}  M^d_{ij}     d^{Int.}_{R_{j}} +
                    \overline{u}^{Int.}_{L_{i}} M^u_{ij}    u^{Int.}_{R_{j}} +
               \overline{\ell}^{Int.}_{L_{i}} M^{\ell}_{ij} {\ell}^{Int.}_{R_{j}}
 ~+ ~H.C.
\label{eq:yuka_masse}
\end{equation}
where $M^u = (v/\sqrt{2}) Y^u$ and $M^d = (v^*/\sqrt{2}) Y^d$.
Physical masses are obtained by finding transformations of the fields
such that the corresponding mass matrices become real and diagonal:
% and we have make a choice of having the mass matrices diagonal
%and real (corresponding to the masses), defining:
\begin{equation}
M^f(diag) = V_L^f M^f (V_R^f)^{\dagger}~.
\label{eq:diago}
\end{equation}
Therefore, the mass eigenstates are 
\begin{eqnarray}
d_{L_{i}} = (V_L^d)_{ij} d^{Int.}_{L_{j}} ~ ; ~ d_{R_{i}} = (V_R^d)_{ij} d^{Int.}_{R_{j}} \\ \nonumber
u_{L_{i}} = (V_L^u)_{ij} u^{Int.}_{L_{j}} ~ ; ~ u_{R_{i}} = (V_R^u)_{ij} u^{Int.}_{R_{j}} \\ \nonumber
{\ell}_{L_{i}}=(V_L^{\ell})_{ij} {\ell}^{Int.}_{L_{j}} ~;~{\ell}_{R_{i}}=(V_R^{\ell})_{ij} 
{\ell}^{Int.}_{R_{j}} \\ \nonumber
{\nu}_{L_{i}} = (V_L^{\nu})_{ij} {\nu}^{Int.}_{L_{j}}
\label{eq:eigenmasse}
\end{eqnarray}
In this basis the Lagrangian for the weak interaction can be written as:
\begin{equation}
{\cal {L}}_W = \frac{g}{2} \overline{u}_{L_{i}} \gamma^{\mu} 
\left [ V_L^u (V_L^d)^{\dagger} \right ] d_{L_{j}} W_{\mu}^{a} + h.c.
\label{eq:smweak_new}
\end{equation}
where 
\begin{equation}
V(CKM) = V_L^u (V_L^d)^{\dagger}~.
\label{eq:ckmsm}
\end{equation}
V(CKM) is the CKM matrix. 
%It is clear that the phenomenon of flavour changing, due to W boson emission, corresponds to 
%the fact that a quark of a given mass can transform into another quark of different mass.\\
The phenomenon of flavour changing can be appreciated in two different ways 
(different basis). If we use the basis in which the mass matrices are diagonal, 
the Lagrangian for the interactions is not anymore family blind. 
The interaction among quarks belonging to different families are possible 
and the couplings are encoded in the CKM matrix. 

If the same procedure is applied in the lepton sector it follows that:
\begin{equation}
V(leptons) = (V_L^{\nu} (V_L^{\ell})^{\dagger}) = (V_L^{\ell} (V_L^{\ell})^{\dagger}) = 1~,
\label{eq:ckmlept}
\end{equation}
since the mass matrix of the neutrinos is arbitrary (the neutrinos are massless in the SM), 
we can always choose $V_L^{\nu}=V_L^{\ell}$.\\
There is freedom in parametrising the CKM matrix:\\
- a permutation between the different generations (it is normally chosen to order the quarks by 
  increasing value of their mass ($u,c,t$ and $d,s,b$), \\
- the presence of phases in the CKM matrix. It is clear that M($diag$) is unchanged if the matrices 
  $V_{L(R)}$ are multiplied by a matrix containing only phases: $\tilde{V}_{L(R)}^f 
  = P^f V_{L(R)}^f$, it follows $V(CKM)=P^uV(CKM)P^{*d}$.

%  The usual choice is 
As long as some of these phases are not observable, one has 
to require that the CKM matrix contains the minimal number of phases, 
all the others being absorbed in the definition of quark wave functions.

The $2 \times 2$ matrix can be used to illustrate  the contribution 
 of these arbitrary phases in the
CKM matrix:
\begin{eqnarray}
V =
\left(\begin{array}{cc} 
             V_{11} & V_{12} \\  
             V_{21} & V_{22}  
\end{array}\right)
\rightarrow
\left(\begin{array}{cc} 
             e^{-i \phi_1} &   0 \\  
               0 &   e^{-i \phi_2}
\end{array}\right)\ 
\left(\begin{array}{cc} 
             V_{11} & V_{12} \\  
             V_{21} & V_{22}  
\end{array}\right)\ 
\left(\begin{array}{cc} 
             e^{+i \chi_1} &   0 \\  
               0 &   e^{+i \chi_2}
\end{array}\right)\  & \\ \nonumber
= \left(\begin{array}{cc} 
             V_{11} e^{-i(\phi_1-\chi_1)}  & V_{12}e^{-i(\phi_1-\chi_2)} \\  
             V_{21} e^{-i(\phi_2-\chi_1)}  & V_{22}e^{-i(\phi_2-\chi_2)}  
\end{array}\right)\ 
\end{eqnarray}
It can be noted that:
\begin{equation}
(\phi_2-\chi_2)=(\phi_2-\chi_1)+(\phi_1-\chi_2)-(\phi_1-\chi_1)
\end{equation}
Among the  four phases, corresponding to the four quark flavours, only three can be chosen 
in an arbitrary way, since one phase difference is obtained as a linear sum of the
other three. In the general case, the number of arbitrary phases is:~~$2n(families)-1$.

The CKM matrix is a rotation matrix and, in a complex plane, can be parametrized in terms
of a given number of angles (real numbers) and phases (complex numbers) as indicated in 
Table \ref{tab:phases}.

\begin{table*}[htb]
\begin{center}
\begin{tabular}{|c|c|c|c|}
 \hline
Family of quarks & num. of Angles  & num. Phases  &     Irreducible Phases  \\ \hline 
     $n$         &    $n(n-1)/2$   &  $n(n+1)/2$  &     $n(n-1)/2-(2n-1)$   \\ 
                 &                 &              &     $=(n-1)(n-2)/2$     \\ \hline
      2          &       1         &     3        &         0               \\
      3          &       3         &     6        &         1               \\
      4          &       6         &     10       &         3               \\
 \hline 
 \end{tabular}
\caption{\it {Numbers of angles and phases parametrising a complex rotation matrix. The last
column gives the number of phases which cannot be reabsorbed into the quark fields. }}
\label{tab:phases}
\end{center}
\end{table*}
It results that a 2$\times$ 2 matrix (the Cabibbo matrix) is parametrized in terms of
one real parameter and contains no phase. The 3$\times$ 3 matrix (CKM) is parametrised in 
terms of three real parameters and one irreducible phase. The presence of this complex number in the
Lagrangian, as explained at the end of Section \ref{sec:storia} (footnote 7), is responsible, 
and it is the only one, of the fact that the CP symmetry is violated in the Standard Model.

\section{The CKM Matrix}
\label{sec:ckmmatrix}
\setcounter{equation}{0}
Many parametrizations of the CKM
matrix have been proposed in the literature.   The most popular are
the standard parametrization \cite{ref:CHAU} recommended by \cite{ref:pdg02} 
and a generalization of the Wolfenstein 
parametrization \cite{ref:WO} as presented in \cite{ref:BLO}.

With
$c_{ij}=\cos\theta_{ij}$ and $s_{ij}=\sin\theta_{ij}$ 
($i,j=1,2,3$), the standard parametrization is
given by:
\begin{equation}\label{2.72}
V_{\rm CKM}=
\left(\begin{array}{ccc}
c_{12}c_{13}&s_{12}c_{13}&s_{13}e^{-i\delta}\\ -s_{12}c_{23}
-c_{12}s_{23}s_{13}e^{i\delta}&c_{12}c_{23}-s_{12}s_{23}s_{13}e^{i\delta}&
s_{23}c_{13}\\ s_{12}s_{23}-c_{12}c_{23}s_{13}e^{i\delta}&-s_{23}c_{12}
-s_{12}c_{23}s_{13}e^{i\delta}&c_{23}c_{13}
\end{array}\right)\ ,
\end{equation}
where $\delta$ is the phase necessary for {\rm CP} violation.
$c_{ij}$ and
$s_{ij}$ can all be chosen to be positive
and  $\delta$ may vary in the
range $0\le\delta\le 2\pi$. However, measurements
of CP violation in $K$ decays force $\delta$ to be in the range
 $0<\delta<\pi$. 
$s_{13}$ and $s_{23}$ are small numbers: ${\cal O}(10^{-3})$ and ${\cal O}(10^{-2})$, 
respectively. Consequently, from phenomenological applications, the four independent 
parameters are taken to be: 
\begin{equation}
s_{12}=| V_{us}|, \quad s_{13}=| V_{ub}|, \quad s_{23}=|
V_{cb}|, \quad \delta.
\label{2.73}
\end{equation}

The first three quantities can be extracted from tree level decays mediated
by the  $s \to u$, $b \to u$ and $b \to c$ transitions, respectively.
The phase $\delta$ can be obtained from CP violating or 
loop processes sensitive to the $V_{td}$ matrix element.

The absolute values of the elements of the CKM matrix show a 
hierarchical pattern with the diagonal elements being close to unity: 
$|V_{us}|$ and $|V_{cd}|$ being of order $0.2$, $|V_{cb}|$ and
$|V_{ts}|$ of order $4\cdot 10^{-2}$ and $|V_{ub}|$ and $|V_{td}|$ of order 
$5\cdot 10^{-3}$.  
The Wolfenstein parametrization is useful to illustrate this structure.
It shows that the matrix is almost diagonal, namely that the coupling between 
quarks of the same family is close to unity, and 
is decreasing as the separation between families increases:

\begin{equation}
\begin{array}{cccc}
V_{CKM} =
&
\left ( \begin{array}{ccc}
1 - \frac{\lambda^{2}}{2} &                \lambda  &                    A \lambda^{3} (\rho - i \eta) \\
   - \lambda              &           1 - \frac{\lambda^{2}}{2}  &          A \lambda^{2}             \\
A \lambda^{3} (1 - \rho -i \eta)  &       -A \lambda^{2}  &                      1
\end{array} \right ) & + O(\lambda^{4}).
\end{array}
\label{eq:eqw}
\end{equation}
The set (\ref{2.73}) is replaced by:
\begin{equation}
\lambda, \quad A, \quad \rho,  \quad \eta ~,
\label{eq:set_wolf}
\end{equation}
known as the Wolfenstein parameters.
%\begin{figure}[hbt!]
%\begin{center}
%\includegraphics[width=12cm]{gerarchia.ps}
%\caption{Pictorial view of the Wolfeinstein hierarchy.}
%\label{fig:gerarchia}
%\end{center}
%\end{figure}
To obtain the exact expression of the CKM parameters in the Wolfenstein parametrization, 
it is convenient to go back to the standard parametrization and to 
make the following change of variables in (\ref{2.72}) \cite{ref:BLO,ref:schubert}:

\begin{equation}\label{2.77} 
s_{12}=\lambda\ ,
\qquad
s_{23}=A \lambda^2\ ,
\qquad
s_{13} e^{-i\delta}=A \lambda^3 (\rho-i \eta).
\end{equation}
%to all orders in $\lambda$.

At order $\lambda^{5}$, the obtained CKM matrix in the extended Wolfenstein parametrization 
is:
{\tiny
\begin{equation}
\begin{array}{cccc}
V_{CKM} =
&
\left ( \begin{array}{ccc}
1 - \frac{\lambda^{2}}{2} - \frac{\lambda^4}{8}  &        \lambda      &  A \lambda^{3} (\rho - i \eta) \\
- \lambda +\frac{A^2 \lambda^5}{2}(1-2 \rho) -i A^2 \lambda^5 \eta   & 
1 - \frac{\lambda^{2}}{2}-\lambda^4(\frac{1}{8}+\frac{A^2}{2}) &
                           A \lambda^{2}       \\
A \lambda^{3} \left [1 - (1-\frac{\lambda^2}{2})(\rho +i \eta)\right ] &
-A \lambda^{2}(1-\frac{\lambda^2}{2})\left [ 1 + \lambda^{2}(\rho +i \eta)\right ] &
                 1-\frac{A^2 \lambda^4}{2}
\end{array} \right )
& 
+ {\cal O}(\lambda^{6}).
\end{array}
\label{eq:eq8}
\end{equation}
}

By definition, the expression for $V_{ub}$ remains unchanged relative 
to the original Wolfenstein parametrization and the
corrections to $V_{us}$ and $V_{cb}$ appear only at ${\cal O}(\lambda^7)$ and
${\cal O}(\lambda^8)$, respectively.
The advantage of this generalization of the Wolfenstein parametrization,
over other generalizations found in the literature, 
is the absence of relevant corrections in $V_{us}$, $V_{cd}$, $V_{ub}$ and 
$V_{cb}$. 
It can be noted that the element $V_{td}$ can be re-expressed as:
$$
V_{td} = A \lambda^{3} (1-\overline{\rho} -i \overline{\eta})
$$
where \cite{ref:BLO}
\begin{equation}\label{2.88d}
\overline{\rho}=\rho (1-\frac{\lambda^2}{2}),
\qquad
\overline{\eta}=\eta (1-\frac{\lambda^2}{2}).
\end{equation}

This elegant change in $V_{td}$, with respect to the original Wolfenstein parametrization, 
allows a simple generalization of the so-called unitarity triangle to higher orders in 
$\lambda$ \cite{ref:BLO} as discussed below.

\subsection{The Unitarity Triangle}
\label{sec:uuuut}
From the unitarity of the CKM matrix ($V V^{\dag} = V^{\dag} V = 1 $), non diagonal elements of the
matrix products corresponding to six equations relating its
elements can be written. In particular, in transitions involving $b$ quarks, the scalar 
product of the third column with the complex conjugate of the first row must vanish:
\begin{equation}
V_{ud}^{\ast} V_{ub}~+~ V_{cd}^{\ast} V_{cb}~+~ V_{td}^{\ast} V_{tb}~=~0
\label{eq:triangle}
\end{equation}

\begin{figure}[h!]
\begin{center}
\includegraphics[width=7cm]{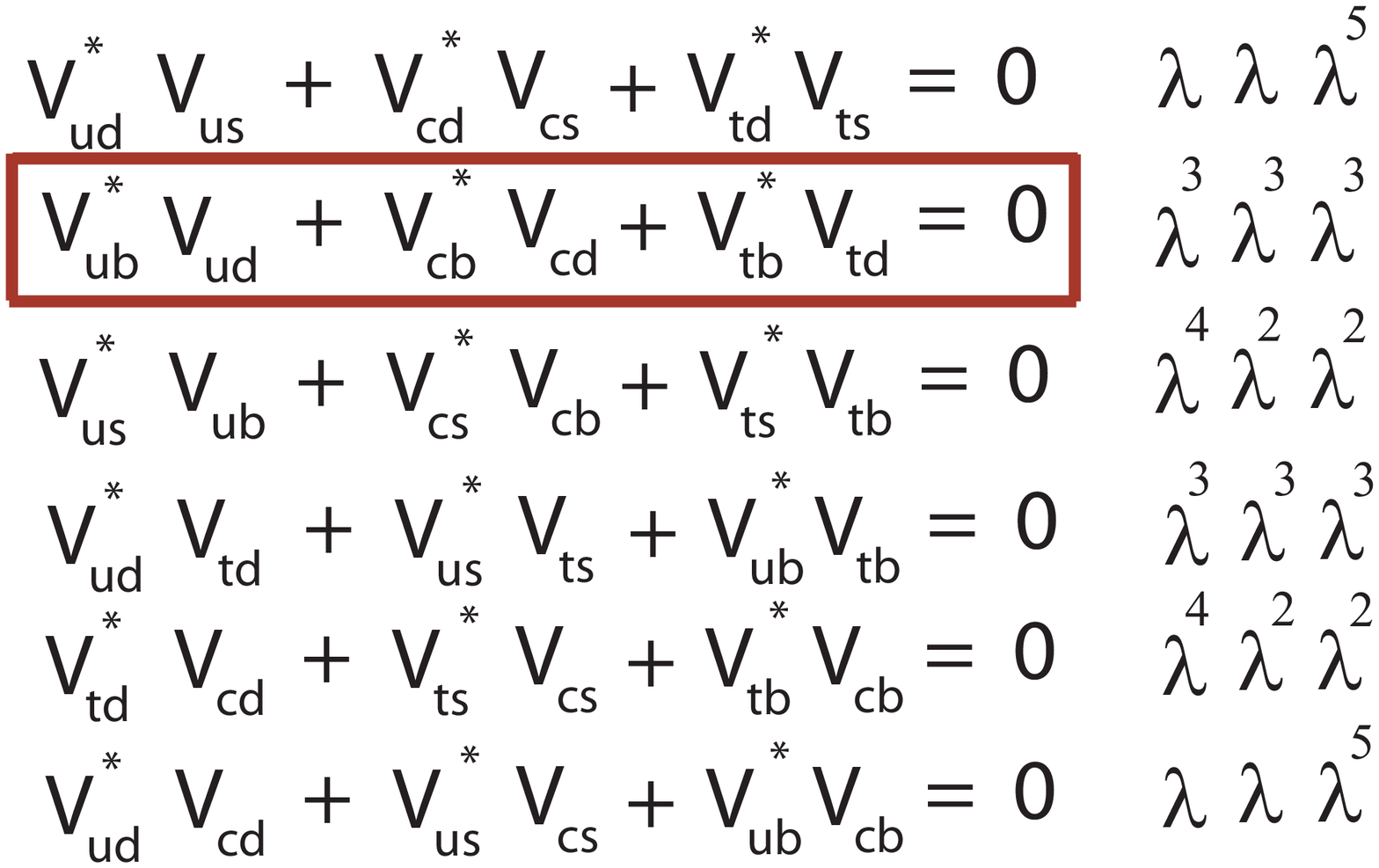}
\caption{The six triangle equations from the unitarity condition of the CKM matrix.}
\label{fig:unitarity}
\end{center}
\end{figure}
Using the parametrization given in Equation (\ref{eq:eq8}), and neglecting contributions 
of order ${\cal O}(\lambda^7)$, the different terms, in this expression,
are respectively:
\begin{eqnarray}
 V_{ud} V_{ub}^{\ast}~=~A \lambda^3(\overline{\rho}+i \overline{\eta}), \\ \nonumber 
 V_{cd} V_{cb}^{\ast}~=~-A \lambda^3,                                   \\ \nonumber 
 V_{td} V_{tb}^{\ast}~=~A \lambda^3(1-\overline{\rho}-i \overline{\eta})
\end{eqnarray}

The three expressions are proportional  to $A \lambda^3$, which can be factored out, and 
the geometrical representation  of Eq. (\ref{eq:triangle}), in the 
($\overline{\rho}$, $\overline{\eta}$) plane, is a triangle with summit at 
C(0, 0), B(1, 0) and A($\overline{\rho}$, $\overline{\eta}$).
%It should be noted that to obtained this triangle in a complex plane we have taken the 
%complex conjugate of the expression (\ref{eq:triangle}).

\begin{itemize}
\item
The lengths $CA$ and $BA$, to be denoted respectively by $R_b$ and $R_t$, are given by
\begin{equation}\label{2.94}
\overline{AC} \equiv R_b \equiv \frac{| V_{ud}^{}V^*_{ub}|}{| V_{cd}^{}V^*_{cb}|}
= \sqrt{\overline\rho^2 +\overline\eta^2}
= (1-\frac{\lambda^2}{2})\frac{1}{\lambda}
\left| \frac{V_{ub}}{V_{cb}} \right|,
\end{equation}
\begin{equation}\label{2.95}
\overline{AB} \equiv R_t \equiv \frac{| V_{td}^{}V^*_{tb}|}{| V_{cd}^{}V^*_{cb}|} =
 \sqrt{(1-\overline\rho)^2 +\overline\eta^2}
=\frac{1}{\lambda} \left| \frac{V_{td}}{V_{cb}} \right|.
\end{equation}
%\item

\begin{figure}[hbt!]
\begin{center}
\includegraphics[width=6cm]{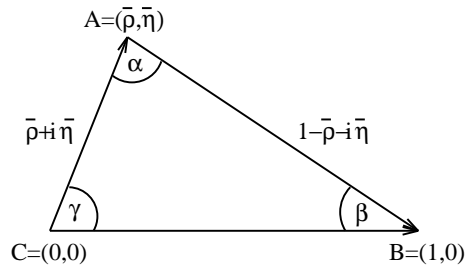}
\caption{The Unitarity Triangle.}
\label{fig:utria}
\end{center}
\end{figure}

The angles $\beta$ and $\gamma=\delta$ of the unitarity triangle are related
directly to the complex phases of the CKM-elements $V_{td}$ and
$V_{ub}$, respectively, through
\begin{equation}\label{e417}
V_{td}=|V_{td}|e^{-i\beta},\qquad V_{ub}=|V_{ub}|e^{-i\gamma}.
\end{equation}

Each of the angles is the relative phase of two adjacent sides (a part for possible
extra $\pi$ and minus sign) so that:
\begin{eqnarray}
\beta  = & arg(\frac{V_{td}V_{tb}^*}{V_{cd}V_{cb}^*}) ~=~ \rm{atan}(\frac{\etabar}{1-\rhobar})  \\
\gamma = & arg(\frac{V_{ud}V_{ub}^*}{V_{cd}V_{cb}^*}) ~=~ \rm{atan}(\frac{\etabar}{\rhobar})    
\label{eq:utangle}
\end{eqnarray}

\item
The unitarity relation (Eq. \ref{eq:triangle}) can be rewritten as
\begin{equation}
\label{RbRt}
R_b e^{i\gamma} +R_t e^{-i\beta}=1
\end{equation}
\item The angle $\alpha$ can be obtained through the relation $\alpha+\beta+\gamma=180^\circ$
      expressing the unitarity of the CKM-matrix.
\end{itemize}

The triangle shown in Figure \ref{fig:utria} 
-which depends on two parameters ($\overline{\rho},~\overline{\eta}$))-, 
plus $|V_{us}|$ and $|V_{cb}|$ give the full description of the CKM matrix. 

The Standard Model, with three families of quarks and leptons, predicts that all measurements 
have to be consistent with the point A($\overline{\rho},~\overline{\eta}$).

\subsection{General Introduction to Oscillation and CP Violation}
\label{sec:generalintro}
In this section we give a general introduction to the oscillation and CP 
violation formalism in view of their impact on the CKM matrix element determination.

A general system which satisfies the coupled Schrodinger equation can be written as:
\begin{eqnarray}
i \frac{d}{dt} \left ( \begin{array}{c} B^0 \\ \overline{B}^0 \end{array} \right ) = 
            H  \left ( \begin{array}{c} B^0 \\ \overline{B}^0 \end{array} \right ) \quad \quad \quad \quad    \\ \nonumber
\quad \quad \quad H = \left ( \begin{array}{cc} H_{11} & H_{12} \\
                              H_{21} & H_{22} \end{array} \right )  ~;~ H_{ij} = M_{ij} -i \Gamma_{ij}/2
\label{eq:general}
\end{eqnarray}

The ${B}^0$ and the $\overline{B}^0$ are the flavour eigenstates. Transitions between $B^0$ and 
$\overline{B}^0$ are then possible and the Hamiltonian has to be diagonalized to find the 
new eigenstates which are:
\begin{eqnarray}
|B^0_L> = p |B^0> + q | \overline{B}^0 > ~~;~~&
|B^0_H> = p |B^0> - q | \overline{B}^0 > 
\label{eq:lightheavy}
\end{eqnarray}
\noindent
where $|q|^2+|p|^2 = 1$. 
Solving the eigenvalues equation (supposing that CPT is conserved) and defining 
$\Delta m=M_H-M_L$ and $\Delta \Gamma =\Gamma_H -\Gamma_L$ it follows:
\begin{eqnarray}
\Delta m^2 - 1/4 \Delta \Gamma^2 = & 4|M_{12}|^2 - |\Gamma_{12}|^2                \\   \nonumber
\Delta m \Delta \Gamma = & 4 Re(M_{12} \Gamma_{12}^*)                             
\label{eq:eigenvalues}
\end{eqnarray}

In the Standard Model,  ${\rm B}^0-\overline{\rm {B}^0}$ transitions occur through a second-order 
process -a box diagram- with a loop that contains W and up-type quarks. The exchange of the top
quark dominates the part relative to the mass difference ($M_{12}$), while light quarks contribute to
the part relative to the decay ($\Gamma_{12}$) (only common states to $B^0$ and $\overline{B}^0$ 
contribute). It results that : $\Gamma_{12}/M_{12} = m_b^2/ m_t^2 <<1$. The relations 
(\ref{eq:eigenvalues}) simplify to:
\begin{eqnarray}
\Delta m       = & 2|M_{12}|                                                         \\ \nonumber 
\Delta \Gamma  = & \frac{2 Re(M_{12} \Gamma{12}^*)}{|M_{12}|}                        \\ \nonumber   
\frac{q}{p}    = & -\frac{|M_{12}|}{M_{12}}                                                       
\label{eq:simply}
\end{eqnarray}

The last expression is valid at leading approximation\footnote{Beyond the leading approximation
the expression become : $\frac{q}{p}= -\frac{\Delta m -i/2 \Delta \Gamma}{2 M_{12} - i\Gamma_{12}}$}.       
We are now interested in the time evolution of the flavour eigenstates in the hypothesis 
($\Delta \Gamma << \Delta m$):
\begin{eqnarray*}
|B^0_{phys.}(t)> = e^{-i m t}~ e^{-\Gamma/2~t}
                  (cos~ \Delta m/2 t ~|B^0>~ +~ i ~\frac{q}{p} ~sin \Delta m ~t/2 ~|\overline{B}^0>)     \\
|\overline{B}^0_{phys.}(t)> = e^{-i m t}~ e^{-\Gamma/2 t}
                  (cos~ \Delta m/2 t ~|\overline{B}^0>~ +~ i~ \frac{q}{p} sin \Delta m t/2 ~|{B}^0>)
\label{eq:time}
\end{eqnarray*}
It follows:
\begin{eqnarray}
 <f|H|B^0_{phys.}(t)>|^2  = \quad \quad \quad \quad \quad
\quad \quad \quad \quad \quad \quad \quad \quad \quad \quad \quad \quad                   \\ \nonumber  
    \frac{e^{-\Gamma t}}{2} [ (1 + cos \Delta m t) |<f|H|B^0(t)>|^2   +    \\ \nonumber
    (1 - cos \Delta m t) |\frac{q}{p}|^2 |<f|H|\overline{B}^0(t)>|^2  +    \\ \nonumber 
        2 \rm{Im}~ (~|\frac{q}{p}|~\rm{sin}~ \Delta m t  ~<f|H|{B}^0(t)><f|H|\overline{B}^0(t)>^*)  ] 
\label{eq:basicCP}
\end{eqnarray}
The probability that a meson $B^0$ produced (by strong interaction) at time $t=0$ transforms (by weak interaction) into a $\overline{B}^0$ (or stays as a $B^0$) at time $t$ is given by:
\begin{eqnarray}
Prob(B^0_{phys.}(t) \rightarrow B^0_{phys.}(t) (\overline{B}^0_{phys.}(t))) 
        = \frac{1}{2} e^{- \Gamma t} ( 1 +(-) cos \Delta m t) 
\label{eq:basicosci}
\end{eqnarray}
Defining:
\begin{equation}
\lambda            =       \frac{q}{p} \frac{<f|H|\overline{B}^0>}{<f|H|{B}^0>} 
                   =       \frac{q}{p} \frac{\overline{A_f}}{A_f}
~~ ; ~~
\overline{\lambda} =      \frac{p}{q} \frac{<\overline{f}|H|{B}^0>}{<\overline{f}|H|\overline{B}^0>} 
                   =      \frac{p}{q} \frac{{A_{\overline{f}}}}{\overline{A}_{\overline{f}}}
\label{eq:deflambda}
\end{equation}
the equation becomes:
\begin{eqnarray}
|<f|H|B^0_{phys.}(t)>|^2 = 
\quad \quad \quad \quad \quad \quad \quad \quad \quad \quad \quad \quad \quad     \\ \nonumber  
                          \frac{e^{-\Gamma t}}{2}|<f|H|B^0(t)>|^2                 \\ \nonumber  
                           [~ (1 + cos \Delta m t)              +                    \\ \nonumber 
                           (1 - cos \Delta m t) |\lambda|^2                       \\  \nonumber
                          - 2 Im(\lambda) sin \Delta m t ~ ]
\label{eq:basicCPlambda}
\end{eqnarray}
and similarly for $|<f|H|\overline{B}^0_{phys.}(t)>|,|<\overline{f}|H|B^0_{phys.}(t)>|$ and 
$|<\overline{f}|H|\overline{B}^0_{phys.}(t)>|$. 
\noindent
We concentrate here on the two cases of CP violation which are the most relevant for CKM physics.\\

\noindent
{\it $B^0$ sector: direct CP Violation and CP Violation in the interference between
mixing and decays.}
\vspace*{0.1cm}

CP violation can occur because $\rm{Im} \lambda \neq \rm{Im} \overline{\lambda}$ and/or when $|\lambda|$,
$|\bar{\lambda}|$ are different from unity.
In this case the four quantities $Prob(B^0_{phys.} \rightarrow f),
Prob(\overline{B}^0_{phys.} \rightarrow f), Prob(B^0_{phys.} \rightarrow \overline{f})$ and 
$Prob(\overline{B}^0_{phys.} \rightarrow \overline{f})$ (see eq. \ref{eq:basicCPlambda}) 
have to be studied and thus $|\lambda|^2, |\overline{\lambda}|^2, \rm{Im} \lambda, 
\rm{Im} \overline{\lambda}$ are determined. 
The simplest case is when the final state $f$ is a specific CP state. 
In this case $\overline{\lambda}=1/\lambda \equiv \lambda_f$ and the previous conditions simplify 
to $\rm{Im} \lambda_f \neq 0$ and/or $|\lambda_f| \neq$ 1.
The following asymmetry can be studied:
\begin{eqnarray}
  A_{CP}(mixing-decay) =& \frac{Prob(B^0_{phys}(\Delta t)\rightarrow {f}) - 
                           Prob(\overline{\rm B}^0_{phys}(\Delta t)\rightarrow {f})}
                          {Prob({\rm B}^0_{phys}(\Delta t)\rightarrow {f}) +
                           Prob(\overline{\rm B}^0_{phys}(\Delta t)\rightarrow {f})}  \\ \nonumber
                    = &C_f~cos \Delta m_d ~\Delta t ~+ S_f ~sin \Delta m_d ~\Delta t ~,  
\label{eq:sin2b1}
\end{eqnarray}
where 
\begin{equation}
C_f = \frac{1-|\lambda_f|^2}{1+|\lambda_f|^2} ~;~ S_f = -\frac{2~ \rm{Im} \lambda_f}{1+|\lambda_f|^2}
\label{eq:sin2b2}
\end{equation}
$C_f$ corresponds to direct CP violation, since it is related to differences in the decay amplitudes,
while $S_f$ is related to the interference between the mixing and decays, involving the imaginary 
parts of $p/q$ and of the decay amplitudes. It is important to note that, also in case $|\lambda_f|$=1,
CP violation is possible if $Im \lambda_f \neq$ 0. This case is particularly interesting. When only 
one amplitude dominates the decay process, $|\lambda_f| =1$, implying $C_f = 0$ and 
$S_f = -\rm{Im} \lambda_f$. We will see in the following that $-\rm{Im} \lambda_f$ is the sine of
twice an angle of the unitarity triangle.\\

\noindent
{\it $B^+$ sector: the direct CP Violation.}
\vspace*{0.1cm}

The transition amplitudes can be written as:
\begin{eqnarray}
|<f|H|{B}^+>| = v_1  A_1 e^{i\theta_1} + v_2 A_2 e^{i\theta_2} \\ \nonumber
|<\overline{f}|H|{B}^->| = v_1^* A_1 e^{i\theta_1} + v_2^* A_2 e^{i\theta_2}
\end{eqnarray}
where $v_{1,2}$ are the weak-CKM couplings and $A_{1,2} (\theta_{1,2})$ are the modulus and the strong 
phase respectively. The weak phase changes sign under CP (the strong phase do not). It follows:
\begin{eqnarray}
a_{CP}(direct) = \frac{|<f|H|{B}^+>|^2 - |<\overline{f}|H|{B}^->|^2}
                 {|<f|H|{B}^+>|^2 + |<\overline{f}|H|{B}^->|^2} \\ \nonumber
               = \frac {2 r_B sin (arg(v_1/v_2)) sin(\theta_1 - \theta_2)}
                       {1 + r_B^2 +2 r_B cos (arg(v_1/v_2)) cos(\theta_1 - \theta_2)} \quad &; \quad r_B = \frac{|v_1| A_1}{|v_2| A_2}
\end{eqnarray}

The basic conditions to have direct CP violation are the presence of two competing amplitudes, $r_B \neq 0$, 
and of nonzero weak phase and strong phase differences.

The weak phase difference is related to one of the unitarity triangle angles. To be more explicit, if
for instance we consider a process which can occur through $V_{ub}$ and $V_{cb}$ mediated transitions, 
$arg(v_1/v_2)=\gamma$.\\

\noindent
{\it CP Violation in the Kaon sector.}
\vspace*{0.1cm}

Historically the parameter characterizing CP violation was defined as $\epsilon$ given in:
\begin{equation}
|K_S> = \frac{|K_1>+\epsilon K_2>}{\sqrt(1+|\epsilon|^2)} ~~;~~
|K_L> = \frac{|K_2>+\epsilon K_1>}{\sqrt(1+|\epsilon|^2)}
\label{eq:klks}
\end{equation}
$K_{1,2}$ are the CP eigenstates and the previous equation can be written in terms of flavour eigenstates:
\begin{eqnarray*}
|K_S> = \frac{1}{\sqrt 2 ~\sqrt(1+|\epsilon|^2)} ~[(1+\epsilon) |K^0> + (1-\epsilon) |\overline{K}^0>] \\
|K_L> = \frac{1}{\sqrt 2 ~\sqrt(1+|\epsilon|^2)} ~[(1+\epsilon) |K^0> - (1-\epsilon) |\overline{K}^0>]
\label{eq:klks2}
\end{eqnarray*}
where $\epsilon$ is related to $p,q$ parameters by: $\epsilon =\frac{p-q}{p+q}$.

Two CP violating quantities are measured in the neutral Kaon sector:
\begin{equation}
\eta_{00}  = \frac{<\pi^0 \pi^0|H|K_L>}{<\pi^0 \pi^0|H|K_S>} ~~;~~
\eta_{\pm} = \frac{<\pi^+ \pi^-|H|K_L>}{<\pi^+ \pi^-|H|K_S>}
\label{eq:eta00pm}
\end{equation}
Defining
\begin{eqnarray*}
A_{00}  = <\pi^0 \pi^0|H|K^0> ~~;~~ \overline{A}_{00}  = <\pi^0 \pi^0|H|\overline{K}^0>  \\
A_{+-}  = <\pi^+ \pi^-|H|K^0> ~~;~~ \overline{A}_{+-}  = <\pi^+ \pi^-|H|\overline{K}^0>  \\
\lambda_{00} = \frac{q}{p} \frac{\overline{A}_{00}}{A_{00}} ~~;~~
\lambda_{+-} = \frac{q}{p} \frac{\overline{A}_{+-}}{A_{+-}} 
\label{eq:defa00pm}
\end{eqnarray*}
which implies
\begin{equation}
\eta_{00} = \frac{1-\lambda_{00}}{1+\lambda_{00}} ~~;~~
\eta_{+-} = \frac{1-\lambda_{+-}}{1+\lambda_{+-}} 
\label{eq:eta00lambda00}
\end{equation}

The $\pi \pi$ final states can have isospin I=0,2. Experimentally it is observed that
$A_{I=2}/A_{I=0} \simeq 1/20$ (known has the $\Delta$ I=1/2 rule). In the approximation
that only the I=0 amplitude contributes -no direct CP violation- it follows:
\begin{equation} 
\epsilon_K = \frac{<\pi^0 \pi^0_{I=0}~ |H|K_L>}{<\pi^0 \pi^0_{I=0}~ |H|K_S>} = \eta_{00}
\label{eq:epsilonkdef}
\end{equation}
and similarly for the $\eta_{+-}$.

Contrarily to B mesons, in case of Kaon physics $\Delta \Gamma \simeq \Delta M$. 
Using the expression of $q/p$ (eq. \ref{eq:eigenvalues}) it follows that:
\begin{equation}
\epsilon_K = \frac{e^{i \pi/4}}{\sqrt 2 \Delta m_K}~(\rm{Im} M_{12} + 2 \zeta \rm{Re} M_{12})
\end{equation}
where $ \zeta = Im(A(K \rightarrow \pi \pi)_{I=0}) / Re(A(K \rightarrow \pi \pi)_{I=0})$.
The contribution, proportional to $\zeta$, which is of about 2$\%$ correction to $\epsilonk$ 
can be neglected. 

\subsection{Standard Model formulae relating \boldmath$\rhobar$ and \boldmath$\etabar$
 to experimental and theoretical inputs}
\label{sec:formulae}

\begin{figure}[h]
\begin{center}
\includegraphics[width=10cm]{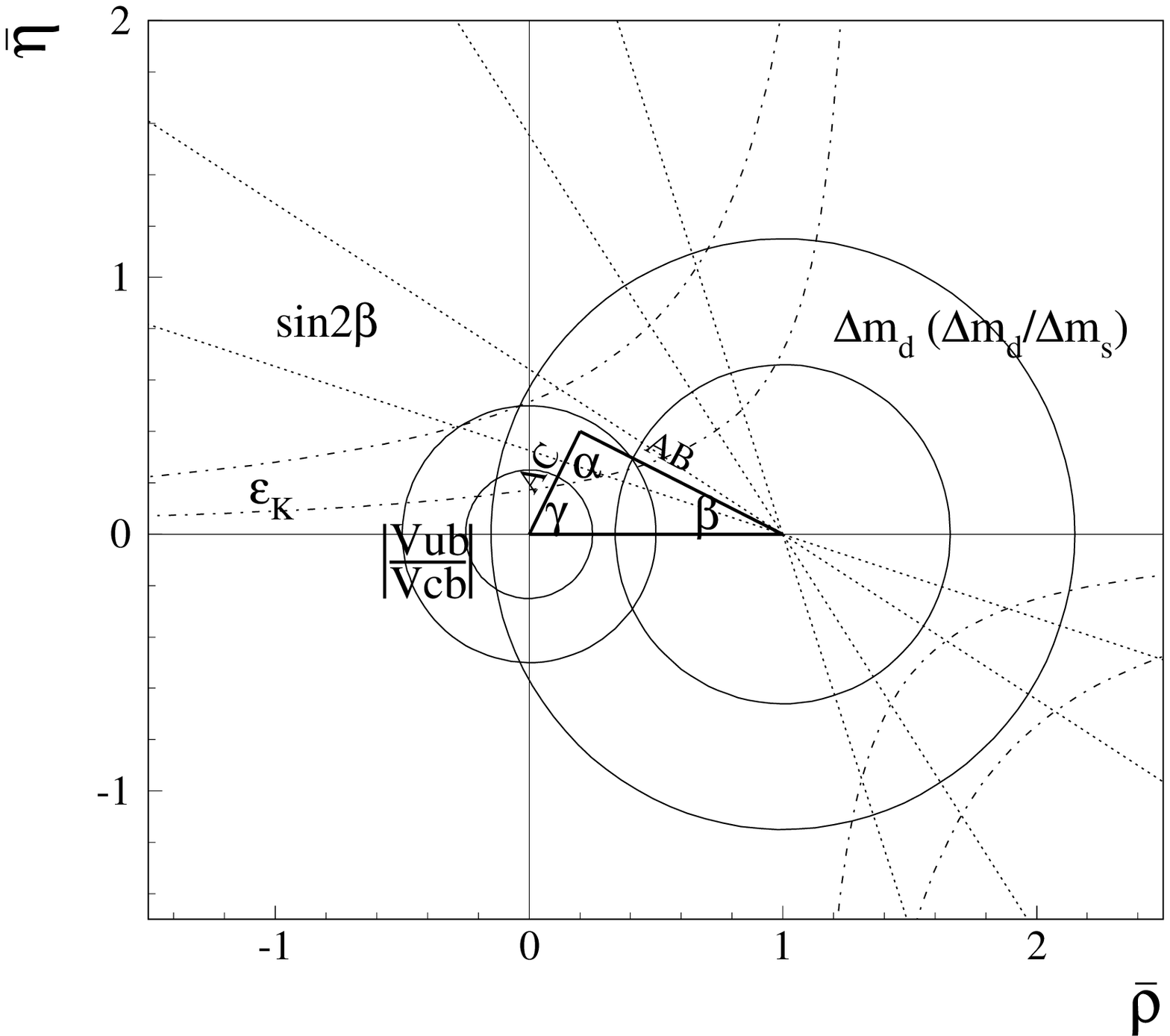}
\caption{Unitarity Triangle. Constraints from
$\left | V_{ub} \right |/\left | V_{cb} \right |$, $\epsilonk$, 
$\Delta m_d$ or $\Delta m_d/\Delta m_s$ and $\snb$ are shown.}
\label{fig:bands}
\end{center}
\end{figure}

Five measurements restrict, at present, the possible range of variation of 
the $\rhobar$ and $\etabar$ parameters:

\begin{itemize}

\item 
B hadrons can decay through the $b \rightarrow c$ and $b \rightarrow u$ transitions.
Semileptonic decays offer a relatively large branching fraction ($\simeq$ 10 $\%$) and 
corresponding measurements
can be interpreted using a well established theoretical framework. 
The relative rate of charmless over 
charmed $b$-hadron semileptonic decays is proportional to the square of the ratio:
\begin{equation}
\vubovcb = \frac{\lambda}{1~-~\frac{\lambda^2}{2}}\sqrt{\rhobar^2+\etabar^2}\ ,
\label{eq:C_vubovcb} 
\end{equation}
and it allows to measure the length of the side AC of the triangle (Figure \ref{fig:bands}).

\item 
 In the Standard Model,  ${\rm B}^0-\overline{\rm {B}^0}$ oscillations occur through a second-order 
process -a box diagram- with a loop that contains W and up-type quarks. The box diagram with 
the exchange of a $top$ quark gives the dominant contribution.
The oscillation probability is given in eq. (\ref{eq:basicosci}) and the time oscillation 
frequency, which can be related to the mass difference between the light and heavy mass 
eigenstates of the ${\rm B}^0_d-\overline{{\rm B}^0_d}$ system
(eq. \ref{eq:simply}, $\Delta m = 2|M_{12}|$ ), is expressed, in the SM, as\footnote{$\Delta {m}_q$ is usually expressed in ps$^{-1}$ unit. 1 ps$^{-1}$ corresponds to 
6.58 10$^{-4}$eV.}:
\begin{equation}
\dmd\ = 
\ {G_F^2\over 6 \pi^2} m_W^2 \ \eta_b S(x_t) \ A^2 \lambda^6 \
[(1-\rhobar)^2+\etabar^2] \ m_{B_d} \ f_{B_d}^2 \hat B_{B_d} \ , 
\label{eq:deltam} 
\end{equation}
where $S(x_t)$ is the Inami-Lim function~\cite{ref:inami} and  $x_t=m_t^2/M^2_W$,
$m_t$ is the $\overline{MS}$ top quark mass, $m_t^{\overline{MS}}(m_t^{\overline{MS}})$, and 
$\eta_b$ is the perturbative QCD short-distance NLO correction.
The value of $\eta_b=0.55 \pm 0.01$ has been obtained in~\cite{ref:bur1} and 
$m_t=(167 \pm 5)\  \GeV$ is used, as deduced from measurements by CDF and D0 Collaborations \cite{ref:top}.
The remaining factor, $f_{B_d}^2 \hat B_{B_d}$, encodes the information 
of non-perturbative QCD. Apart for $\rhobar$ and $\etabar$, the most uncertain parameter in this 
expression is $f_{B_d} \sqrt{\hat B_{B_d}}$ (\ref{sec:parth}).\\
In the vacuum saturation approximation the matrix element of the V-A current
is calculated between the vacuum and the pseudoscalar meson and only the axial current contributes. 
The constant $f_{B_d}$ translates the probability that the quark and the antiquark meet to decay or the size of the B
meson wave function at the origin. Another parameter is also introduced: the bag factor ${\hat B_{B_d}}$ which is
inserted to take into account all possible deviation from vacuum saturation approximation. The values of the bag factors
are expected to be close of the unity.\\
The measurement of $\Delta m_d$ gives a constraint on the length of the side AB of the triangle
(Figure \ref{fig:bands}).
\item
The ${\rm B}^0_s-\overline{\rm {B}^0_s}$ time oscillation frequency, which can be related to the mass 
difference between the light and heavy mass eigenstates of the ${\rm B}^0_s-\overline{{\rm B}^0_s}$ system,
is proportional to the square of the $|V_{ts}|$ element. Neglecting terms,
having a small contribution, $\mid V_{ts} \mid$ is independent of $\rhobar$ and $\etabar$.
The measurement of $\Delta m_s$ would then give a strong constraint on the non-perturbative QCD 
parameter $f_{B_s}^2 \hat B_{B_s}$. In any case, the ratio between the values of the 
mass difference between the mass-eigenstates, measured in the $\Bd$ and in the 
$\Bs$ systems can be used:
\begin{equation}
\frac{\dmd}{\dms}~=~\frac{m_{B_d}f_{B_d}^2 \hat B_{B_d}}{m_{B_s}f_{B_s}^2 \hat B_{B_s} }\
\left( \frac{\lambda}{1-\frac{\lambda^2}{2}} \right )^2 \frac{(1-\rhobar)^2+\etabar^2}{\left ( 1 + \frac{\lambda^2}{1-\frac{\lambda^2}{2}}\rhobar \right )^2
+\lambda^4 \etabar^2}\ .
\label{eq:dms} 
\end{equation}
The advantage in using the ratio $\frac{\dmd}{\dms}$, instead of only $\Delta m_d$, is that the 
ratio $\xi=f_{B_s}\sqrt{\hat B_{B_s}}/f_{B_d}\sqrt{ \hat B_{B_d}}$ is expected to be better 
determined from theory than the individual quantities entering into its expression. 
The measurement of the ratio $\dmd/\dms$ gives a similar type of constraint as $\dmd$, on the length of the 
side AB of the triangle.

\item 
Indirect CP violation in the $\mbox{K}^0-\overline{\rm{K}^0}$ system is usually expressed
in terms of the $\epsilonk$ parameter (as defined in Section \ref{sec:generalintro}) 
which is the fraction of CP violating component in the mass eigenstates. 
In the SM, the following equation is obtained
\begin{eqnarray}
\epsilonk\  =  C_\varepsilon \ A^2 \lambda^6 \ \etabar  \times
\quad \quad \quad \quad \quad \quad \quad \quad \quad \quad \quad \quad \quad \quad \quad \quad \quad \quad
 \\ \nonumber
\left[ -\eta_1 S(x_c) \left ( 1 -\frac{\lambda^2}{2}\right ) 
+ \eta_2 S(x_t) A^2 \lambda ^4 \left(1-\rhobar\right)
+ \eta_3 S(x_c,x_t)  \right] \hat B_K 
\label{eq:epskdef}
\end{eqnarray}
where $C_{\varepsilon} = \frac{G_F^2 f_K^2 m_K m_W^2}{6 \sqrt{2} \pi^2 \Delta m_K}$.\\
$S(x_i)$ and $S(x_i,x_j)$ are the appropriate  
Inami-Lim functions~\cite{ref:inami} depending on $x_q=m_q^2/m_W^2$, including the
next-to-leading order QCD corrections~\cite{ref:bur1,ref:basics}.
The most uncertain parameter is $\hat \BK$ (\ref{sec:parth}). \\
The constraint brought by the measurement of $\epsilonk$ corresponds to
an hyperbola in the $(\rhobar$,~$\etabar)$ plane (Figure \ref{fig:bands}).

\item The measurement of CP violation in the B sector.

The mixing induced CP asymmetry, $a_{J\psi K_S}$, in $\Bd \to J/\psi K_S$ or 
($\to J/\psi K_L$) decays allows to determine the angle $\beta$ of the Unitarity 
Triangle essentially without any hadronic uncertainties.
As explained before, a possible manifestation of the CP asymmetry could appear 
in the interference between amplitudes describing decays with and without mixing. 
The process $B^0 \rightarrow J/\Psi K^0$ is dominated by tree diagram\footnote
{The same process could be described by a Penguin diagram (with a $ts$ transition) and a J/$\Psi$
emitted from gluons. This process is proportional to $V_{ts}V^*_{tb}$. It is important 
to note that the amplitude associated to this process has the same phase, at order 
$\cal{O}$$(\lambda^2)$, as the dominant tree-level one. At order $\cal{O}$$(\lambda^4)$, 
$V_{ts}$ is complex and differs from $V_{cb}$. Thus the correction to $\beta$ 
is suppressed by a factor $\cal{O}$($\lambda^4$) and by an extra factor because the J/$\Psi$ must be 
emitted by at least three gluons.
}
and it follows that:
\begin{eqnarray*}
\frac{q}{p} &=& \left( \frac{V^*_{tb}V_{td}}{V_{tb}V^*_{td}} \right )  ~~~ \rm{from~B~mixing}   \\ 
\frac{<J/\Psi K^0|H|\overline{B^0}>}{<J/\Psi K^0|H|{B^0}>} &=&                       
     \left( \frac{V^*_{cs}V_{cb}}{V_{cs}V^*_{cb}} \right )   ~~~\rm{from~B~decay~amplitudes}   \\ 
     & & \left( \frac{V^*_{cd}V_{cs}}{V_{cd}V^*_{cs}} \right )  ~~~ \rm{from~K~mixing}         \\ 
                |\lambda_f|^2 = 1 ~~~~&;& ~~~~   \rm{Im}~\lambda_f = \eta_{CP} sin2 \beta
\end{eqnarray*} 
where $\eta_{CP}$ is the CP eigenvalue of the final state. The asymmetry defined in eq. \ref{eq:sin2b1} 
gives:
\begin{equation}
  A_{CP}(J/\Psi K_S)  = - 2~ \snb ~sin \Delta m_d ~\Delta t ~, 
\label{eq:sin2bdef}
\end{equation}

The measurement of $  A_{CP}(J/\Psi K^0)$ gives a constraint corresponding to $\snb$, 
in the $(\rhobar$,~$\etabar)$ plane (Figure \ref{fig:bands}).

\end{itemize}

Constraints on $\rhobar$ and $\etabar$ are obtained by comparing present measurements with 
theoretical expectations using the expressions given above 
and taking into account the different sources of uncertainties. In addition
to $\rhobar$ and $\etabar$, these expressions depend on other quantities.
Additional measurements or theoretical determinations have been used to provide
information on the values of these parameters; details are given 
in the next sections.

To illustrate the different constraints described in the present section, 
in Figure~\ref{fig:bands}, 
the uncertainty bands for the quantities, obtained using 
Eqs.~(\ref{eq:C_vubovcb})--(\ref{eq:sin2bdef}), are presented. Each band, 
corresponds to only  one of  the constraints and contains 95\%  
of the events obtained by varying  the input parameters.

In  the first column of Table \ref{tab:rhoeta} the different measured 
quantities are listed, with their explicit dependence on $\rhobar$ and $\etabar$ 
given in the third column. 

\begin{table*}[htb]
\begin{center}
\begin{tabular}{|c|c|c|}
 \hline
 Measurement & ${CKM}\times$other & Constraint \\
 \hline 
 $Br(b\rightarrow u \ell \overline{\nu})$/ $Br(b \rightarrow c \ell \overline{\nu})$ 
 & {$|V_{ub} / V_{cb}|^2$}  & ${{\overline \rho}^2 + {\overline \eta}^2}$ \\
 {$\Delta m_d$} &  $|V_{td}|^2 {f^2_{B_d}B_{B_d}}f(m_t)$ & 
 ${(1-{\overline \rho})^2 + {\overline \eta}^2}$ \\
 {$\Delta m_d \over \Delta m_s$} &  
 ${\left| V_{td}\over V_{ts} \right|^2}
 {f^2_{B_d}B_{B_d}\over f^2_{B_s}B_{B_s}}$ & 
 ${(1-{\overline \rho})^2 + {\overline \eta}^2}$ \\
 {$\epsilonk$} & $f(A,{\overline \eta},{\overline \rho},{B_K})$ & 
 $\propto {{\overline\eta}(1-{\overline \rho})}$ \\
 {$A_{CP}(J/\Psi K^0$)} & $\snb$ & 
 $ 2{\overline \eta}{\overline \rho}/\left [(1-{\overline \rho})^2+{\overline \eta}^2\right ]$ \\
 \hline
 \end{tabular}
\caption{ \it Different measurements contributing in the determination
of $\rhobar$ and $\etabar$, with their functional dependences.}
\label{tab:rhoeta}
\vspace*{-0.5cm}
\end{center}
\end{table*}
%%%%%%%%%%%%%%%%%%%%%%%%%%%%%%%%%%%%%%%%%%%%%%%%%%%%%%%%%%%

The values and errors of the relevant quantities used in the fit of the CKM parameters are summarized 
in Table \ref{tab:inputs}.

For the extraction of the CKM parameters we use the Bayesian approach \cite{ref:ciuchinietal}.
In these lectures we do not enter into any details related to the statistical method. Here we want just to
explain the splitting of the errors as given in Table \ref{tab:inputs}.
We take a Gaussian distribution (${\cal{G}}(x-x_{0})$)
when the uncertainty is dominated by statistical effects, or when there are several
contributions of similar importance to systematic errors, so that the 
central limit theorem applies. We take a uniform p.d.f. if the parameter value is believed 
to be (almost) certainly within a given interval, and the points inside this interval are 
considered as equally probable. The second model is used for theoretical uncertainties.
${\cal{U}}(x) = 1/2 \sigma_{\rm theo}$ for $x \in [x_{0}-\sigma_{\rm theo},x_{0}+\sigma_{\rm theo}]$
and ${\cal{U}}(x) = 0$ elsewhere. The combined p.d.f. $(\cal{P})$ is obtained by convoluting 
the Gaussian p.d.f. $(\cal{G})$ with the uniform p.d.f. $(\cal{U})$: $\cal{P} = \cal{G} \otimes \cal{U}$.
When several determinations of the same quantity are available the final p.d.f, in the Bayesian approach, 
is obtained by taking the product of individual p.d.f.s (and normalizing the obtained distribution 
to unity). For more details on statistical methods see \cite{ref:ckm1}.

\begin{table*}[h]
{%\footnotesize
\begin{center}
\begin{tabular}{@{}lllll}
\hline
\\
         Parameter               &  Value     & Gaussian ($\sigma$)    &   Uniform             &   Ref.   \\
                                 &            &                        & (half-width)          &          \\
\\
\hline
         $\lambda$               &  0.2241    &  0.0036                &    -                   & 
                                                                                \cite{ref:ckm2}   \\
\hline
$\left |V_{cb} \right |$(excl.) & $ 42.1 \times 10^{-3}$  & $2.2 \times 10^{-3}$ & - &
                                                                                Section \ref{sec:vcb}  \\
$\left |V_{cb} \right |$(incl.) & $ 41.4 \times 10^{-3}$  & $0.7 \times 10^{-3}$ & $0.6 \times 10^{-3}$ &
                                                                                Section \ref{sec:vcb} \\
$\left |V_{ub} \right |$(excl.) & $ 33.0  \times 10^{-4}$ & $2.4 \times 10^{-4}$ & $4.6 \times 10^{-4}$ & 
                                                                                Section \ref{sec:vub} \\
$\left |V_{ub} \right |$(incl.) & $ 40.9  \times 10^{-4}$ & $4.6 \times 10^{-4}$ & $3.6 \times 10^{-4}$ & 
                                                                                Section \ref{sec:vub} \\
 \hline
$\Delta m_d$                    & $0.502~\mbox{ps}^{-1}$  & $0.007~\mbox{ps}^{-1}$ & - & 
                                                                                Section \ref{sec:osci} \\
$\Delta m_s$       & $>$ 14.5 ps$^{-1}$ at 95\% C.L.      & \multicolumn{2}{c}{sensitivity 18.3 ps$^{-1}$}    & 
                                                                                Section \ref{sec:osci} \\
$m_t$              & $167~GeV$                            & $ 5~GeV$   & - &          \cite{ref:top}     \\
$f_{B_d} \sqrt{\hat B_{B_d}}$ & $223~MeV$  & $33~MeV$ &  $\pm 12~MeV$  &        Section \ref{sec:parth}   \\
$\xi=\frac{ f_{B_s}\sqrt{\hat B_{B_s}}}{ f_{B_d}\sqrt{\hat B_{B_d}}}$ 
                                  & 1.24   & 0.04 & $\pm 0.06$ &                Section \ref{sec:parth}   \\
$\eta_b$                          & 0.55   & 0.01 &      -     &                     \cite{ref:bur1}       \\
 \hline
$\hat B_K$                        &   0.86 & 0.06 &     0.14   &                Section \ref{sec:parth}   \\
$\epsilonk$   & $2.280 \times 10^{-3}$    & $0.019 \times 10^{-3}$  & -     &        \cite{ref:pdg02}      \\
$\eta_1$      & 1.38                      & 0.53                    & -     &        \cite{ref:basics}     \\
$\eta_2$      & 0.574                     & 0.004                   & -     &        \cite{ref:bur1}       \\
$\eta_3$      & 0.47                      & 0.04                    & -     &        \cite{ref:basics}     \\
$f_K$         & 0.159 GeV                 & \multicolumn{2}{c}{fixed}     &        \cite{ref:pdg02}      \\
$\Delta m_K$  & 0.5301 $\times 10^{-2} ~\mbox{ps}^{-1}$ & \multicolumn{2}{c}{fixed} & \cite{ref:pdg02}   \\
 \hline
        $\snb$             & 0.739   & 0.048 & - &                       Section \ref{sec:sin2beta} \\
\hline
$m_b$         & 4.21 GeV                  & 0.08 GeV                & --     &  Section \ref{sec:vcb}  \\
$m_c$         & 1.3 GeV                   & 0.1 GeV                 & --     &  Section \ref{sec:vcb}  \\
$\alpha_s$    & 0.119                     & 0.03                    & --     &          \cite{ref:basics}  \\
$G_F $        & 1.16639 $\times 10^{-5} \GeV^{-2}$      & \multicolumn{2}{c}{fixed} & \cite{ref:pdg02}  \\
$ m_{W}$      & 80.23 GeV                 & \multicolumn{2}{c}{fixed}      &          \cite{ref:pdg02}  \\
$ m_{B^0_d}$  & 5.2794 GeV                & \multicolumn{2}{c}{fixed}      &          \cite{ref:pdg02}  \\
$ m_{B^0_s}$  & 5.3696 GeV                & \multicolumn{2}{c}{fixed}      &          \cite{ref:pdg02}  \\
$ m_K$        & 0.493677 GeV              & \multicolumn{2}{c}{fixed}      &          \cite{ref:pdg02}  \\
\hline
\end{tabular} 
\end{center}
}
\caption {\it {Values of the relevant quantities used in the fit of the CKM parameters.
In the third and fourth columns the Gaussian and the flat parts of the uncertainty are given (see text), 
respectively. The central values and errors are those adopted at the end of the ``CKM
Unitarity Triangle'' Workshops (\cite{ref:ckm1},\cite{ref:ckm2}) and by HFAG \cite{ref:hfag} and 
are given and explained in the following sections (as indicated in the last column). The averages for
the non perturbative QCD parameters are made by the CKM-LDG group \cite{ref:ldg}}}
\label{tab:inputs} 
\end{table*}

\section{B Physics at different facilities}

In this chapter we will discuss B physics at different machines.
The main contributors in B hadron studies are:
\begin{itemize}
\item the $e^+ e^-$ colliders 
       \begin{itemize}
       \item the symmetric-B factories operating at $\Upsilon(4S)$ (ARGUS/CRYSTAL BALL and CLEO/CUSB
             experiments running at DORIS and CESR, respectively, from 1979 to 2002) 
       \item the asymmetric-B factories operating at $\Upsilon(4S)$ 
             (Belle at KEK and BaBar at PEP experiments running from 1999)
       \item the $\Zz$ resonance experiments (the LEP collaborations which run 
             from 1989 to 1995 and the SLD collaboration at SLC which run from 1989 to 1998). 
       \end{itemize}  
\item the $p \overline{p}$ collider
       \begin{itemize}
       \item the TeVatron collider, operating at $\sqrt s$ = 1.8 TeV - phase I 
             (D0 and CDF experiment from 1987 to 2000). They are presently running with an improved 
             luminosity
             at $\sqrt s \simeq$ 1.9 TeV -phase II. 
       \end{itemize}
\end{itemize}

An overview of these experiments, operating at different facilities, is given in Table \ref{tab:stat}.

At the  $\Upsilon(4S)$,  pairs of ${\rm B}^{\pm}$ and $\Bd \left (\Bdb \right)$ mesons are 
produced  on top of the hadronic background continuum from lighter
$q \overline{q}$ pairs. The two B mesons are created simultaneously in 
a L=1 coherent state, such that before the first decay the final state contains
a ${\rm B}$ and a ${\rm \overline{B}}$; at the time of the decay of the first B meson, 
the second one is in the opposite flavour eigenstate.
The production cross section is about\linebreak 1.2 nb. 
Because of the energy available, only ${\rm B}^{\pm}$ and ${\Bd}$ mesons are emitted. 
In symmetric B-factories B particles are produced almost at rest while at the
asymmetric factories they have a boost of $\beta \gamma$ = 0.56 (0.44) (for BaBar (Belle)).
It is important to note that, in both cases, the average B momentum in the $\Upsilon(4S)$ rest
frame is of the order of about 350 MeV/c.

Considering that the B lifetime is of the order of 1.6 ps, the flight distance of a 
B hadron, defined as $L = \gamma \beta c \tau$ is, on average, at asymmetric B-factories, 
of the order of 250 $\mu$m.
This distance is measurable and highlights the greatest advantage of asymmetric B-factories 
where time dependent analyses, necessary for CP violation studies, are possible.

The B decay products are the only tracks produced in the events, there is no accompanying additional 
hadron. As a consequence the energy taken by each B meson is equal to the half the total energy
in the $e^+ e^-$ center-of-mass frame;
this constraint is, for instance, very important in rejecting the non-B events.
The decay products of the two B particles are spread isotropically over the space
and such events can be distinguished from the continuum which are more jetty-like.

At the $\Zz$ resonance, B hadrons are produced from the coupling of the $\Zz$ to a 
$b \overline{b}$ quark pair. The production cross section is of $\sim$ 6 nb, 
which is five times larger than at the $\Upsilon(4S)$.
Hadronic events account for about  70 $\%$ of the total production rate; among 
these, the fraction of $b \overline{b}$ events is $\sim 22\%$\footnote{whereas the fraction of $c \overline{c}$ events is $\sim 17\%$.},
which is rather similar to the one observed when running at the $\Upsilon(4S)$ energy ($\sim$ 25 \%). 
B hadrons are thus copiously produced\footnote { In the intermediate energy region (``continuum'') where 
the annihilation through one photon is dominant (V-coupling) the cross 
section scales with the energy available in the center of mass (squared), 
being of the order of 30 pb at 30 GeV and of about 10 pb at 60 GeV. 
In this energy range the fraction of 
$b \overline{b}$ events is  $\sim  9\%$ whereas the fraction of 
$c \overline{c}$ events is $\sim 35\%$ (being the coupling proportional to the square of the
electric charge.}. 
The produced $\bb$ pair picks up from the 
vacuum other quark-antiquarks pairs and hadronizes into B hadrons plus 
few other particles.
 Therefore, not only  ${\rm B}^{\pm}$ and $\Bd$ mesons are produced, but also 
 $\Bs$ mesons or $b$-baryons can be present in 
the final state. 
The $b$ and $\overline{b}$ quarks hadronize independently. $b$ quarks 
fragment differently from light quarks, because of their high mass as
compared with $\Lambda_{QCD}$. As a result, B hadrons carry,  on average,  
about 70$\%$ of the available beam energy, whereas the rest of the energy is 
distributed among the other particles emitted in the fragmentation process. 
As a consequence, 
the two B hadrons fly in opposite directions and their decay products 
belong to jets situated in two different hemispheres.

The hard fragmentation and the long lifetime of the $b$ quark make that the 
flight distance of a B hadron at the Z pole, defined 
as $L = \gamma \beta c \tau$, is on average of the order of 3 mm. 

At $p \overline{p}$ colliders, the situation is rather different. 
Here $b$ quarks
are produced mainly through the gluon-gluon fusion process 
$gg \rightarrow b \overline{b}$. At the Fermilab 
Collider ($\sqrt s$ = 1.8 TeV), the differential $b$-production cross section 
depends on the rapidity 
and on the transverse momentum. In total, it is typically 
of the order of 50$\mu b$, which is large. 
B decay
products are situated inside events having a average multiplicity 
which is much larger
than the multiplicity  at the Z pole. Furthermore the ratio $\sigma_{b \overline{b}}/\sigma_{tot}$
is of the order of a few per mill. As a consequence,  only 
specific channels e.g. with fully reconstructed final states, 
or semileptonic decays, can be studied with a reasonable
signal to background ratio.

Registered data sets from experiments operating at different facilities 
are summarized in Table \ref{tab:stat}.

{\tiny
\begin{table}[htb!]
\begin{center}
\begin{tabular}{cccccccc} \hline
 Experiments & Number of $\bb$ events &       Environment            &      Characteristics      \\ 
  & ($\times$ 1000000) &                   &           \\ \hline
  LEP Coll.  & $\sim 1 $ per expt. &       Z$^0$ decays           &  back-to-back 45 GeV b-jets,   \\
             &  (4 expts.)  & ($\sigma_{\bb} \sim~$ 6nb)   &    all B hadron produced.     \\ 
\hline
    SLD      & $\sim 0.1 $           &       Z$^0$ decays           &  back-to-back 45 GeV b-jets,   \\
             &                        & ($\sigma_{\bb} \sim~$ 6nb)   &    all B hadron produced,     \\
             &                        &                              &        beam polarized.         \\  
\hline
   ARGUS     & $\sim 0.2 $           &  $\Upsilon(4S)$ decays       &    mesons produced at rest,    \\
             &                        & ($\sigma_{\bb} \sim~$ 1.2nb) &       $\Bd$ and $\Bp$.       \\
\hline
   CLEO      & $\sim  9  $           &  $\Upsilon(4S)$ decays       &    mesons produced at rest,    \\
             &                        & ($\sigma_{\bb} \sim~$ 1.2nb) &       $\Bd$ and $\Bp$.       \\
\hline
   BaBar     & $\sim  130$           &  $\Upsilon(4S)$ decays       &     asymmetric B-factories         \\
   Belle     &                       & ($\sigma_{\bb} \sim~$ 1.2nb) &        $\Bd$ and $\Bp$.            \\
\hline
    CDF      & $\sim~ \rm{several} $ & $p \overline{p}$ collider-Run I    & events triggered with leptons, \\
             &                        &    $\sqrt s$ = 1.8 TeV       &    all B hadron produced.     \\  
             &                       & (($\sigma_{\bb} \sim~$ 50$\mu$b) &                               \\
\hline
\end{tabular}
\caption{\it { Summary of recorded statistics by experiments operating
               at different facilities and main characteristics.}}
\label{tab:stat}
\end{center}
\end{table}
}
%\begin{figure}[tbp]
%\begin{center}
%\includegraphics[angle=-90,width=13cm]{chap1_aleph_event_bs.eps}
%\caption[]{A LEP $b \overline{b}$ event. In the upper part the ALEPH detector and a 
%zoom on the charged tracks seen by the silicon detectors are displayed. In the lower part 
%the reconstructed event is shown. The event is constituted of two jets
%which define two separate hemispheres. In one of this hemisphere a $\overline{B}_s^0$
%decays semileptonically : $\overline{B}^0_d \rightarrow D_s^+ e^- \nu X$ (secondary
%vertex), followed by the decay : $ D_s^+ \rightarrow K^+ K^- \pi^+$ (tertiary vertex).
%The primary vertex (marked with IP) is also shown.}
%\label{fig:chap1_aleph_event}
%\end{center}
%\end{figure}

\section{Evaluation of the parameters entering in the determination of the CKM parameters.}
This section gives a short summary on the determination
of the quantities entering in Unitarity Triangle fits.
The discussion on the central values and attributed 
errors for these quantities has been extensively done and agreed
values were adopted during the First Workshop on the ``Unitarity Triangle Parameters 
Determination'' held at CERN from the 12-15 February 2002 \cite{ref:ckm1}.
More recent values are taken from the updates done during the Second Workshop on 
the ``Unitarity Triangle Parameters Determination'' held at Durahm from the 5-9 April 2003  
\cite{ref:ckm2}. Many of the experimental averages have been calculated by the 
HFAG (Heavy Flavour Averaging Group) and can be found in \cite{ref:hfag}.

\subsection{Determination of $|V_{cb}|$}
\label{sec:vcb}

The $|V_{cb}|$ element of the CKM matrix can be accessed by studying the decay 
rate of inclusive and exclusive semileptonic $b$-decays.

\subsubsection{Determination of $|V_{cb}|$ using inclusive analyses}
%\label{sec:vcbincl}

The first method to extract $|V_{cb}|$ makes use of B-hadrons inclusive
semileptonic decays and of the theoretical calculations
done in the framework of the OPE (Operator Product Expansion).
The inclusive semileptonic width $\Gamma_{s.l.}$ is expressed as:
\begin{eqnarray} 
 \Gamma_{s.l.} =  \frac{BR(b \rightarrow c l \nu)}{\tau_b}  = 
                  \gamma_{theory} |V_{cb}|^2 ; & \nonumber \\
  \gamma_{theory} = f(\alpha_s,m_b,\mu_{\pi}^2,1/m_b^3...).
\label{eq:vcbtheo}
\end{eqnarray} 

From the experimental point of view the semileptonic width has been measured by the LEP/SLD and 
$\Upsilon(4S)$ experiments with a relative precision of about 2$\%$:
\begin{eqnarray}
   \Gamma_{sl} = & (0.431 \pm 0.008 \pm 0.007) 10^{-10} MeV  &  \small{\Upsilon(4S)}    \nonumber \\
   \Gamma_{sl} = & (0.439 \pm 0.010 \pm 0.007) 10^{-10} MeV  &  \small{\rm{LEP/SLD}}    \nonumber \\
   \Gamma_{sl} = & (0.434 \times (1 \pm 0.018)) 10^{-10} MeV  &  \small{\rm average}
\label{eq:gammsl}
\end{eqnarray}

Using the theoretical determinations of
the parameters entering into the expression of $\gamma_{theory}$ in 
Eq. (\ref{eq:vcbtheo}),
the uncertainty on $|V_{cb}|$ comes out to be of the order of 5$\% (2.0~10^{-3})$. Thus the
precision on the determination of $|V_{cb}|$ is limited by theoretical uncertainties
which are mainly related to the non perturbative QCD parameters. 

%\begin{table}[h]
%\begin{center}
%\begin{tabular}{|l|c c c c|}
%\hline
%Fit & Fit & Fit & Syst. & \\
%Parameter & Values & Uncertainty & Uncertainty & \\ \hline
%$m_b(1~{\rm{GeV}})$ & 4.59 & $\pm$ 0.08 & $\pm$ 0.01 & GeV~\\
%$m_c(1~{\rm{GeV}})$ & 1.13 & $\pm$ 0.13 & $\pm$ 0.03 & GeV~\\
%$\mu_{\pi}^2(1~{\rm{GeV}})$ & 0.31 & $\pm$ 0.07 & $\pm$ 0.02 & GeV$^2$\\
%$\rho_D^3$    & 0.05 & $\pm$ 0.04 & $\pm$ 0.01 & GeV$^3$ \\ \hline
%%$\rho_{LS}^3$ & -0.20 & $\pm$ 0.15 & $\pm$ 0.06 & GeV$^3$ \\  \hline
%\end{tabular}
%\end{center}
%\caption{\it Results of fit for the $m_b(\mu)$, $m_c(\mu)$ and 
%$\mu_{\pi}^2(\mu)$ formalism. The value of $m_b$, in the $\overline{\rm{MS}}$ 
%corresponds to $m_b(m_b) = 4.23 \pm 0.08 \pm 0.01$}
%\label{tab:5}
%\end{table}

These parameters can be experimentally determined
using the fact that OPE gives expressions in terms of operators whose
averaged values are universal when considering different aspects of the same reaction.

Moments of the hadronic mass spectrum, of the lepton energy spectrum and of 
the photon energy in the $b \rightarrow s \gamma$ decay are sensitive to the 
same non
perturbative QCD parameters contained in the factor $\gamma_{theory}$ of 
Eq. (\ref{eq:vcbtheo}) and, in particular, to the mass of the $b$ and $c$ quarks 
and to the Fermi motion of the heavy quark inside the hadron, $\mu_{\pi}^2$\footnote{In another formalism, based on pole quark masses, the $\overline{\Lambda}$ 
and $\lambda_1$ parameters are used, which can be related to the difference between 
hadron and quark masses and to $\mu_{\pi}^2$, respectively.}.
For more details, see for instance \cite{ref:moments},\cite{ref:ligetimom}.

First measurements have been done by CLEO and preliminary results
have been  obtained by BaBar and DELPHI.

\begin{figure}[h]
\centering
\includegraphics[width=120mm]{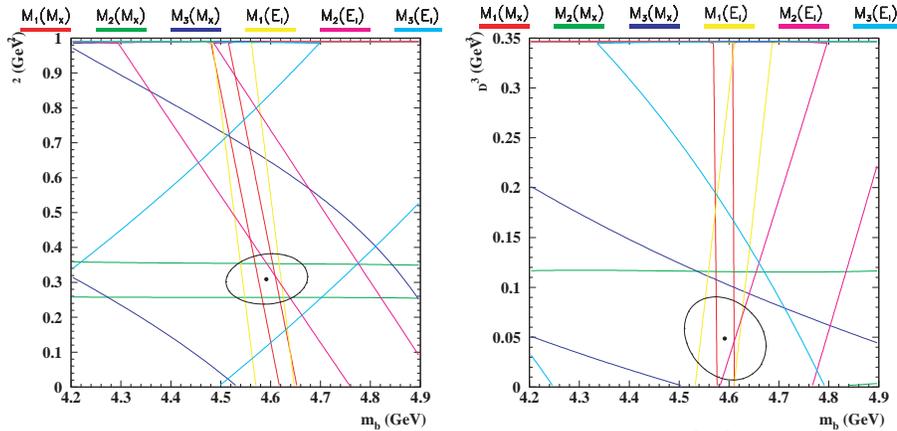}
\vspace*{-0.5cm}
\caption{The moments analysis performed by DELPHI Collaboration \cite{ref:moments}.
The projection of the constraints, brought by six measured moments, over the $m_b-\mu^2_{\pi}$ (left) and $m_b-\rho^3_D$
(right) planes ($\rho_D$ being related to the corrections corresponding to
$1/m_b^3$ terms). The bands correspond to the 
total measurement accuracy and are given by keeping all other parameters fixed at 
their central values. The ellipses represent the 1$\sigma$ contour.}
\label{fig:moments}
\end{figure}

As an example, DELPHI data have been  used for the determination of these non perturbative 
QCD parameters and an illustration of the obtained results is given in Figure \ref{fig:moments}.

%The results are collected in Tables \ref{tab:5}.\\
Using the experimental results on  $\Gamma_{sl}$, Eq. (\ref{eq:gammsl}), and 
on the determination of the non 
perturbative QCD parameters, the following value for $|V_{cb}|$ is obtained:
\begin{equation}
 |V_{cb}| = \left ( 41.4 \pm 0.7 \pm 0.6_{\rm{theo.}} \right ) 10^{-3} \rm{(inclusive)}
\label{eq:vcbinclres}
\end{equation}

This result brings an important improvement in the determination of the $|V_{cb}|$ element. 
The dominant part of the initial theoretical errors is now accounted for as  
experimental uncertainties, using the fitted non perturbative quantities 
($m_b$, $m_c$, $\mu_{\pi}^2$ and $1/m_b^3$ contributions) and the 
remaining theoretical error has been reduced by more than a factor three (previously the quoted 
theoretical error was $\pm\ 2.0~10^{-3}$). 
%In the remaining theoretical error have been included contributions
%from the scale uncertainty at which perturbative QCD corrections have been evaluated, 
%which was varied between
%$2m_b$ and $m_b/2$, and from estimates of higher order (${\cal O}(1/m_b^4)$)
%terms.

\subsubsection{Determination of $|V_{cb}|$ using $\rm {B} \rightarrow \rm{D^* \ell \nu}$ analyses}
%\label{sec:vcbexcl}

An alternative method to determine $|V_{cb}|$ is based on exclusive $\overline{\rm{ B}^0_d} \rightarrow 
{\rm D}^{*+} \ell^- 
\overline{\nu_l}$ decays. Using HQET (Heavy Quark Effective Theory), 
an expression for the differential decay rate can be derived:
\begin{equation}
\frac{d\Gamma}{dw} = \frac{G_F^2}{48 \pi^2} |V_{cb}^2| |F(w)|^2 G(w) ~;~ w = v_B.v_D 
\label{eq:exclvcb}
\end{equation}
%\begin{figure}[htb]
%\includegraphics[width=120mm]{/exp/delphi/stocchi/HABILITATION/CKMSTATUS/vcb_ave.eps}
%\caption{Summary of the measurements of $F(1) \times |V_{cb}|$ (see Paper $num.~ 2$ as listed in 
%Section \ref{sec:published} and \cite{ref:vcbWG}).}
%\label{fig:vcb_ave}
%\end{figure}
%\noindent
$w$ is the 4-product of the B ($v_B$) and the D meson ($v_D$) velocities. G($w$) is a kinematical
factor and F($w$) is the form factor describing the transition. 
At zero recoil ($w$=1) and for infinite quark masses,
F(1) goes to unity. The strategy is then to measure $d\Gamma/dw$, to extrapolate 
at zero recoil and to determine $F(1) \times |V_{cb}|$.

The world average result (as given in PDG 2004) \cite{ref:hfag} is:
\begin{equation}
 |V_{cb}| = ( 42.1 \pm 1.1 \pm 1.9_{F(1)})~ 10^{-3} ~=~  
            ( 42.1 \pm 2.2) ~10^{-3} ~~~\rm{(exclusive)}
\label{eq:vcbexclres}
\end{equation}
To evaluate $|V_{cb}|$, the value of F(1) = 0.91 $\pm$ 0.04 have been 
used \cite{ref:lellouch,ref:latticeF1}.

\subsubsection{Determination of $|V_{cb}|$ using inclusive and exclusive methods}

%\label{sec:vcbcomb}
\noindent
Combining these two determinations of $|V_{cb}|$ gives:
\begin{eqnarray}
 |V_{cb}| = (41.5 \pm 0.8) 10^{-3}   ~~~{\rm{(exclusive+inclusive)}}
\label{eq:vcbave}
\end{eqnarray}

The average has been obtained neglecting possible correlations between the two
methods to determine of $|V_{cb}|$. This assumption is safe from the 
experimental point of view, whereas detailed studies are still missing from 
theory side. It should be noted that the inclusive method is dominating the 
final precision on $|V_{cb}|$.

To conclude, it is important to remind that, as $|V_{cb}| = A \lambda^2$,
the measurement of $|V_{cb}|$
allows the determination of $A$ one of the four free parameters of the 
CKM matrix. Furthermore $\vcb$ gives the scale of the Unitarity 
Triangle.

It is important to note also that $\vcb$, today, is known 
with 2$\%$ 
accuracy. This achievement has to be considered as a legacy 
from LEP and CLEO experiments.

\subsection{Determination of $|V_{ub}|$}
\label{sec:vub}

The measurement of $|V_{ub}|$ is rather difficult because one has to suppress 
the large background coming from the more abundant semileptonic $b$ to $c$ 
quark transitions. 

Several new determinations of the CKM element $|V_{ub}|$ are now available \cite{ref:hfag}

\subsubsection{Determination of $|V_{ub}|$ using inclusive analyses} 
%\label{sec:vubincl}

As for $|V_{cb}|$, the extraction of $|V_{ub}|$ from inclusive semileptonic 
decays is based on HQET implemented through OPE. 

By using kinematical and topological variables, it is possible to 
select samples enriched in $b \rightarrow u \ell^- \overline{\nu_{\ell}}$ transitions. 
There are, schematically, three main regions in the semileptonic 
decay phase space to be considered:
\begin{itemize}
\item  the lepton energy end-point region: 
       $E_{\ell}>\frac{M^2_B-M^2_D}{2M_B}$ (which was at the origin 
       for the first evidence of $b \rightarrow u$ transitions)
\item  the low hadronic mass region: $M_{X} < M_{D}$  
       (pioneered by the DELPHI Coll. \cite{ref:ichep02})
\item  the high $q^2$ region: $M^2_{\ell \nu}=q^2>(M_B-M_D)^2$. 
\end{itemize}
in which the background from $b \rightarrow c \ell^-  \overline{\nu_{\ell}}$ decays
is small.

\begin{figure}[h]
\begin{center}
\includegraphics[width=100mm]{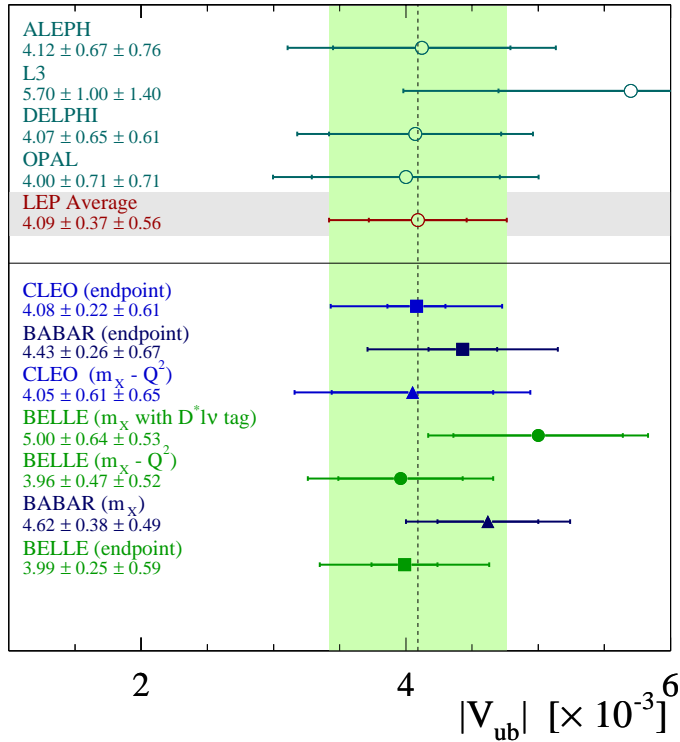}
\caption{Summary of $|V_{ub}|$ inclusive measurements \cite{ref:hfag}. 
For the extraction of the CKM parameters we use the average calculated
in \cite{ref:ckm1}, presented in \cite{ref:ichep02} and given in 
Table \ref{tab:inputs}.}
\label{fig:vubincl}
\end{center}
\end{figure}

A summary of the different determinations of $|V_{ub}|$ is given in 
Figure \ref{fig:vubincl}. 
For the extraction of the CKM parameters we use the average calculated in 
\cite{ref:ckm1}, presented in \cite{ref:ichep02} and given in Table \ref{tab:inputs} :
\begin{eqnarray}
 \vub &=& ( 40.9 \pm 4.6 \pm 3.6) \times 10^{-4}  \quad \mbox{LEP-CLEO}~(inclusive)
\label{eq:vubincl}
 \end{eqnarray}

\subsubsection{Determination of $|V_{ub}|$ using exclusive analyses}
%\label{sec:vubexcl}

%\begin{figure}[htb]
%\includegraphics[width=100mm]{vubexcl.eps}
%\caption{Summary of $|V_{ub}|$ exclusive measurements (\cite{ref:ckm1}.}
%\label{fig:vubexcl}
%\end{figure}

\begin{figure}[h]
\begin{center}
\includegraphics[width=80mm]{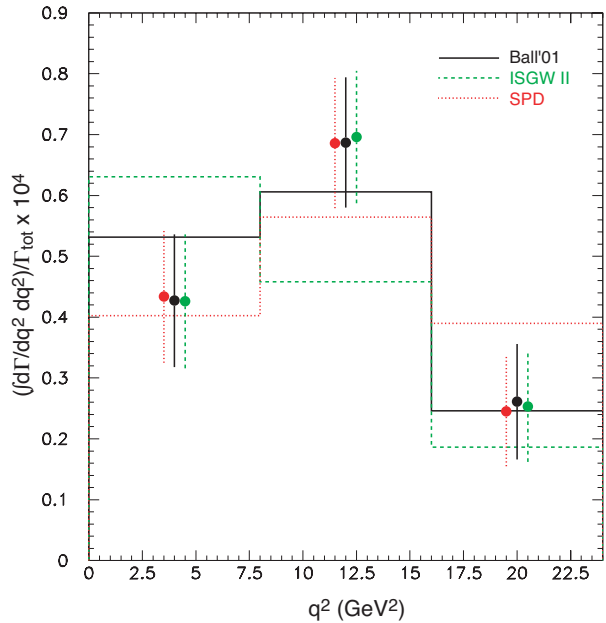}
\caption{Differential branching fraction for 
$B^0 \to \pi^- \ell^+ \nu$  measured as a function of $q^2$, by the CLEO Coll., 
and compared with predicted values (histograms) for three models 
used to extract $|V_{ub}|$.}
\label{fig:cleovub}
\end{center}
\end{figure}

The second method to determine $|V_{ub}|$ consists in the reconstruction 
of charmless semileptonic B decays: ${\rm \overline{B}} \rightarrow \pi (\rho) \ell \overline{\nu}$.

Experimentally, the use of exclusive final states provides extra kinematical constraints for
background suppression. Theoretically, the uncertainties are of a
different nature as those already described in the inclusive analysis. 
%The extraction of $|V_{ub}|$ is complicated by the
%fact that the quarks are not free, but bound inside mesons. 
The probability that the final state quarks  form a given meson is
described by form factors and, to extract $|V_{ub}|$ from actual measurements, 
the main problem rests in the determination of these hadronic form factors.
As there is no heavy quark in the final state, symmetry arguments which
were helpful to determine the form factor in $\overline{B} \rightarrow \Dstar \ell \overline{\nu}$
decays cannot be invoked. 
%which enter in the expressions of the decay rates. 
Light-Cone Sum Rules can provide an evaluation at the 15-20$\%$ accuracy level. 
Lattice QCD calculations give a similar precision 
but these uncertainties are expected to be reduced in the near future. 
The main limitation in lattice calculations is that, at present, they 
can be used only in the high $q^2$ region. 

%is given in Figure \ref{fig:vubexcl}.
A summary of the different determinations of $|V_{ub}|$ can be found in \cite{ref:ckm1}
and \cite{ref:ckm2}.
The combined value of $|V_{ub}|$ is obtained by assuming that systematic uncertainties,
attached to individual measurements, can be 
composed quadratically, for their uncorrelated components, and have
correlated contributions, of
similar size. This correlated part of the systematics arises mainly from the
modelling of the $b\to u $ background. 
%The experimental error of the
%combined value includes this correlated contribution. 
The relative
theoretical error is similar for all measurements and, for the time being, the
error from the BaBar measurement is used. The result is
\begin{eqnarray}
 \vub &=& (33.8 \pm 2.4 ^{+3.7}_{-5.4}) \times 10^{-4}   \\ \nonumber
      &=& (33.0 \pm 2.4 \pm 4.6) \times 10^{-4}
\label{eq:vubexcl}
 \end{eqnarray}

The accuracy on the determination of $\vub$ using exclusive decays is limited
by the theoretical uncertainty on hadronic form factor determination.
%The corresponding
%statistical distribution results from the convolution
%of a Gaussian ($\sigma=2.9  \times 10^{-4}$) and of a flat distribution
%whose variance is given by the second quoted error.\\
%The final p.d.f. for $\vub$ would than be 
%a convolution of a Gaussian and a flat distribution.\\
An interesting analysis has been presented by the CLEO Collaboration 
at ICHEP02 \cite{ref:ichep02}, using the $\Bd \to \pi^- \ell^+ \nu_{\ell}$ decay mode, 
which consists in extracting the signal 
rates in three independent regions of $q^2$. In this way it is possible 
to discriminate between models. The fit shows that the ISGW~II model 
is compatible with data at only  1$\%$ probability level. This approach
could be used, in future, to reduce the importance of theoretical errors, 
considering that the ISGW ~II gave, at present, the further apart $V_{ub}$ determination
\cite{ref:ichep02}.

\subsubsection{Determination of $|V_{ub}|$ using inclusive and exclusive methods}

Combining the two determinations of $\vub$ (\ref{eq:vubincl},\ref{eq:vubexcl}),  
 we obtain, in practice, almost a Gaussian p.d.f. corresponding to:  
\begin{equation}
\vub = (35.7 \pm 3.1) \times 10^{-4}.
\label{eq:vub_corr}
\end{equation} 

New and more precise results from Belle and Babar Collaborations will much improve
the present situation.

\subsection{Measurements of ${\rm B}^0-\overline{{\rm B}^0}$ oscillations}
\label{sec:osci}

\subsubsection{Measurements of the ${\rm B}^0_d-\overline{{\rm B}^0_d}$ oscillation frequency: $\Delta m_d$} 
%\label{sec:oscidmd}

The probability that a $\rm B^0$ meson oscillates into a $\overline{\rm B}^0$ or
remains as a $\rm{B}^0$ is given in Eq. \ref{eq:basicosci}.

The measurement of $\Delta m_d$ has been the subject of an intense experimental activity
during the last ten years. Results are available which correspond to 
the combination of 27 analyses, 
using different event samples, performed by the LEP Coll./SLC/CDF/B-Factories experiments. 

\begin{figure}[htb]
\begin{center}
\includegraphics[width=90mm]{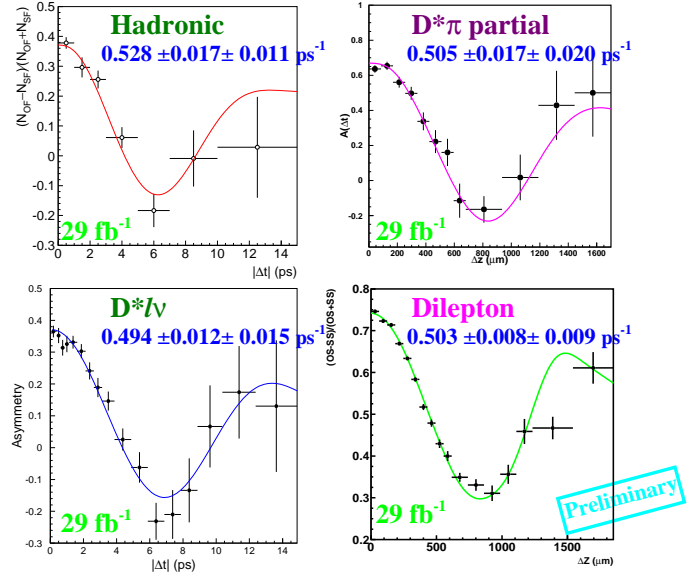}
\caption{The plots show the ${B}^0_d - \overline{{B}}^0_d$ oscillations (Belle Coll.). 
The points with error bars are the data. The result of the fit gives the value 
for $\Delta m_d$.}
\label{fig:dmdbfactory}
\end{center}
\end{figure}

A typical proper time distribution is shown in Figure \ref{fig:dmdbfactory}. The 
oscillating behaviour is clearly visible.

\begin{figure}[h]
\begin{center}
\includegraphics[width=120mm]{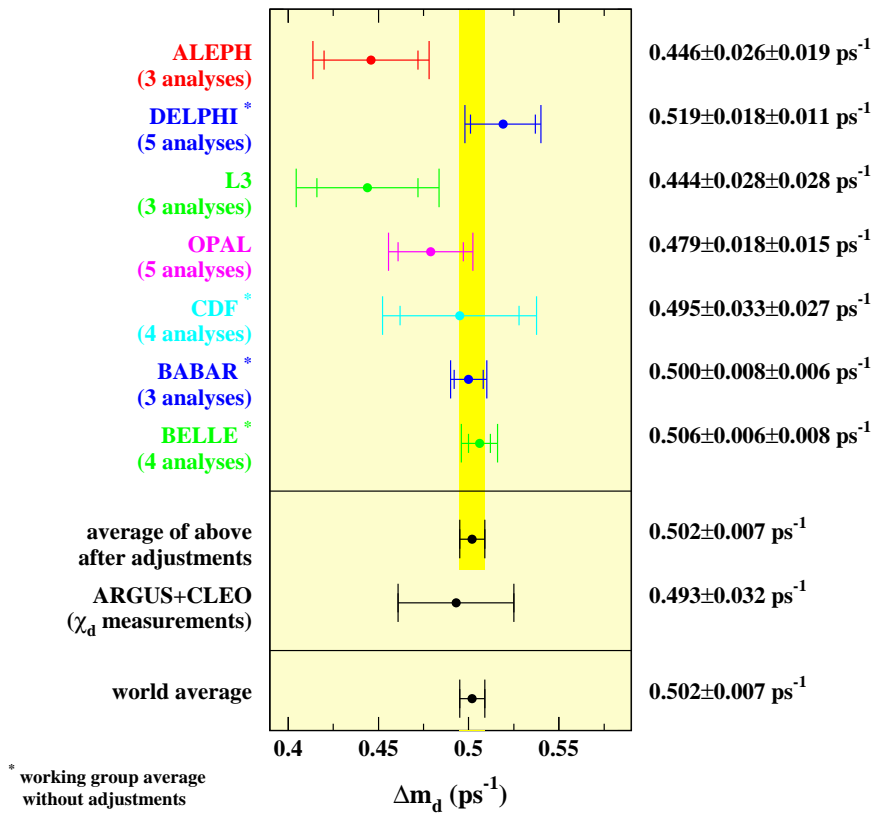}
\caption{Summary  of $\Delta m_d$  measurements \cite{ref:hfag}.}
\label{fig:dmd_ave}
\end{center}
\end{figure}

Figure \ref{fig:dmd_ave} gives the results for $\Delta m_d$, obtained by each experiment and
the overall average \cite{ref:hfag}:
\begin{eqnarray}
\Delta {m}_d &=& (0.502 \pm 0.007)~{ps}^{-1}.
\label{eq:dmdresults}
\end{eqnarray}

The accuracy is of about 1$\%$. The B-factories have the main contribution to this accuracy.
Improvements can still be expected from these facilities and they are expected to reach 
a few per mill precision.

\noindent
\subsubsection{Search for ${\rm B^0_s}-\overline{\rm B^0_s}$ oscillations}
%\label{sec:oscidms}

As the $\Bs$ meson is expected to oscillate more than 20 times faster than the $\Bd$ 
($\dms/\dmd \propto 1/\lambda^2$) and as $\Bs$ mesons are less abundantly produced, the search for 
${\rm B}^0_s-\overline{\rm {B}^0_s}$ oscillations is more difficult. 
The observation of fast oscillations requires the highest resolution on the proper time and 
thus on the $\Bs$ decay length. 
\begin{figure}[h]
\begin{center}
\includegraphics[width=125mm]{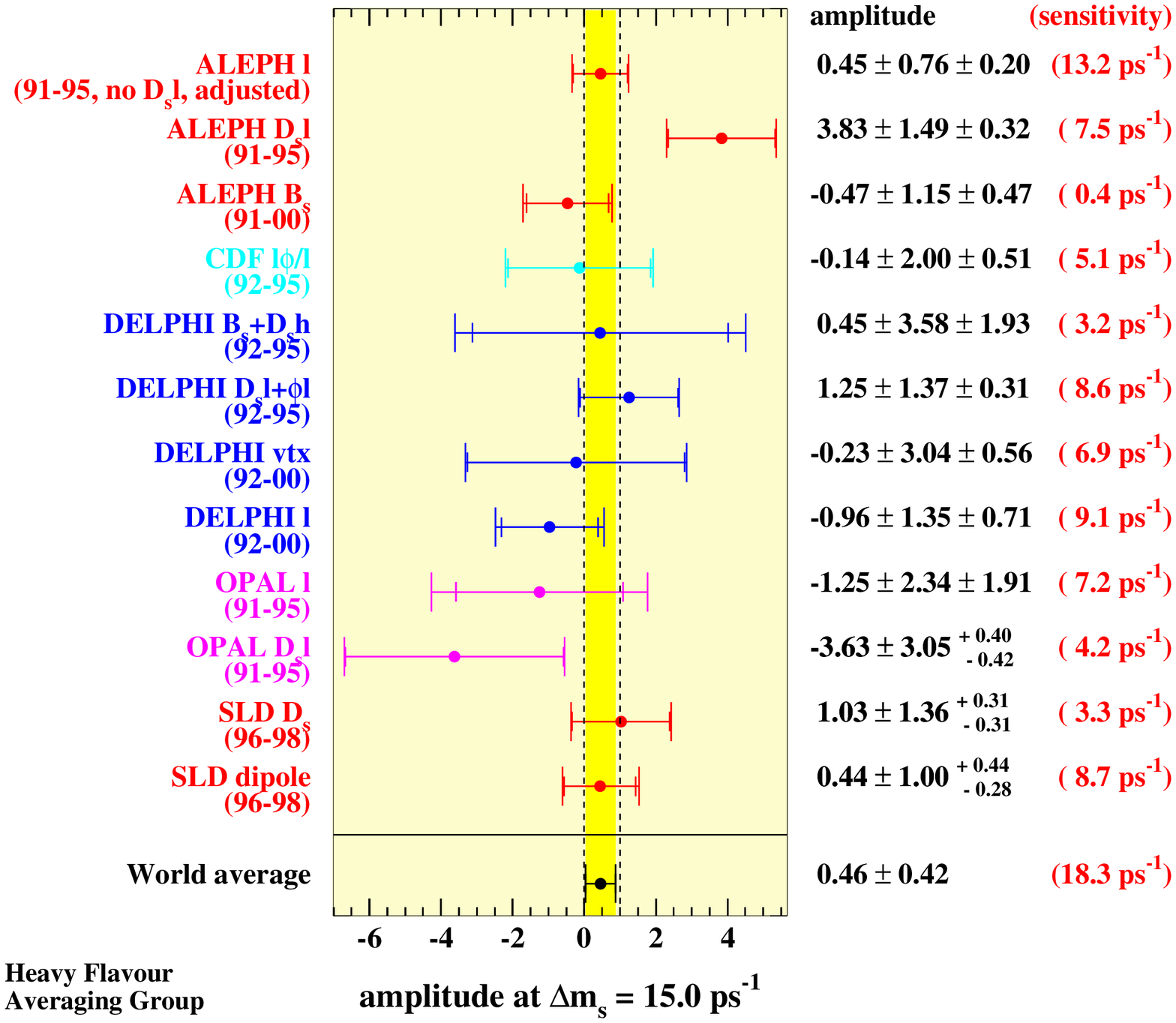}
\caption{${\rm B}_s^0$ oscillation results. Values of the fitted amplitude at 
$\Delta {m}_s = 15~\rm{ps}^{-1}$ and of  the sensitivity obtained by each experiment 
\cite{ref:hfag}.}
\label{fig:dms_ave}
\end{center}
\end{figure}

No signal for ${\rm B}^0_s-\overline{\rm B}^0_s$ oscillations has been observed so far. 

The method used to measure or to put a limit on $\Delta {m}_s$ consists
in modifying Eq. (\ref{eq:basicosci}) in the following way \cite{ref:ampli}:
\begin{equation}
 1 \pm \cos{ (\Delta {m}_s t)} \rightarrow 1 \pm {\cal A} \cos{( \Delta {m}_s t)}. 
\label{eq:amplitude}
\end{equation}
${\cal A}$ and its error, $\sigma_{{\cal A}}$, are measured at fixed values  
of $\Delta {m}_s$, instead of $\Delta {m}_s$ itself. 
In case of a clear oscillation signal, at a given frequency, the amplitude should be 
compatible with ${\cal A} = 1$ at this frequency.
With this method it is easy to set a limit. 
The values of $\Delta
{m}_s$ excluded at 95\% C.L. are those satisfying the condition ${\cal A}(\Delta{m}_s) 
~+~ 1.645  ~\sigma_{{\cal A}(\Delta {m}_s)} < 1$. 

With this method, it is easy also to combine results from different experiments and to
treat systematic uncertainties in the usual way since, for each value of $\Delta {m}_s$, 
a value for $\cal A$ with a Gaussian error $\sigma_{\cal A}$, is measured. 
Furthermore, the sensitivity of a given analysis can be defined as the value of
$\Delta {m}_s$ corresponding to 1.645 $\sigma_{\cal A} (\Delta {m}_s) = 1$ (using 
$\cal A$($\Delta {m}_s) = 0$), namely supposing that the ``true'' value of
$\Delta {m}_s$ is well above the measurable value.

\begin{figure}[h]
\begin{center}
\includegraphics[width=120mm]{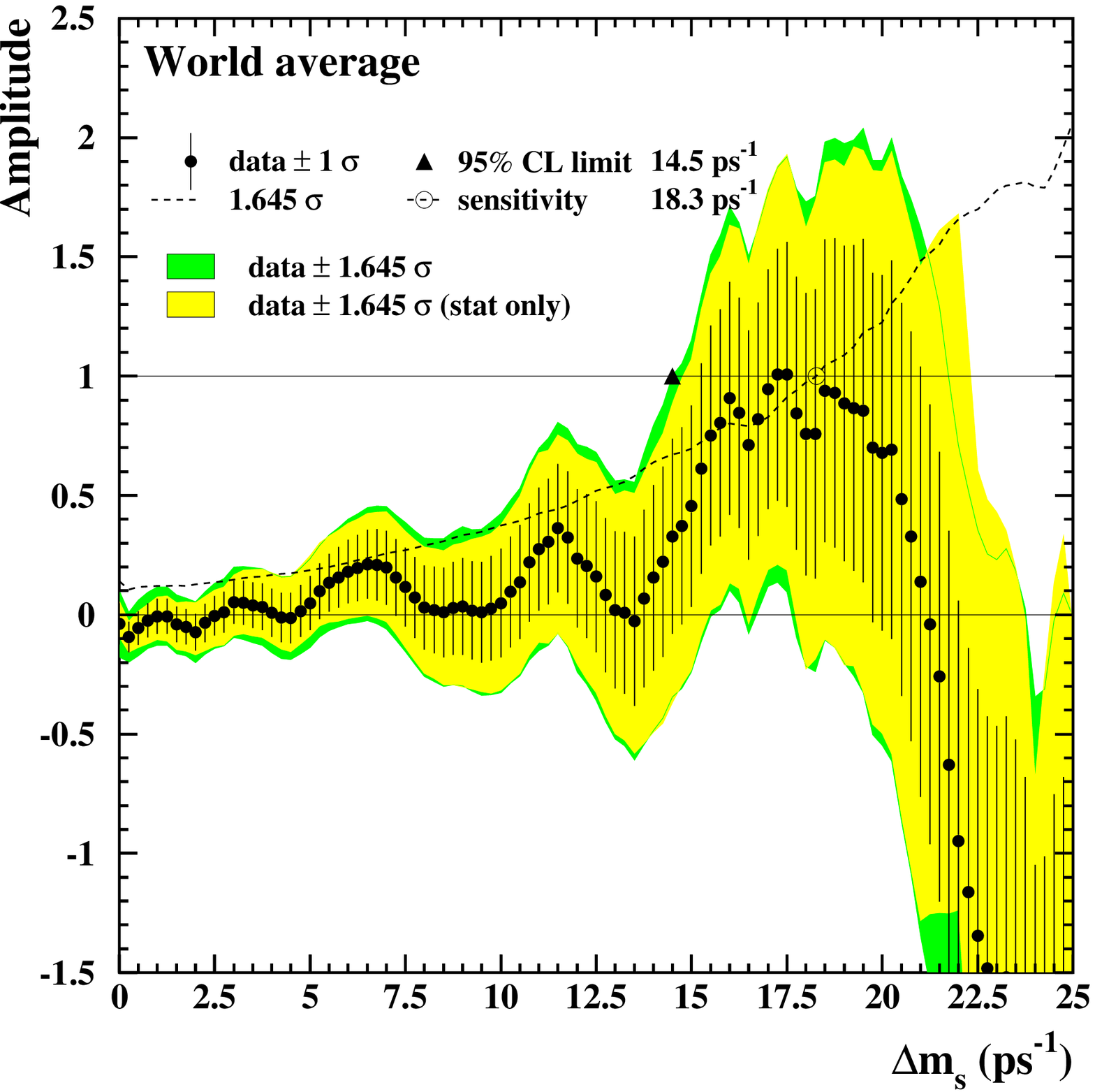}
\caption{The plot \cite{ref:hfag} gives combined $\Delta {m}_s$ results from 
LEP/SLD/CDF analyses shown as an amplitude versus $\Delta {m}_s$ plot. 
The points with error bars are the data; the lines show the 95\% C.L. curves 
(darker regions correspond to the inclusion of systematics). 
The dotted curve corresponds to the sensitivity.}
\label{fig:dms_ampli}
\end{center}
\end{figure}

During last years, impressive improvements in the analysis techniques 
allowed to increase the sensitivity of the search for ${\rm B}^0_s-\overline{\rm B}^0_s$ oscillations. 
Figure \ref{fig:dms_ave} gives details of the different $\Delta m_s$ analyses.
The combined result of LEP/SLD/CDF analyses \cite{ref:hfag} (Figure \ref{fig:dms_ampli}) 
corresponds to:
\begin{eqnarray}
\quad \Delta {m}_s & > 14.5~{ps}^{-1}~~{at}~~95\%~~{C.L.} & \nonumber \\
 ~\rm{with~a ~ sensitivity:} &  \Delta m_s = 18.3 ~{ps}^{-1}.   &
\label{eq:dmsresults}
\end{eqnarray}

The present combined limit implies that $\Bs$ oscillate at least 30 
times faster than $\Bd$ mesons. Taking into account only the $\lambda$
dependence of the ratio $\dmd/\dms$ (eq. \ref{eq:dms}),
this factor would be about 20. The present limit 
gives strong constraints on the $\rhobar$ parameter whose value ends up to be about 0.2.

The significance of the ``bump'' appearing around 17 ps$^{-1}$ is about 2.2 $\sigma$ and
no claim can be made for the observation of ${\rm B}^0_s-\overline{{\rm B}^0_s}$ oscillations.\\
Tevatron experiments are expected to measure soon these oscillations.

\subsection{Some theoretical inputs: \boldmath$\BK$, \boldmath$f_B \sqrt{\hat B_B}$ and  
\boldmath$\xi$}
\label{sec:parth}

Constraints on $\rhobar$ and $\etabar$ depend also upon three parameters which are related to the 
strong interaction operating in the non-perturbative regime: $f_B \sqrt{\hat B_B}$, $\xi$ and $\BK$. 

Expressions for these constraints have been given, respectively,
in Eqs. (\ref{eq:deltam}), (\ref{eq:dms}) and (\ref{eq:epskdef}).
Important improvements have been achieved during the last few years in the evaluation 
of these parameters in the framework of Lattice QCD and a world-wide effort
is organized in view of having precise determinations
of these parameters. As a consequence, in this phenomenological
analysis, only  most recent results from Lattice QCD are used. 

\subsubsection{Brief introduction to Lattice QCD (LQCD) } 
%\label{sec:lqcd}

Lattice QCD (LQCD) was invented about 25 years ago by K. Wilson \cite{ref:Wilson}.

Perturbation theory can be seen as a tool to perform functional integrals
by which all vacuum expectation values of the quantum fields can be expressed. 
LQCD approach consists in a numerical evaluation of the functional integrals. 
It needs a discretization of the four-dimensional space-time by introducing a
basic length, the lattice spacing (often indicated as $a$). 
So LQCD does not introduce new parameters or field variables in the discretization 
and it retains the same properties as QCD. In this sense, it is correct to say, 
that LQCD is not a model, as quark models for example, and therefore physical 
quantities can be computed from first principles without arbitrary assumptions. 
The only input parameters are the strong coupling constant and the six quark 
current masses.
\vspace*{0.1cm}

\noindent
{\it Statistical errors.}\\
Considering N points in each direction, the lattice will have a volume (N a)$^4$ 
( having so two natural cutoffs: a finite space resolution and a finite volume ). 
The standard integrals are sampled over a finite net of points, whereas the 
functional integrals are sampled over a finite set of functions (or configurations).
The vacuum expectation values are obtained by ``averaging'' over all the configurations. 
Those evaluations are done using MonteCarlo techniques. In this spirit, LQCD 
simulations are theoretical experiments carried out by numerical integration of the 
functional integral by MonteCarlo techniques. 
In this respect uncertainties on output quantities are evaluated following criteria 
which are very close to those used in experimental measurements. 
Results are obtained with  ``statistical errors", i.e. uncertainties 
originated by stochastic fluctuations, which may be reduced
by increasing the sample of gluon-field configurations on which averages are performed.  
It is very reasonable to assume that the statistical fluctuations have a Gaussian
distribution. \\
For several quantities statistical errors have been reduced to the percent level (or even less).
However most of the  results are affected by systematic effects.\\
\vspace*{0.1cm}

\noindent
{\it Systematic errors.}\\
Systematic uncertainties come from discretization effects, finite volume effects, 
the treatment of heavy quarks, chiral extrapolation and quenching.
Errors coming from the discretization and from the finite volume can be addressed by 
brute-force improvements of numerical simulations or by improvements in the discretization
procedures. \\
The quenched approximation is obtained by  turning off virtual quark 
loops. An important consequence of this approximation is that the potential between
a quark/antiquark pair depends on this approximation. In the full theory,
at large distance, there is a screened potential between two hadrons because the
string breaks by the creation of a $q \overline{q}$ pair. In quenched LQCD the string
does not couple to such pairs and the long distance behaviour of the two theories is rather different.
This problem is not so important since,
for the long distance scale 
which matters in hadronic physics, and in which we are interested,
there is a ``natural'' cutoff 
of about one Fermi due to confinement. \\
It is reasonable to expect that quenching corrections are lying between 10-20$\%$ 
for most of evaluated
physical quantities. \\
Because of computing limitations, most numbers have been obtained 
in the quenched approximation. 
Theoretical estimates and some preliminary results in the (partially) 
unquenched case  are also available and are
used to estimate the corresponding systematic error of quenched results. \\
These calculations are usually performed with two light quarks in the 
fermion loops, at values of the light-quark  masses larger than the 
physical values and an extrapolation in these masses 
is required. Calculations are generally made at few values of 
the lattice spacing and thus contain discretization errors.
An estimate of quenching errors is obtained by comparing quenched
and unquenched  results  at similar  values of the lattice spacing. \\
Another important issue is related to the chiral extrapolation. In fact it
is difficult to simulate realistically light quarks, with their physical masses, 
and calculations are 
usually made for a set of (valence) quark masses, ranging from about $m_s$/2 to 2 $m_s$.
The results need then to be interpolated or extrapolated. Similar extrapolation
needs to be done, in partially quenched calculations, considering the range of sea quark
masses used. The problem arises since there are logarithmic dependences in physical 
quantities as the valence and/or the sea quark mass are extrapolated to their physical values
(divergences in some cases if masses vanish).
In practice different extrapolations can be performed if one considers or not these terms.
The JLQCD collaboration finds \cite{ref:chiral} that these different extrapolations 
tend to decrease the value of $f_{B_d}$ relative to $f_{B_s}$.  
At present a reasonable view \cite{ref:kron,ref:lellouch} is to allow a decrease of
$f_{B_d}$ by -10$\%$ and a negligible change in $f_{B_s}$.\\

For the present phenomenological analysis, the following values and errors have been used
\begin{eqnarray} 
f_{B_d} \sqrt{\hat B_{B_d}} = (223 \pm 33 \pm 12)~MeV    \\ \nonumber
\xi=\frac{ f_{B_s}\sqrt{\hat B_{B_s}}}{ f_{B_d}\sqrt{\hat B_{B_d}}} 
                            = 1.18 \pm 0.04 \pm 0.06 \\ \nonumber
\hat B_K                    = 0.86 \pm 0.06 \pm 0.14 
\end{eqnarray}

These estimates have to be considered as conservative, since they assume a 
maximal effect due to chiral extrapolation, reflected in the last error.
These last errors are taken as flat distributions.

A detailed description on how these values have been obtained can be found in 
\cite{ref:ckm1}. The CKM-LDG Group \cite{ref:ldg} is taking care of these averages.

\subsection{Determination of $\snb$ from CP asymmetry in  $J/\psi K^0$ decays.}
\label{sec:sin2beta}

BaBar and Belle collaborations have recently updated their measurements.
The world average is \cite{ref:hfag}:
\begin{equation}
         \snb = 0.739 \pm 0.048 
\label{eq:sin2beta_ave}
\end{equation}

All details concerning the analyses techniques are described in these proceeding by
the seminar corresponding to U. Mallik.

\section{Determination of the Unitarity Triangle parameters}
\label{sec:results}

In this section we give the results for the quantities defining the Unitarity
Triangle, assuming the validity of the Standard Model: $\rhobar$, $\etabar$, $\snb$, 
$\sna$ and $\gamma$ as well as for other quantities as $\dms$, $f_B$ and $\hat{B}_K$.
The inputs used are summarised in Table \ref{tab:inputs} (see Section \ref{sec:formulae}).\\
For more details and concerning latest results see \cite{ref:pageweb}.

\subsection{Fundamental test of the Standard Model in the fermion sector}

\begin{figure}[h]
\begin{center}
\includegraphics[width=14cm]{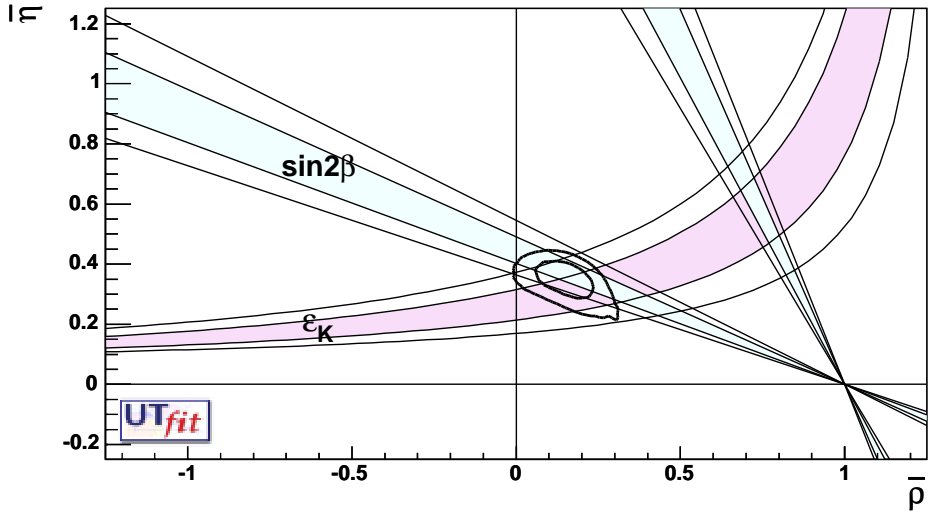}
\caption{The allowed regions for $\overline{\rho}$ and $\overline{\eta}$
(contours at 68\%, 95\%) as selected by the measurements of 
$\left | V_{ub} \right |/\left | V_{cb} \right |$, 
$\Delta {M_d}$, and by the limit on $\Delta {M_s}/\Delta {M_d} $ are compared with the 
bands (at 68\% and 95\% C.L.) from the measurements of CP violating quantities in the 
kaon ($\epsilonk$) and in the B ($\snb$) sectors.}
\label{fig:testcp}
\end{center}
\end{figure}

The most crucial test consists in the comparison between the region selected by the measurements
which are sensitive only to the sides of the Unitarity Triangle (semileptonic B decays and
${\rm B}^0-\overline{{\rm B}^0}$ oscillations) and the regions selected by the direct measurements of 
CP violation in the kaon ($\epsilonk$) or in the B ($\snb$) sectors. 
This test is shown in Figure \ref{fig:testcp}. 
It can be translated quantitatively through the comparison between the values of 
$\snb$ obtained from the measurement of the CP asymmetry in $J/\psi K^0$ 
decays and the one determined from ``sides`` measurements:
\begin{eqnarray}
\snb = & 0.685 \pm 0.047 ~[0.547-0.770]~{\rm at} ~95\%~ C.L.  & ~~\rm {sides~ only}     \nonumber \\
\snb = & 0.739 \pm 0.048 ~[0.681-0.787]~{\rm at} ~95\%~ C.L.  & \rm J/\psi K^0. 
\label{eq:sin2beta}
\end{eqnarray}

The spectacular agreement between these values illustrates the consistency of the Standard 
Model in describing CP violation phenomenon in terms of one single parameter $\etabar$.
It is also an important test of the OPE, HQET and LQCD theories which have been used to 
extract the CKM parameters.

It has to be noted that this test is significant provided the errors on $\snb$ 
from the two determinations are comparable.

%The 1-dim p.d.f and the results for the unitarity triangle parameters are given 
%in Table \ref{tab:1dim_sine} and Figures \ref{fig:1dim_sine}.
Corresponding results, for the unitarity triangle parameters, are given 
in Table \ref{tab:1dim_sine}.

\begin{table*}[h]
\begin{center}
\begin{tabular}{@{}llllll}
\hline
          Parameter           &     ~~~~~68$\%$            &  ~~~~~95$\%$    &    ~~~~~99$\%$   \\ \hline
~~~~~$\overline {\eta}$       & 0.346 $^{+0.039}_{-0.043}$ & (0.227-0.416)   & (0.099-0.437)    \\
~~~~~$\overline {\rho}$       & 0.153 $\pm$ 0.061          & (0.030-0.325)   & (-0.012-0.368)   \\
         ~$\snb$              & 0.685 $\pm$ 0.047          & (0.547-0.770)   & (0.280-0.806)    \\
         ~$\sna$              & -0.01 $\pm$ 0.35           & (-0.85-0.83)    &    ~~~~~-        \\
~~~$\gamma[^{\circ}$]         &  65.3 $\pm$ 9.5            & ( 38.9-84.8)    &   (15.8-90.0)    \\
\hline
\end{tabular} 
\end{center}
\caption {\it Values and probability ranges for the unitarity triangle parameters 
when the constraints from $\epsilonk$ and $\snb$ measurements are not used.}
\label{tab:1dim_sine} 
\end{table*}

\subsection{Determination of the Unitarity Triangle parameters : \boldmath${\overline{\eta}}$,
\boldmath${\overline{\rho}}$, \boldmath${\snb}$, \boldmath${\sna}$, \boldmath${\gamma}$}
\label{sec:allconstr}

By using all five available constraints ($\left | V_{ub} \right |/\left | V_{cb} \right |$, 
$\Delta {m_d}$, $\Delta {m_s}/\Delta {m_d} $, $\epsilonk$ and $\snb$), 
the results given in Table \ref{tab:1dim} are obtained.

\begin{table*}[h]
\begin{center}
\begin{tabular}{@{}llllll}
\hline
    Parameter  &     ~~~~~68$\%$   &      ~~~~~95$\%$     &    ~~~~~99$\%$      \\ 
\hline 
~~~~~$\overline {\eta}$  & 0.342  $\pm$ 0.026          & (0.291-0.396)   & (0.272-0.415) \\
~~~~~$\overline {\rho}$  & 0.174  $\pm$ 0.047          & (0.076-0.260)   & (0.045-0.293) \\
          ~$\snb$        & 0.697  $\pm$ 0.035          & (0.637-0.761)   & (0.619-0.781) \\
          ~$\sna$        & -0.15  $\pm$ 0.25           & (-0.62-0.34)    & (-0.73-0.50)  \\
~~~~$\gamma[^{\circ}$]   & 61.1   $\pm$ 7.8            & (48.6-76.0)     & (43.2-82.9)   \\
\hline
\end{tabular} 
\end{center}
\caption {\it Values and probability ranges for the unitarity triangle parameters obtained by using all 
five available constraints: $\left | V_{ub} \right |/\left | V_{cb} \right |$, 
$\Delta {m_d}$, $\Delta {m_s}/\Delta {m_d} $, $\epsilonk$ and $\snb$.}
\label{tab:1dim} 
\end{table*}

\begin{figure}[h]
\vspace*{-0.3cm}
\begin{center}
\includegraphics[width=14cm]{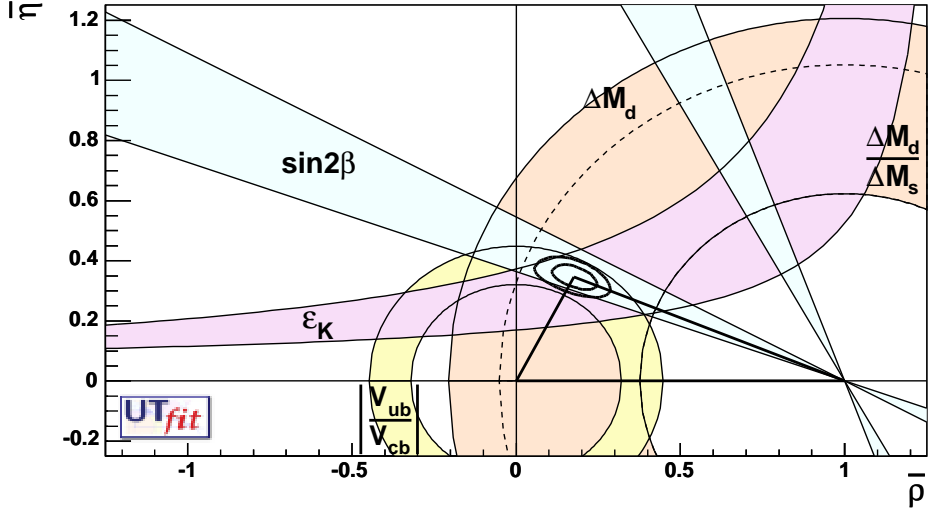}
\caption{ {Allowed regions for $\rhobar$ and $\etabar$ using the parameters listed in Table~\ref{tab:inputs}. The closed contours at 68\% and 95\% probability are shown. The full lines correspond to  95\% probability regions for the constraints, given by the measurements of $\left | V_{ub} \right |/\left | V_{cb} \right |$, $\epsilonk$, $\Delta m_d$, $\Delta m_s$ and $\snb$. 
%from the measurement of the CP asymmetry in the $J/\psi K^0$ decays.
The dotted curve corresponds to the 95\% upper limit obtained from the experimental study of $\Bs-\Bsb$ oscillations.}}
\label{fig:rhoeta}
\end{center}
\vspace*{-0.5cm}
\end{figure}

\begin{figure}[h]
\begin{center}
{\includegraphics[height=4.5cm]{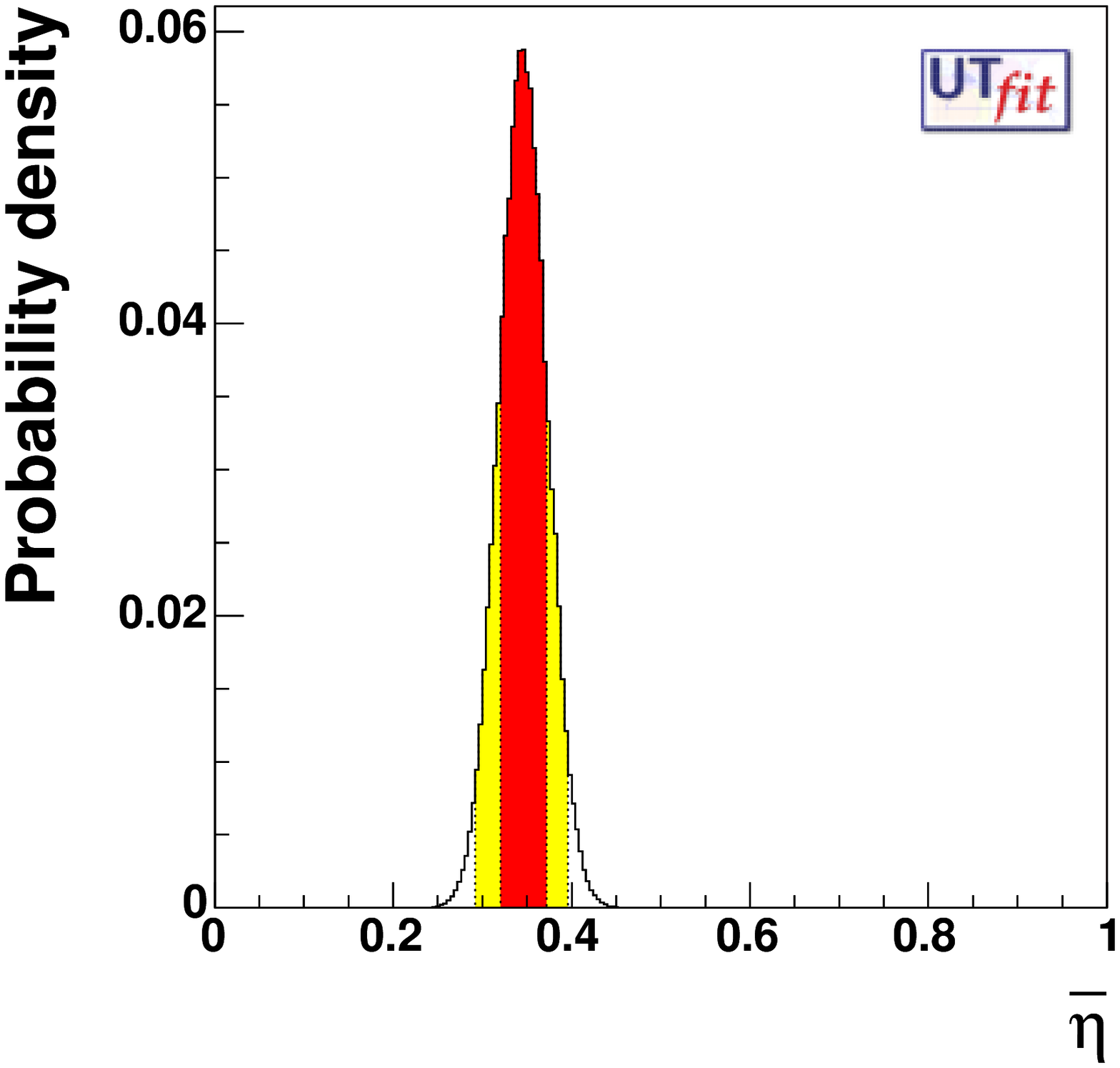}}
{\includegraphics[height=4.5cm]{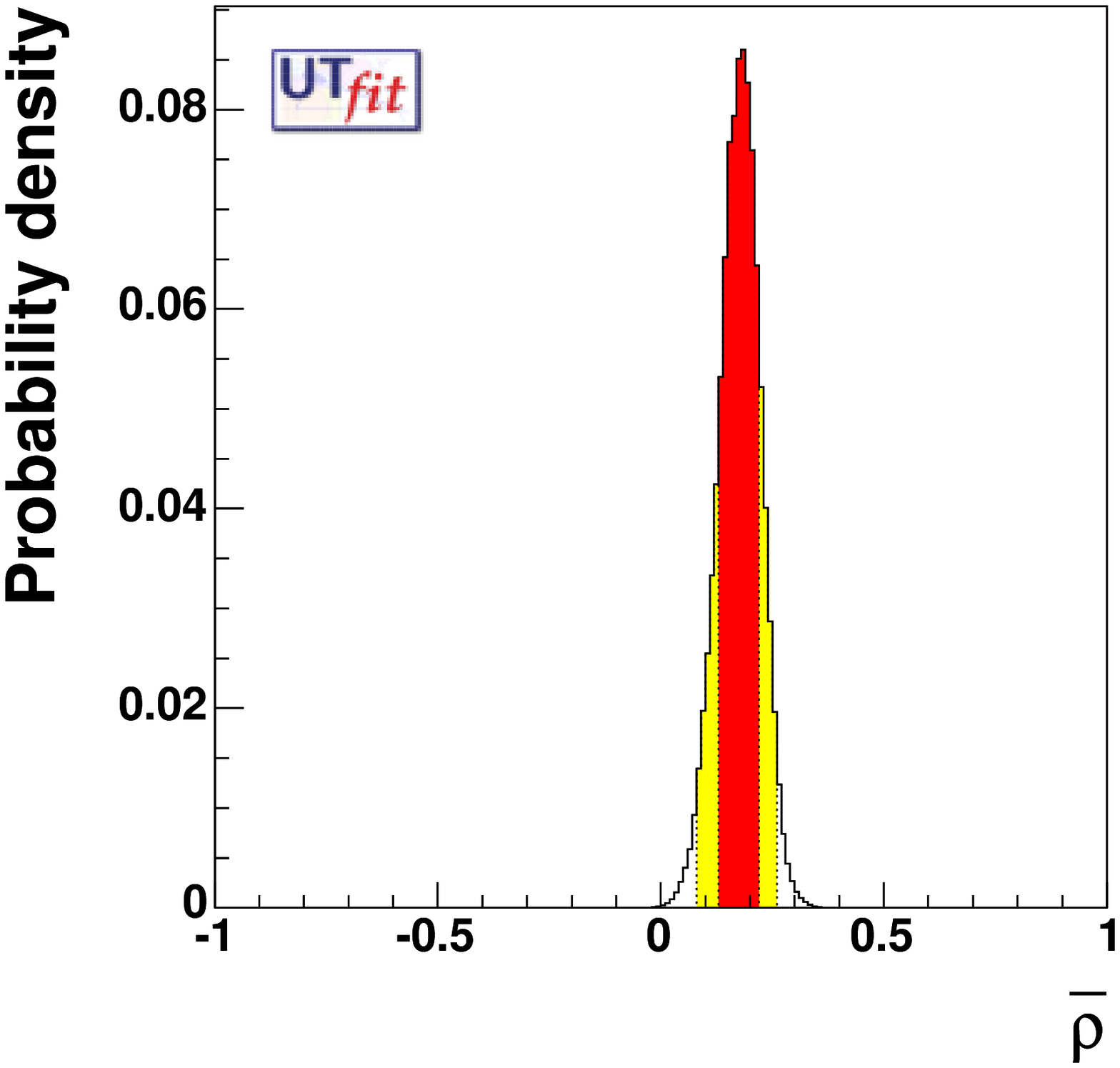}} \\
{\includegraphics[height=4.5cm]{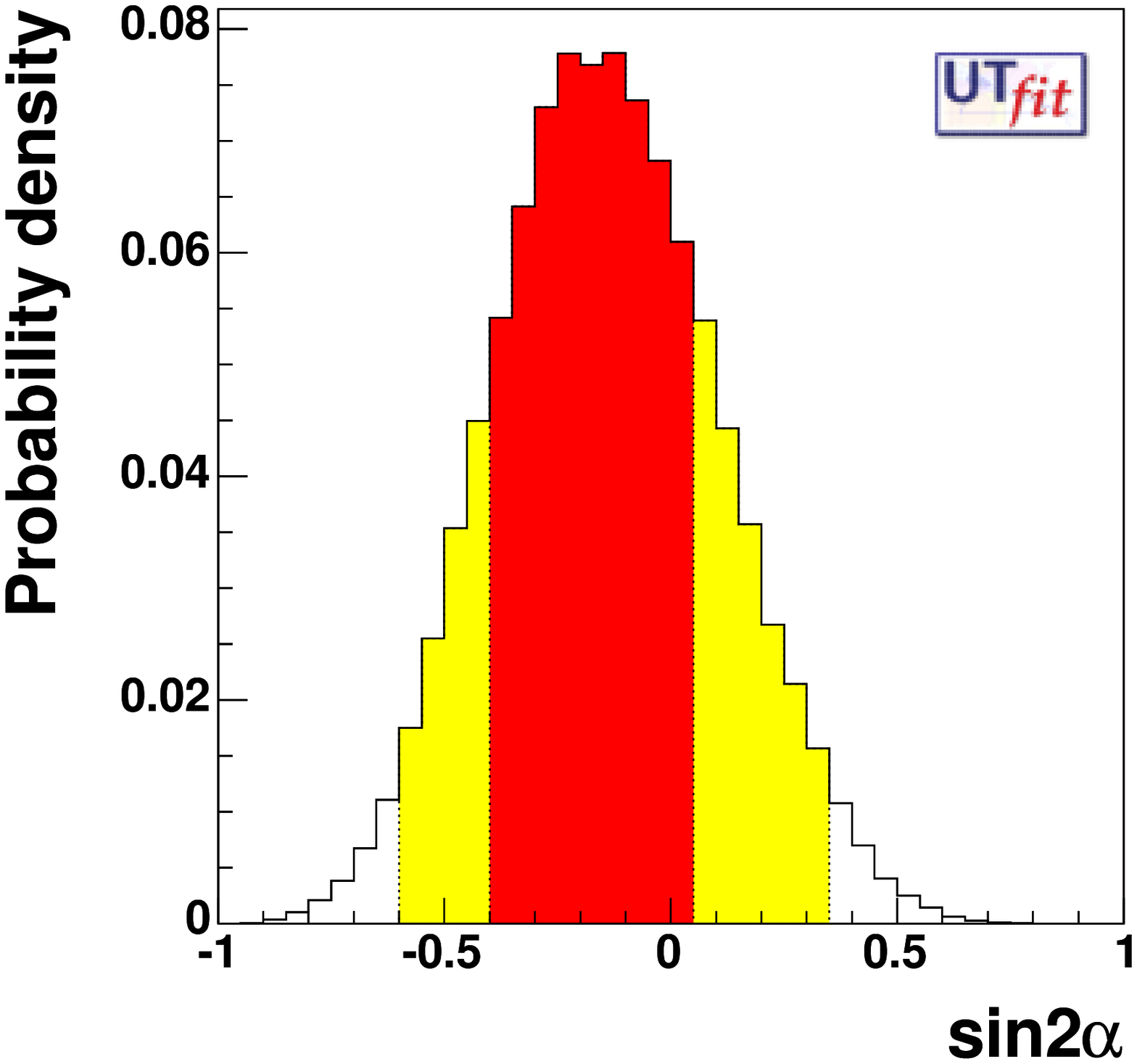}}
{\includegraphics[height=4.5cm]{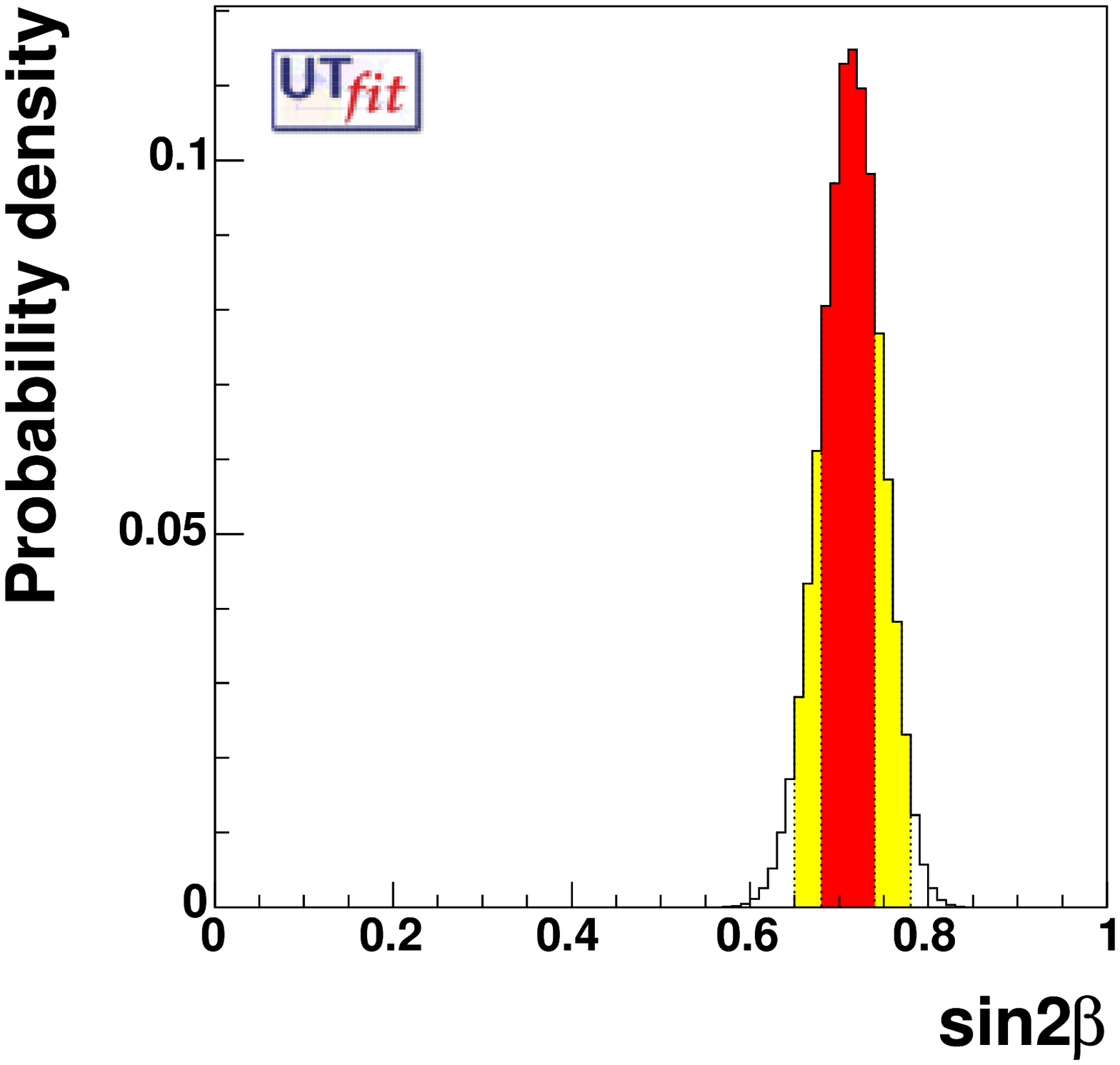}}
{\includegraphics[height=4.5cm]{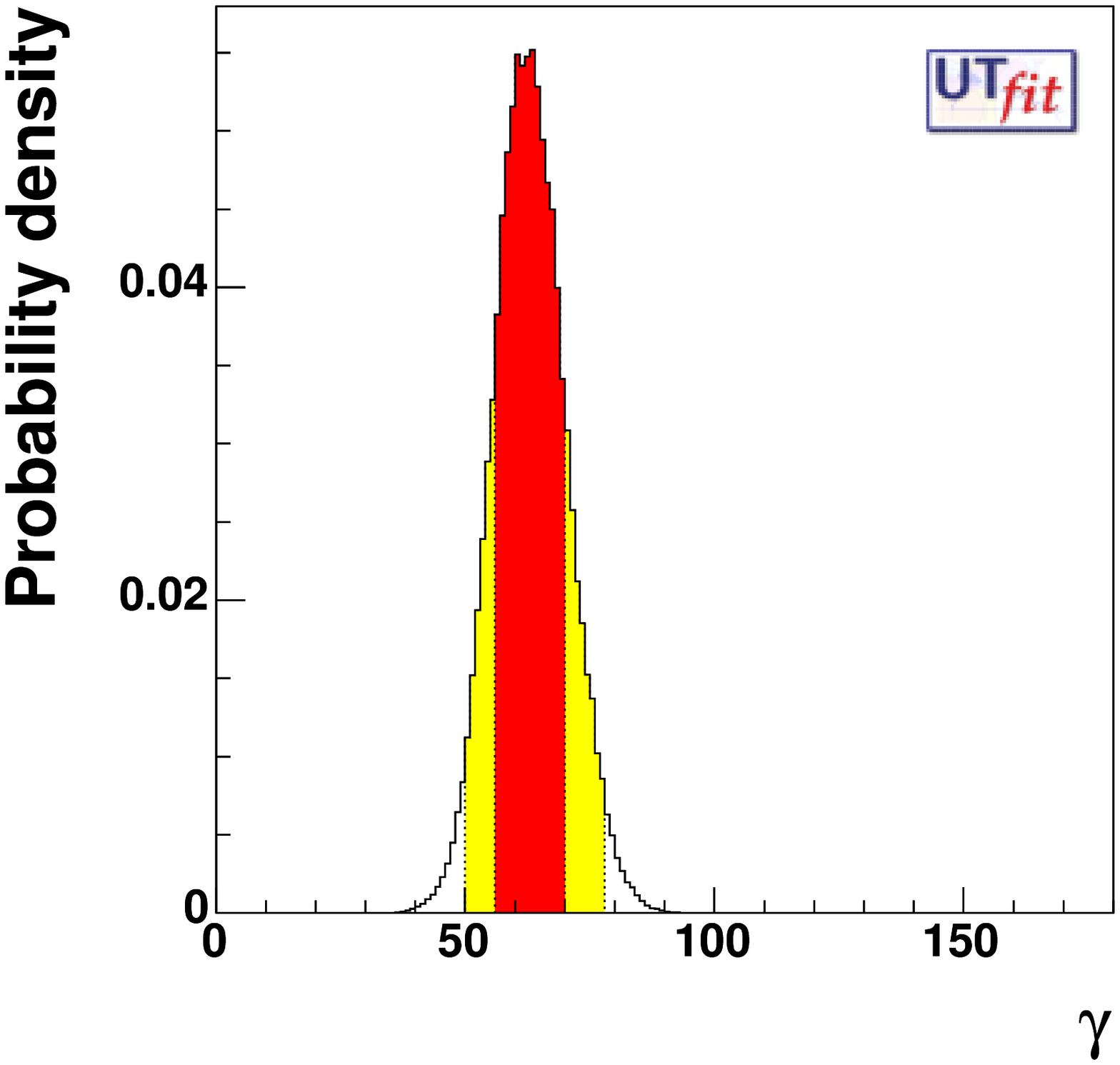}}
\caption{{From top left to bottom, the p.d.f. for $\etabar$, $\rhobar$, $\sna$, 
$\snb$ and $\gamma$. The red (darker) and the yellow (clearer)  zones correspond 
respectively to 68\% and 95\% of the normalised area. All available constraints have been used. }}
\label{fig:1dim}
\end{center}
\end{figure}

Figures \ref{fig:rhoeta} and \ref{fig:1dim} show, respectively, the corresponding selected region in the ($\rhobar,~\etabar)$ plane and the p.d.f. for the Unitarity Triangle parameters.

\subsubsection{Indirect versus direct determination of the Unitarity Triangle angles}

The value of $\snb$ was predicted, before 
its first direct measurement was obtained, by using all other available constraints, 
($\left | V_{ub} \right |/\left | V_{cb} \right |$, $\epsilonk$, $\Delta m_d$ and
$\Delta m_s$). The ``indirect''
\footnote{in the following, for simplicity, we will note as ``direct''(``indirect''), 
the determination of $\snb$ from $A_{CP}(J/\psi K^0)$ (other constraints).}
determination has improved regularly over the years. 
Figure \ref{fig:storiasin2beta} shows this evolution for the ``indirect''
determination of sin2$\beta$ which is compared with the recent determinations of 
$\snb$ from direct measurements.

This test should be repeated with other constraints. 

The values for $\gamma$ and sin2$\alpha$ given in Table \ref{tab:1dim} has to be taken as
predictions for future measurements. A strong message is given for instance for the angle
$\gamma$. The indirect determination of the angle $\gamma$ is known with an accuracy of about 
10\%. It has to be stressed that, with present measurements, the probability that $\gamma$ 
is greater than 90$^{\circ}$ is only 0.003\%.

\subsection{Determination of other important quantities}

In previous sections we have seen that we can get distributions 
for the different unitarity triangle parameters and how it can be instructive
to remove from the fitting procedure the external information on the value of one 
(or more) of the constraints.

In this section we get the distributions for the values of other quantities, 
entering into the Standard Model expressions for the constraints, such as the 
hadronic parameters, or of a constraint as $\dms$. 
In case of the hadronic parameters, for instance, it is instructive to remove, 
from the fit, in turn, their external information.
The idea is to compare the uncertainty on a given quantity, determined in this way,
to its present experimental or theoretical error. This comparison allows to quantify 
the importance of present determinations of the different quantities to define the limits
of the allowed region for the unitarity triangle parameters.

\subsubsection{The expected distribution for \boldmath$\dms$}

\begin{figure}
\begin{center}
{\includegraphics[height=9cm]{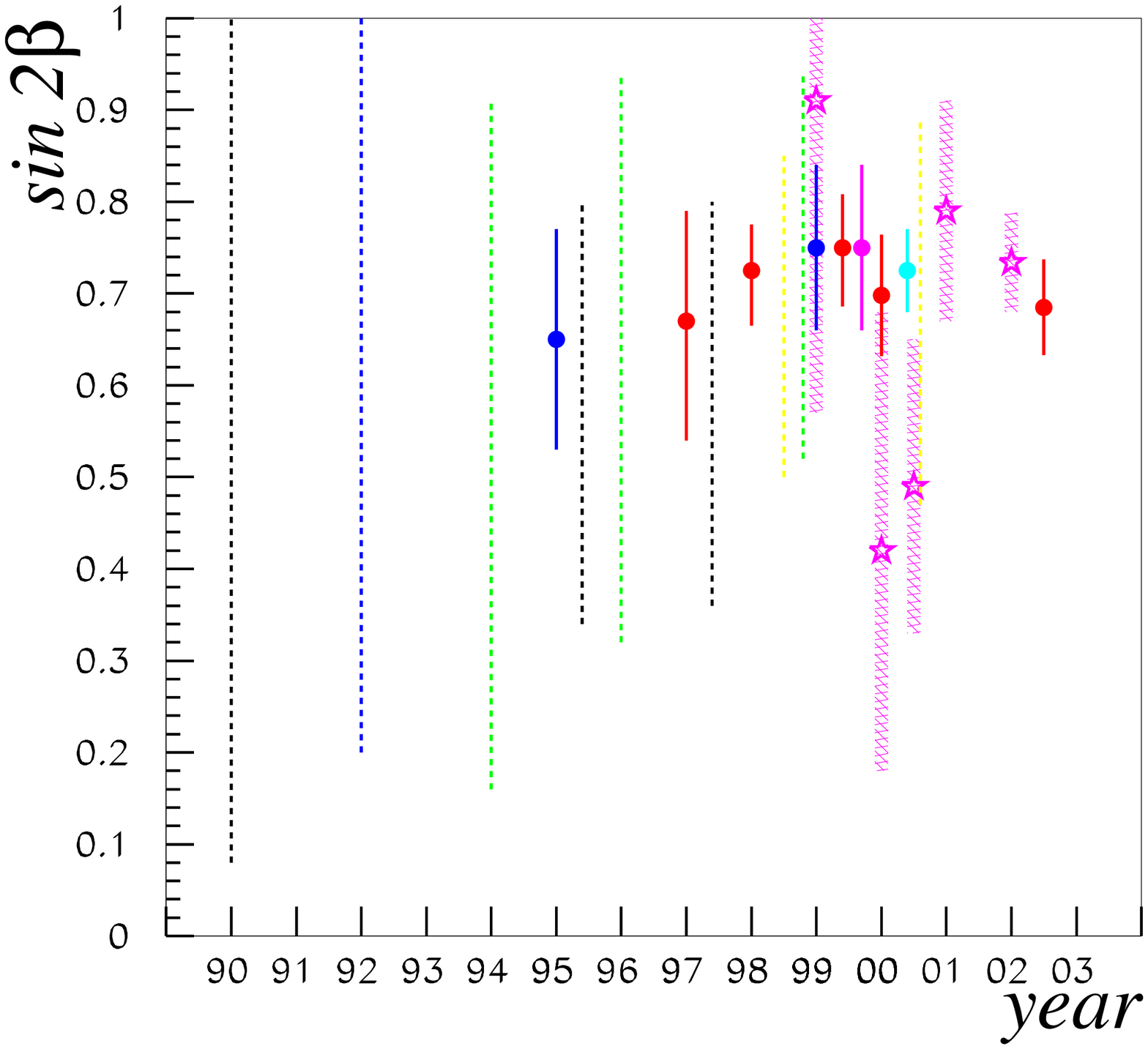}}
\caption{{Evolution of the ``indirect'' determination of $\snb$ over the years.
From left to right, they correspond to the following papers \cite{ref:allrhoeta}: DDGN90, LMMR92, AL94, CFMRS95, BBL95, AL96, 
PPRS97, BF97, BPS98, PS98, AL99, CFGLM99, CPRS99, M99, CDFLMPRS00, B.et.al.00, HLLL00 and the value presented in this document. 
The dotted lines correspond to the 95$\%$ C.L. regions (the only information given in those papers). The larger bands (from year '99)
correspond to values of $\snb$ from direct measurements ($\pm 1 \sigma$).}}
\label{fig:storiasin2beta}
\end{center}
%\end{figure}
%\begin{figure}
\begin{center}
{\includegraphics[height=6cm]{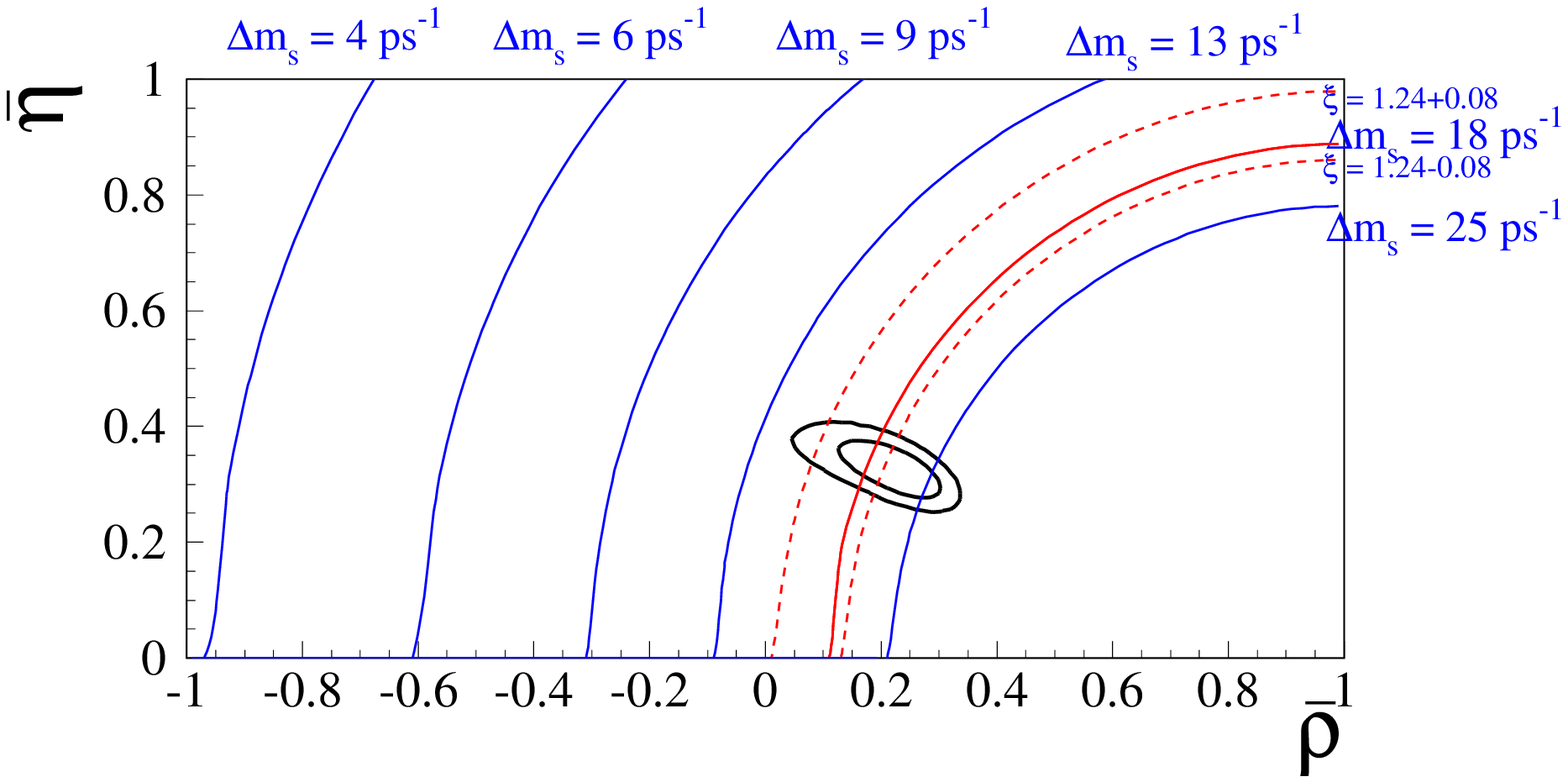}}
\caption{{The allowed regions for $\rhobar$ and $\etabar$ using the constraints given by the measurements of $\epsilonk$, $\left | V_{ub} \right |/\left | V_{cb} \right |$, $\Delta m_d$ and $\snb$ at 68\% and 95\% probability are shown by the closed contour lines.The different continuous circles correspond to fixed values of $\dms$. Dashed circles, drawn on each side of the curve corresponding to $\dms = 18.0~ps^{-1}$, indicate the effect of a variation by $\pm 0.08$ on $\xi$.}}
\label{fig:dmsdemo}
\end{center}
\end{figure}

Figure \ref{fig:dmsdemo} shows the allowed region for $\rhobar$ and $\etabar$ obtained with all the constraints and how the constraint coming from the study of $\Bs$--$\Bsb$ mixing acts in this plane. 
A lower limit at 95$\%$ C.L. on $\dms$ will exclude, at that degree
of confidence, the $\rhobar$-$\etabar$ region situated on the left of the corresponding curve.

\begin{figure}[h]
\begin{center}
%\begin{tabular}{cc}
\includegraphics[height=8.3cm]{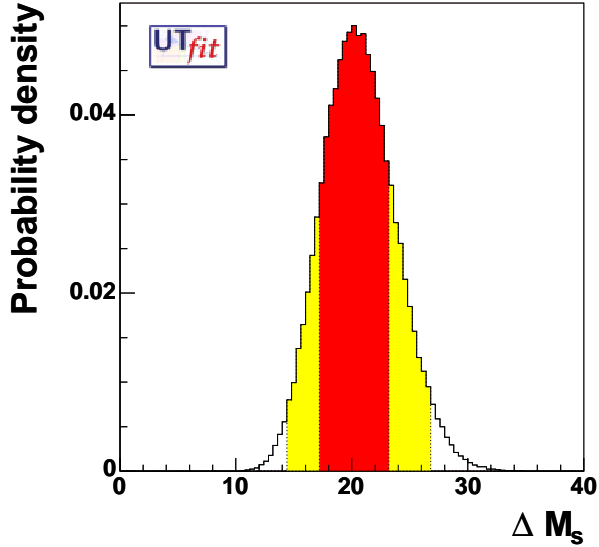} 
%\end{tabular}
\end{center}
\caption{ {$\Delta m_s$ probability distributions. The information from 
${\rm B}^0_s-\overline{{\rm B}^0_s}$ oscillations is not used.}}
\label{fig:dms}
\end{figure}

It is also possible to extract the probability distribution for $\dms$, which is shown in 
Figure \ref{fig:dms}. Corresponding results are given in Table \ref{tab:dmsresults}.
Present analyses at LEP/SLD, with a sensitivity at 19.2~ps$^{-1}$ are situated in a 
high probability region for a positive signal (as the ``signal bump'' appearing 
around 17.5 ps$^{-1}$). 

Accurate measurements of $\dms$ are thus expected soon from the TeVatron.

\begin{table*}[h]
\begin{center}
\begin{tabular}{@{}lllll}
\hline
      ~~~~~~~~~~~~~~Parameter                              & ~~68$\%$             & ~~~~95$\%$   & ~~~~99$\%$  \\ 
\hline
~~~~$\dms$(~\rm{ including ~$\dms$~}) [{\rm ps}$^{-1}$]    &    18.4$\pm 1.6$     & (15.4-21.2)  & (14.6-25.4) \\
$\dms$(~\rm{without~including $\dms$~}) [{\rm ps}$^{-1}$]  &      20.2 $\pm$ 3.0  & (14.4-26.8)  & (13.2-29.2)  \\
\hline
\end{tabular} 
\end{center}
\caption {\it $\dms$ central values and ranges corresponding to defined
levels of probability, obtained when including or not the information from the experimental amplitude spectrum ${\cal A}(\dms)$.}
\label{tab:dmsresults} 
\end{table*}

%%%%%%%%%%%%%%%%%%%%%%%%%%%%%%%%%%%%%%%%%%%%%%%%%%%%%%%%%%%
\subsubsection{Determination of \boldmath$\fbdsqbd$ and $\hat{B}_K$}
The value of  $\fbdsqbd$ can be obtained by removing 
\footnote{Technically we assume a uniform distribution in a range 
which is much larger than the possible values taken by the parameters.}
the theoretical constraint coming from this parameter in the expression of the $\Bd$--$\Bdb$ oscillation frequency $\dmd$.
The main conclusion of this study is that $\fbdsqbd$ is measured with an accuracy
which is better than the current evaluation from lattice QCD, given
in Section~\ref{sec:parth}. Results are summarized in Table \ref{tab:nonptsumm}.
This shows that the present CKM fit, when all the available constraints are used, is, 
in practice, weakly dependent on the  exact value assumed for the uncertainty on $\fbdsqbd$.

\begin{table}[h]
\begin{center}
\begin{tabular}{|c|c|c|c|}
\hline
    Parameter         & 68$\%$                  &    95$\%$    &  99$\%$        \\ \hline
$\fbdsqbd$(MeV)       & 217 $\pm$ 12            &  (196-245)   & (190-258)      \\
$\hat{B}_K$            & 0.69$^{+0.13}_{-0.08}$  &  (0.53-0.96) & (0.49-1.09)   \\
\hline
\end{tabular} 
\caption {\it Values and probability ranges for the non perturbative QCD parameters, if the external
information (input) coming from the theoretical calculation of these parameters is not used in the CKM fits}
\label{tab:nonptsumm} 
\end{center}
\end{table}

The parameter $\hat{B}_K$ can be also determined. Results are also summarized in Table \ref{tab:nonptsumm}.
They indicates that values of $\hat{B}_K$ smaller than 0.5 (0.3) correspond to 0.6$\%$
(5 $\times$ 10$^{-6}$) probability while
large values of $\hat{B}_K$ are compatible with the other constraints over a large domain.
The present estimate of $\hat{B}_K$, from lattice QCD, with a 15$\%$ relative error 
(Table \ref{tab:inputs}) has thus a large impact in the present analysis.

\subsection{Evolution on the precision on $\rhobar$ and $\etabar$ over the last 15 years}

\begin{figure}[htbp!]
\begin{center}
\includegraphics[height=4cm]{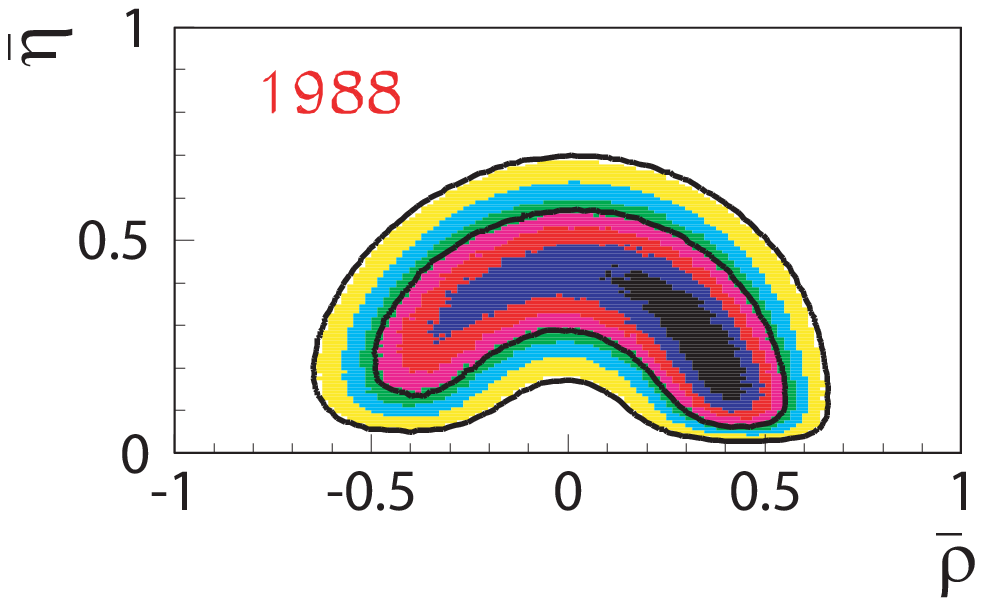}  \\
\includegraphics[height=4cm]{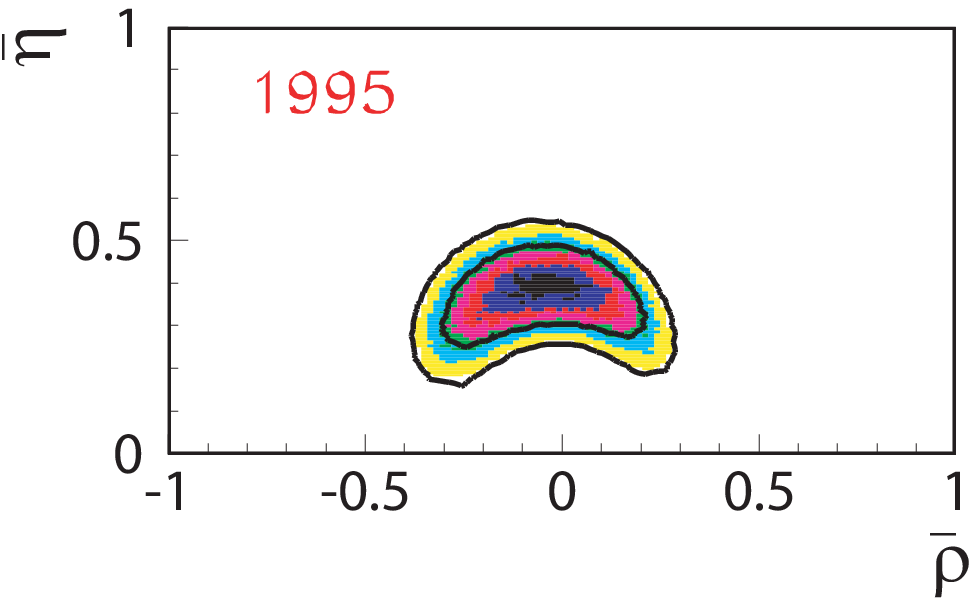}   \\
\includegraphics[height=4cm]{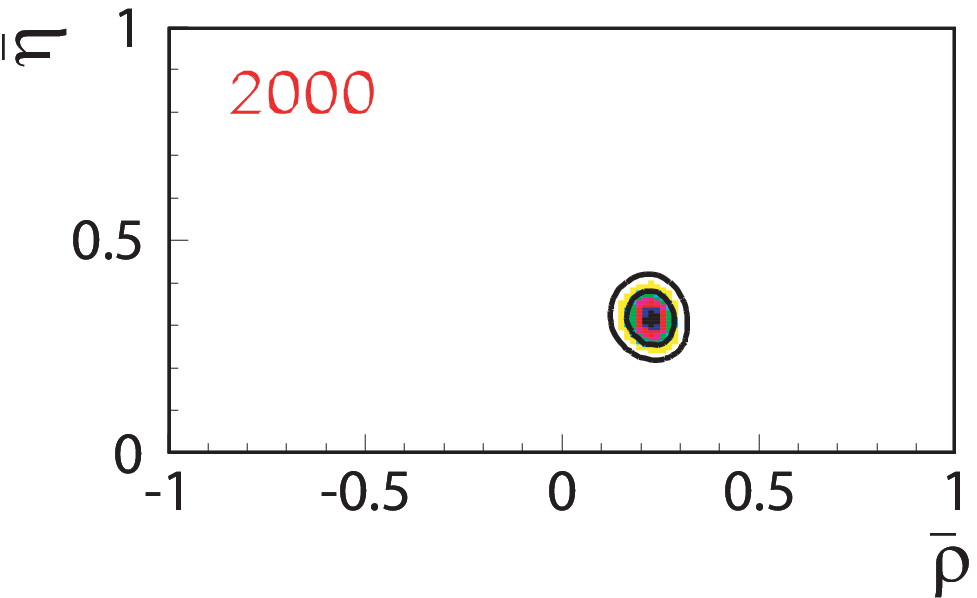} \\
\includegraphics[height=4cm]{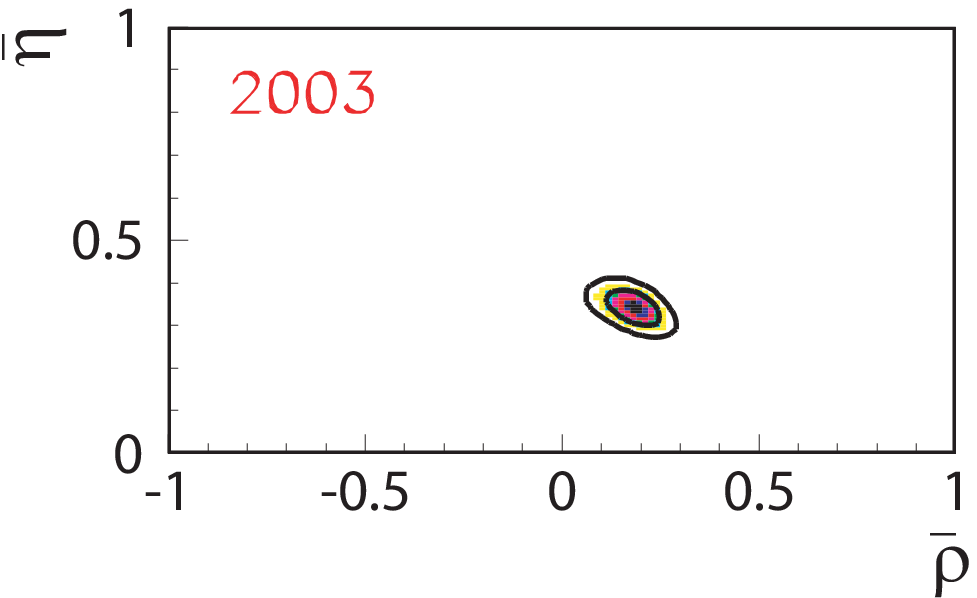}
\caption{{Evolution during the last 15 years of the allowed regions for 
$\overline{\rho}$ and $\overline{\eta}$ (contours at 68\% and 95\% probability are indicated). 
The very last results (updated till Winter 2004) are shown in Figure 
\ref{fig:rhoeta} and in Table \ref{tab:1dim}. }}
\label{fig:storia}
\end{center}
\end{figure}

The evolution of our knowledge concerning the allowed region in the 
($\overline{\rho}$, $\overline{\eta}$) plane is shown in Figure \ref{fig:storia}.

\newpage

The reduction of the size of these regions, from years 1995 to 2000, is essentially due 
to the measurements of the sides of the Unitarity Triangle and to the progress in
OPE, HQET and lattice QCD theoretical parameters determinations. 
The additional reduction, from
years 2000 to 2003, which mainly concerns $\etabar$, is essentially driven by 
the measurement of $\snb$ through the CP violation asymmetry in J$/\psi$ K$^0$
decays.

\subsection{Dulcis in fundo : the new-comers}
The huge statistics collected the B-factories allow the measurements of new CP-violating quantities.
Direct measurements of $\gamma$, $\sin(2\beta + \gamma)$ and sin2$\alpha$ are now available :
\begin{itemize}
\item determination of sin2$\alpha$ using charmless $\pi \pi$ events,
\item determination of $\sin(2\beta + \gamma)$ using D$^{(*)} \pi$ events,
\item determination of $\gamma$ using DK events. 
\end{itemize}
We do not enter in any details for these analyses which are described in U. Mallik lectures. 

\begin{figure}[htb!]
\begin{center}
\includegraphics[height=5cm]{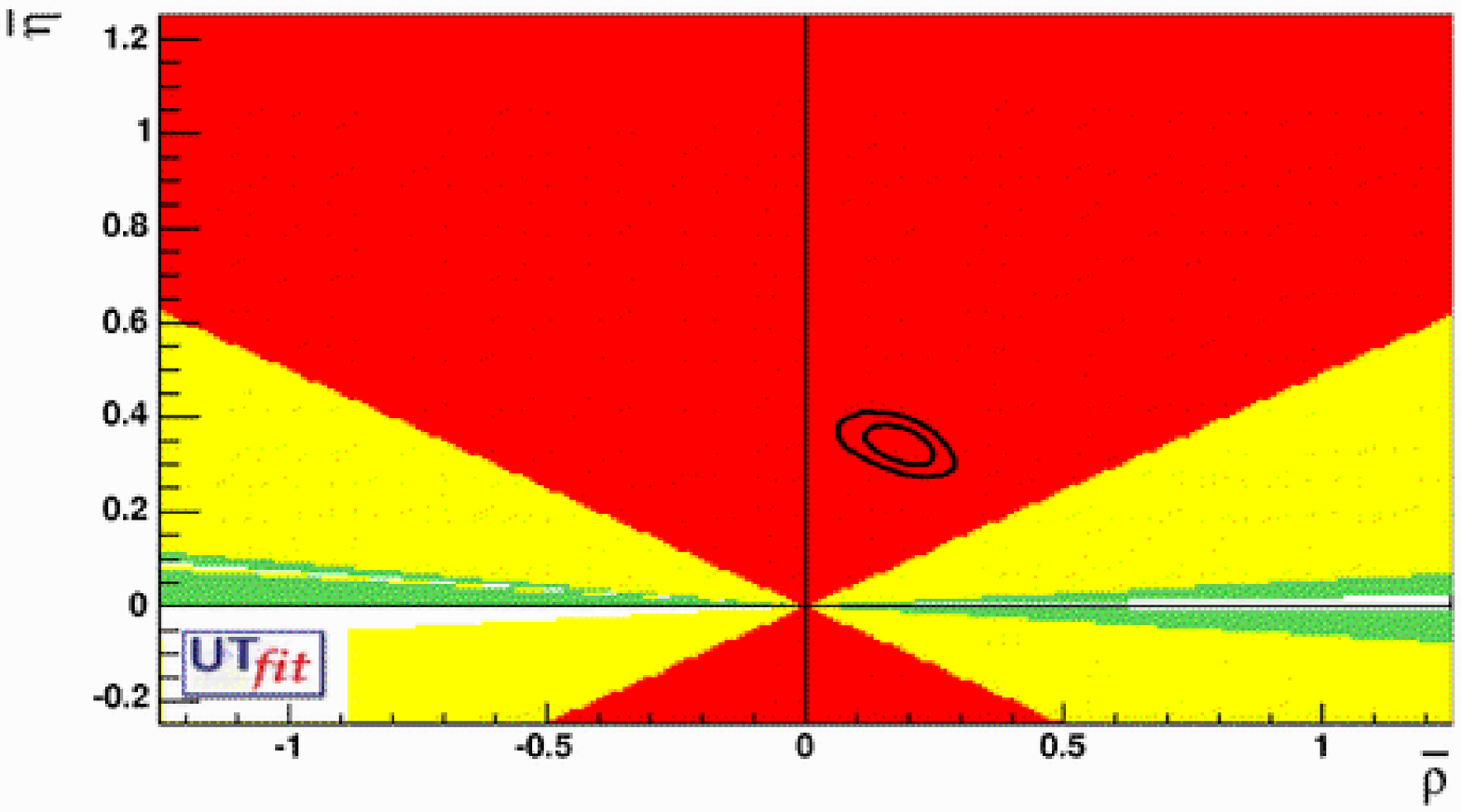}  \\
\includegraphics[height=5cm]{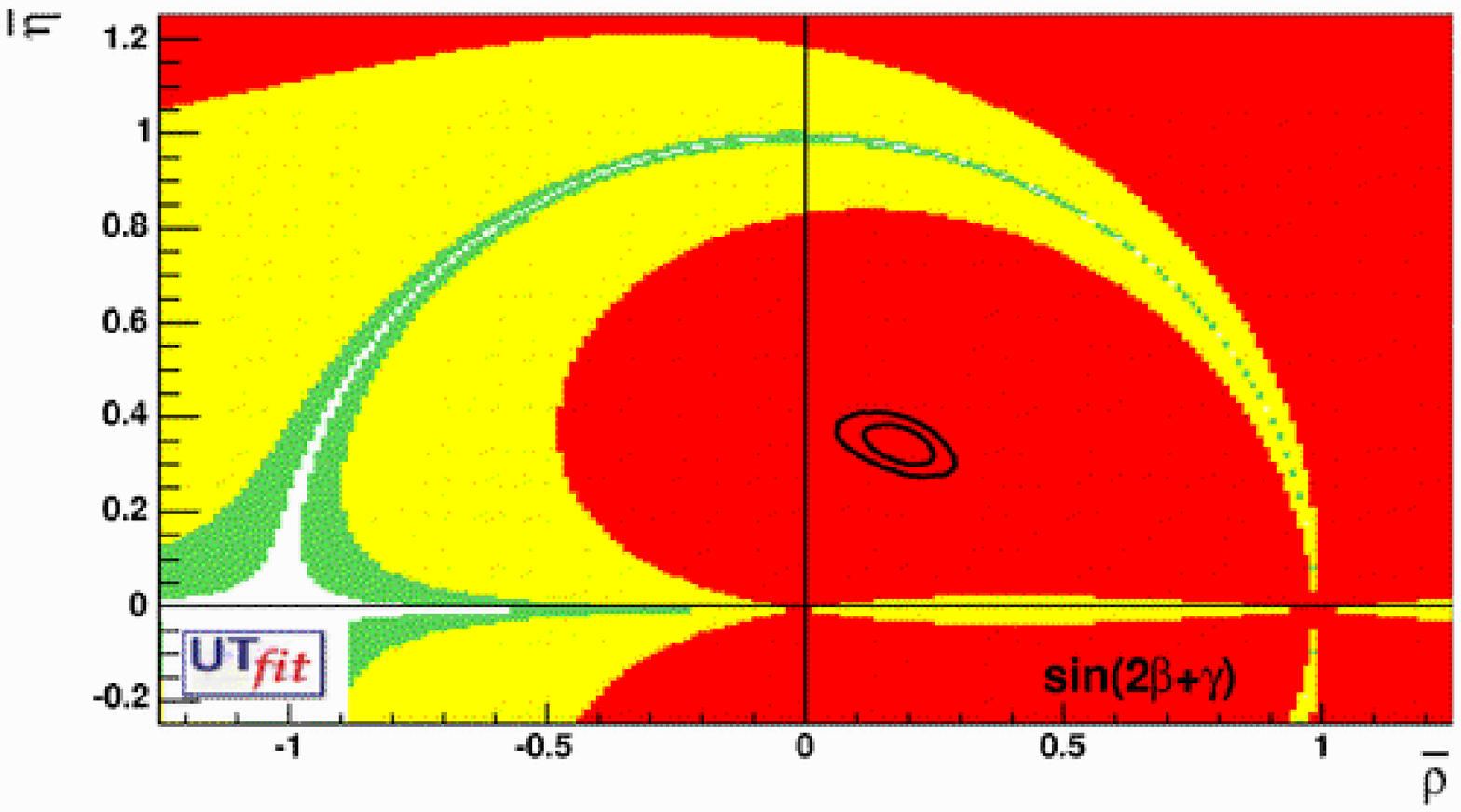}   \\
\includegraphics[height=5cm]{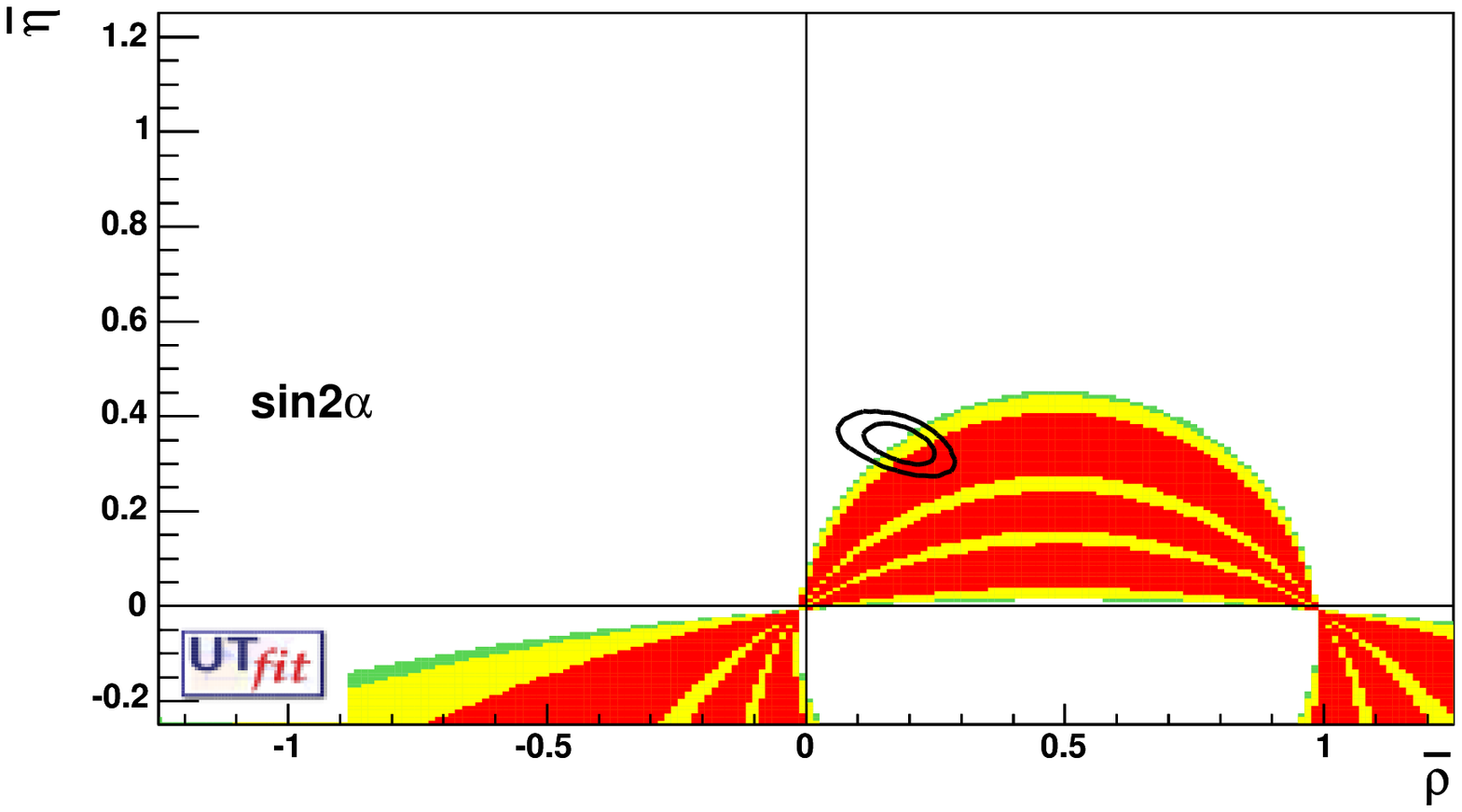} 
\caption{{ From top  to bottom, the allowed region on the $\overline{\rho}-\overline{\eta}$ plane
as selected by the direct measurement of $\gamma$, $\sin(2\beta + \gamma)$ and sin2$\alpha$ 
The red (darker), the yellow (clear) and green (clearer) zones correspond 
respectively to 68\%, 95\% and 99\% of the normalised area. Contours at 68\% and 95\% probability 
selected using all the other available constraints are also shown.}}
\label{fig:nuovi}
\end{center}
\end{figure}

The Figure \ref{fig:nuovi} shows the impact of these new measurements to provide additional constraints in 
the $\overline{\rho}-\overline{\eta}$ plane. More details are given in \cite{ref:pageweb}.

These plots show the potentialities of B-factories, considering that additional measurements 
will be available, in a near future (about 2 years), with more than four times the statistics.

\section{Conclusions}

Flavour physics in the quark sector is entered in its mature age. 
Many and interesting results have been produced during the last 15 years.
Traditional main players (LEP/SLD/CLEO) delivered results until this year, while B 
factories are moving B studies into the era of precision physics.

Many quantities have already been measured with a good precision. 
$|V_{cb}|$ is today known with a relative precision better than 2$\%$. In this case, not only, 
the decay width has been measured, but also some of the non-perturbative QCD parameters entering 
into its theoretical expressions. It is a great experimental
achievement and a success of the theory description of the non-perturbative QCD phenomena in the framework 
of the OPE. Many different methods, more and more reliable, are now available 
for determining the CKM element $|V_{ub}|$. The relative precision, today, is of 
about 10$\%$ and will be certainly improved in a near future at B-factories. 
The time dependence behaviour of $B^0-\bar{B^0}$ oscillations has been 
studied and precisely measured in the $B_d^0$ sector. The oscillation frequency $\Delta m_d$ is 
known with a precision of about 1$\%$. $B_s^0-\bar{B_s^0}$ oscillations have not been measured sofar,
but this search has pushed the experimental limit on the oscillation frequency $\Delta m_s$ 
well beyond any initial prediction. Today we know that $B_s^0$ oscillate at least 30 
times faster than $B_d^0$ mesons. The frequency of $B_s^0-\bar{B_s^0}$ oscillations should be soon 
measured at the TeVatron. 
Nevertheless, the impact of the actual limit on $\Delta m_s$ for the determination of the unitarity 
triangle parameters is crucial.

Many B decay branching fractions and relative CP asymmetries have been measured at B-factories. 
The outstanding result is the determination of sin 2$\beta$ from B hadron decays into charmonium-$K^0$
final states.
On the other hand many other exclusive hadronic rare B decays have been measured and constitute 
a gold mine for weak and hadronic physics, allowing to already extract different combinations
of the unitarity triangle angles. 

The unitarity triangle parameters are today known with a good precision. 
A crucial test has been already done: the comparison between the unitarity triangle 
parameters, as determined with quantities sensitive to the 
sides of the triangle (semileptonic B decays and oscillations), and the measurements of  
CP violation in the kaon ($\epsilon_K$) and in the B (sin2$\beta$) sectors. The agreement is ``unfortunately'' 
excellent. The Standard Model is ``Standardissimo'': it is also working in the flavour sector. 
This agreement is also an important test of the OPE, HQET and LQCD theories which have been 
used to extract the CKM parameters.

The good news is that all these tests are at best at about 10$\%$ level. 
The current and the next facilities can surely push these tests to a 1$\%$ accuracy. 
It is important to note that charm physics can play an important role in this respect 
(providing a laboratory for LQCD) and the Charm-factory (CLEO-C) 
will play a central role for these issues.

\section{Acknowledgements}
I would like to thank the organizers, and especially Gerard Smadja, of the Carg\`ese School 
for the invitation and for having set up a very interesting school in a nice atmosphere.

I really would like to thank Vittorio Lubicz, Patrick Roudeau, Marie-H\'el\`ene Schune and 
Luca Silvestrini for the enlighting discussions and suggestions both in preparing these 
lectures and the document. Many thanks to Marcella Bona, Maurizio Pierini and Fabrizio Parodi, 
the ``UT fitters'' for providing me all the material for the determination of the CKM parameters.\\
Merci \`a Catherine Bourge pour son aide.
%%%%%%%%%%%%%%%%%%%%%%%%%%%%%%%%%%%%%%%%%%%%%%%%%%%%%%%%%%%

\end{document}